\documentclass{aa}  

\usepackage{graphicx}
\usepackage{natbib,twoopt}
\usepackage{placeins}
\usepackage{ulem}  
\usepackage{txfonts}
\usepackage{booktabs, multirow}
\usepackage{orcidlink}
\usepackage{hyperref}
\hypersetup{colorlinks=true,linkcolor=[rgb]{1.,0.2,0.2},citecolor=[rgb]{0.1,0.4,1.},filecolor=[rgb]{0.7,0.2,0.2},urlcolor=[rgb]{0.7,0.2,0.2}}

\def\gsim{\mathrel{\rlap{\raise0.7pt\hbox{$>$}}{\lower 0.8ex\hbox{$\sim$}}}}
\def\lsim{\mathrel{\rlap{\lower3.0pt\hbox{$\sim$}}
\raise0.7pt\hbox{$<$}}}

\def\re{\mbox{$r_{\rm 1/2}$}}
\def\ms{\mbox{$M_\star$}}
\def\msre{\mbox{$M_\star(\re)$}}

\def\mdre{\mbox{$M_{\rm DM}(\re)$}}
\def\mtotre{\mbox{$M_{\rm tot}(\re)$}}
\def\mdyre{\mbox{$M_{\rm dyn}(\re)$}}
\def\sig{\mbox{$\sigma$}}
\def\sigre{\mbox{$\sigma_{\rm e}$}}

\def\mla{{\sc Mela}}
\def\mlaall{{\sc Mela}\_{\sc all}}
\def\mlaetg{{\sc Mela}\_{n\sc etg}}
\def\mladw{{\sc Mela}\_{\sc dw}}
\def\mlaltg{{\sc Mela}\_{\sc ltg}}

\begin{document}

\title{Total and dark mass from observations of galaxy centers with machine learning}
\titlerunning{Total and dark galaxy masses with ML}
\authorrunning{Wu et al.}
\author{Sirui~Wu\inst{\ref{inst1}, \ref{inst2}}\orcidlink{0009-0003-4675-3622}
\and 
Nicola~R.~Napolitano\inst{\ref{inst1}, \ref{inst2}, \ref{inst3}}\thanks{E-mail: napolitano@mail.sysu.edu.cn}\orcidlink{0000-0003-0911-8884}
\and
Crescenzo~Tortora\inst{\ref{inst4}}\orcidlink{0000-0001-7958-6531}
\and
Rodrigo~von~Marttens\inst{\ref{inst5}, \ref{inst9}}\orcidlink{0000-0003-3954-5756}
\and
Luciano~Casarini\inst{\ref{inst6}}\orcidlink{0000-0002-0869-9405}
\and
Rui~Li\inst{\ref{inst7}, \ref{inst8}}\orcidlink{0000-0002-3490-4089}
\and
Weipeng~Lin\inst{\ref{inst1}, \ref{inst2}}\orcidlink{0000-0003-2204-2474}
}
\institute{
School of Physics and Astronomy, Sun Yat-sen University, Zhuhai Campus, 2 Daxue Road, Tangjia, Zhuhai, Guangdong 519082, P.R.~China\label{inst1}
\and
CSST Science Center for Guangdong-Hong Kong-Macau Great Bay Area, Zhuhai, Guangdong 519082, P.R.~China\label{inst2}
\and
Department of Physics E. Pancini, University Federico II, Via Cinthia 6, I-80126, Naples, Italy\label{inst3}
\and
INAF -- Osservatorio Astronomico di Capodimonte, Salita Moiariello 16, I-80131 Naples, Italy\label{inst4}
\and
Instituto de F\'isica, Universidade Federal da Bahia, 40210-340, Salvador-BA, Brazil\label{inst5}
\and
PPGCosmo, Universidade Federal do Esp\'irito Santo, 29075-910, Vit\'oria, ES, Brazil\label{inst9}
\and
Department of Physics, Federal University of Sergipe, Avenida Marechal Rondon s/n, Jardim Rosa Elze, São Cristovão, SE, 49100-000, Brazil\label{inst6}
\and
National Astronomical Observatories, Chinese Academy of Sciences, 20A Datun Road, Chaoyang District, Beijing 100012, P.R.~China\label{inst7}
\and
School of Astronomy and Space Science, University of Chinese Academy of Sciences, Beijing 100049, P.R.~China\label{inst8}
}

\date{Received 4 October 2023; accepted 15 February 2024}

\abstract
{The galaxy total mass inside the 
effective radius is a proxy of the galaxy dark matter content and the star formation efficiency. As such, it encodes important information on the dark matter and baryonic physics.}
{Total central masses can be inferred via galaxy dynamics or gravitational lensing, but these methods have limitations. We propose a novel approach based on machine learning to make predictions on total and dark matter content using simple observables from imaging and spectroscopic surveys.}
{We used catalogs of multiband photometry, sizes, stellar mass, kinematic measurements (features), and dark matter (targets) of simulated galaxies from the  Illustris-TNG100 hydrodynamical simulation to train a Mass Estimate machine Learning Algorithm ({\sc Mela}) based on random forests.}
{We separated the simulated sample into passive early-type galaxies (ETGs), both normal and dwarf, and active late-type galaxies (LTGs) and showed that the mass estimator can accurately predict the galaxy dark masses inside the effective radius in all samples. We finally tested the mass estimator against the central mass estimates of a series of low-redshift (z$\lsim$0.1) datasets, including SPIDER, MaNGA/DynPop, and SAMI dwarf galaxies, derived with standard dynamical methods based on the  Jeans equations. We find that {\sc Mela} predictions are fully consistent with the total dynamical mass of the real samples of ETGs, LTGs, and dwarf galaxies.}
{\mla\ learns from hydro-simulations how to predict the dark and total mass content of galaxies, provided that the real galaxy samples overlap with the training sample or show similar scaling relations in the feature and target parameter space. In this case, dynamical masses are reproduced within 0.30 dex ($\sim2\sigma$), with a limited fraction of outliers and almost no bias. This is independent of the sophistication of the kinematical data collected (fiber vs. 3D spectroscopy) and the dynamical analysis adopted (radial vs. axisymmetric Jeans equations, virial theorem). This makes \mla\ a powerful alternative to predict the mass of galaxies of massive stage IV survey datasets using basic data, such as aperture photometry, stellar masses, fiber spectroscopy, and sizes. We finally discuss how to generalize these results to account for the variance of cosmological parameters and baryon physics using a more extensive variety of simulations and the further option of reverse engineering this approach and using model-free dark matter measurements  (e.g., via strong lensing), plus visual observables, to predict the cosmology and the galaxy formation model.}


\maketitle
%
\section{Introduction}
\label{sec:intro}

Galaxies originate within the gravitational confines of dark matter halos. They consist of baryonic matter, mostly in the form of stars and gas, as well as dark matter (DM). The spatial distribution and interplay of   these two components play a major role in shaping the process of galaxy formation and evolution. Due to the elusiveness of the DM, a compelling characterization of its properties, from the very basic, such as the total dark mass \citep[e.g., using virial theorem,][]{virial-theorem-FP-1997A&A...320..415B}, to the more complex, such as their mass density profiles \citep{NFW-Burkert1995, 1996ApJ...NFW}, remains unattainable. 
 
The only approach to constrain DM distribution in galaxies relies on gravitational effects. Ever since the initial discoveries indicating the presence of DM in galaxies, rotation curves have been extensively utilized to investigate the mass distribution of (rotation-supported) spiral systems \citep[e.g.,][]{Rubin1970ApJ...159..379R,SPARC-Mass-model-rotation-curve-Lelli-2016AJ....152..157L}. For ellipticals, instead, typical probes are stellar velocity dispersion profiles and higher-order velocity moments (\citealt{Kronawitter+00,Gerhard-dynamical-properties-2001AJ....121.1936G,ThomasJ+07_coma,Romanowsky+03,C06-cappellari2006,Napolitano09-NGC4494-planetary-nebulae,T09-CentralDM-2009MNRAS.396.1132Tortora09}), and strong gravitational lensing \citep{TK04,Koopmans+06_SLACSIII, Auger+10_SLACSX, Tortora+10lensing, Sonnenfeld+13_SL2S_IV}.

Our comprehension of the physical processes contributing to the assembling of baryons and DM received a burst thanks to the advent of new techniques, allowing us to collect data for large galaxy samples or resolve their kinematics in detail. Multiobject spectrographs have been used to simultaneously obtain spectra from up to hundreds of objects in a single observation \citep[e.g., SDSS,][]{Blanton-SDSS-multiobject-spectro-2003AJ....125.2276B} and Integral Field Spectrographs (IFS) have provided full two-dimensional kinematical maps, including galaxy rotation, velocity dispersion and even higher-velocity moments, (see, e.g., ATLTAS$^{\rm 3D}$, \citealt{ATLAS3D-I-2011MNRAS.413..813C}; MaNGA, \citealt{2015OverviewMaNGA}; SAMI, \citealt{SAMI-multi-object-integral-field-spectrograph-2012MNRAS.421..872C})

The next-generation (photometric and spectroscopic) sky surveys (or stage IV surveys), such as the {\it Chinese Survey Space Telescope} \citep[CSST,][]{ZhanHu-CSST-2011SSPMA..41.1441Z}, {\it Vera-Rubin/Large Synoptic Survey Telescope} \citep[VR-LSST,][]{LSST-2019}, {\it Euclid mission} \citep{Euclid-2011},   {\it Dark Energy Spectroscopic Instrument} \citep[DESI;][]{DESI-whitepaper-2013arXiv1308.0847L,DESI-Scienc-target-survey-design-2016arXiv161100036D}, and {\it 4-metre Multi-Object Spectrograph Telescope} \citep[4MOST;][]{4MOST2011Msngr.145...14D,4MOST-project-overview-2019Msngr.175....3D} will provide us with even more massive amounts of data, posing tremendous challenges for data modeling. Hence, finding methods that can swiftly obtain reliable results (e.g., to serve as crucial reference values for further analysis and more complex observations) without the need for complex modeling is very necessary.

\begin{table*}
\centering
\small
\begin{tabular}{lll}
\hline
\hline
\multicolumn{3}{c}{Training/Testing Features}\\
\hline
Parameter & Description & Unit  \\
\hline
$g$ & TNG: 3D SDSS $g$-band ($\lambda=469nm$) & mag \\
 & OBS: total 2D $g$-band magnitude ($\lambda=469nm$) & mag \\
$r$ & TNG: 3D SDSS $r$-band ($\lambda=617nm$) of all stellar particles & mag \\
 & OBS: total 2D $r$-band magnitude ($\lambda=617nm$) & mag \\
\re   & TNG: 3D radius containing the half of the stellar mass particles 
& kpc\\
 & OBS: 3D radius from modeling of the galaxy light density profiles & kpc \\
\ms & TNG:  3D Sum of masses of all stellar particles of a galaxy, using Chabrier IMF&    $M_\odot$\\
& OBS: 3D total stellar mass from stellar population models, using  Chabrier IMF  & $M_\odot$ \\
\sig & TNG: 1D velocity dispersion of all the member particles (3D dispersion divided by $\sqrt{3}$) & km/s \\
& OBS: line-of-sight velocity dispersion: within the fiber or (aperture corrected) at $\re$ & km/s\\
\hline
\hline
\multicolumn{3}{c}{Accessory Features}\\
\hline
Parameter & Description & Unit  \\
\hline
\msre & TNG: Sum of masses of stellar particles within the stellar half-mass radius, i.e., $\msre=\ms/2$ &    $M_\odot$\\
 & OBS: Stellar mass inside the 3D half-light radius &  $M_\odot$\\
$M_{\rm g}(\re)$& TNG: Sum of masses of gas particles within the stellar half-mass radius&    $M_\odot$\\
\hline
\hline
\multicolumn{3}{c}{Targets}\\
\hline
Parameter & Description & Unit  \\
\hline
\mtotre & TNG: 3D total mass as the sum of masses of all the particle within stellar half-mass radius & $M_\odot$\\
$M_{\rm dyn}(\re)$ & {OBS:} 3D total mass inside $\re$ derived from (Jeans) dynamical analysis  & $M_\odot$\\
$M_{\rm DM}(\re)$ & TNG: 3D total mass as the sum of masses of all DM particle within stellar half-mass radius & $M_\odot$\\
& OBS: \mdre=\mtotre-\msre\ -- unless gas mass, $M_{\rm g}(\re)$, is measured & $M_\odot$\\
\hline
\end{tabular}
\caption{Features and targets. The differences in physical quantities between the TNG100 simulation and observations are shown face-to-face. We show the features that are used in the training process and other features that are used in the definition of the targets.}
\label{tab: features and targets}
\end{table*}

Machine learning (ML) has become, in the last decade, an efficient solution in various astronomical applications. For example, it has been successfully applied in complex tasks where conventional analytical methods often proved to be challenging or computationally expensive. It has been used for predicting photometric redshifts \citep[e.g.,][]{amaro+19,Photometric-BASS-10.1093/mnras/stab3165,2022A&A...666A..85L}; deriving structural parameters of galaxies from light profiles \citep{Li-CNN-gal-structure-params}; performing star, galaxy, and quasar separation in both images and catalogs \citep{BASS-identification-10.1093/mnras/stab1650,2024MNRAS.527.3347V,2021A&A...645A..87B}; identifying and modeling strong gravitational lensing events \citep{Li+20_KiDS,2022MNRAS.510..500G}; determining DM distribution of galaxies \citep{DM_distribution_CNN_2023MNRAS.525.6015D}; and  connecting properties of galaxies and DM halos \citep{GalaxyNet_2021MNRAS.507.2115M}.

Recently, we have proposed using ML algorithms, trained on cosmological simulations, to estimate the DM content of galaxies \citep[][vM+22 hereafter]{von2022inferring}. Cosmological simulations are based on physical principles and realistic recipes for feedback processes, producing seemingly realistic galaxy distributions and scaling relations. Therefore they can be used to train ML algorithms to estimate the DM content of galaxies. 
In the first paper in this series, we   used a Tree-based Pipeline Optimization Tool \citep[TPOT;][]{TPOT-OlsonGECCO2016} to find the optimal ML pipeline. We demonstrated that we can infer the DM properties of galaxies (e.g., central DM mass, DM half-mass radius) starting from general catalog properties, such as luminosity, size, kinematics,  colors, and stellar masses in the IllustrisTNG simulation \citep[][N+19 hereafter]{TNG-data-nelson2019illustristng}. There are other studies following a similar strategy:  using the {Cosmology and Astrophysics with MachinE Learning Simulation} \citep[CAMELS;][VN+23 hereafter]{CAMLES-public-data-release-2023ApJS..265...54V}. For example, \citet{2021halo-galaxy-GNN}   infer halo mass given the positions, velocities, stellar masses, and radii of the hosted galaxies using graph neural  networks. \citet{2022subhalo-galaxy-MLP}   predict the total mass of a subhalo from their internal properties, such as velocity dispersion, radius, or star formation rate, using neural network and symbolic regression. However, these studies are solely conducted on simulations and have not been tested using real observational data.\footnote{After the publication of the preprint of this article and during the referee process, the preprint of another paper was announced \citep{jiani2023arXiv231110351C}, also aiming at predicting the total and dark mass in galaxies, using convolutional neural networks (CNNs). Unlike this one, this work uses 2D kinematics as input and is explicitly meant not to be applied to real galaxies as it is based on idealistic simulated velocity fields. Although this approach is still at the proof-of-concept stage, it illustrates that deep learning is a promising method to explore further to infer galaxy masses.} On the other hand, utilizing cosmological simulations for replicating and comprehending observational data via ML represents a novel strategy imbued with unforeseeable potentialities. 

For instance, we can constrain cosmological parameters and feedback processes comparing DM-related scaling relations from cosmological simulations with galaxy data, which has been envisioned within the CAMELS project (see, e.g., \citealt{2022_one_galaxy_camels}), and has   recently been shown to be feasible using classical statistical methods \citep[e.g.,][]{busillo2023casco}.  
In this paper, we apply {a random forest algorithm} to predict central total mass and DM mass of {galaxies from both simulations and observations. The Mass Estimate machine Learning Algorithm ({\sc Mela}) is trained with simulation data to make predictions for the galaxies from observations} that are compared with results from literature dynamical analyses.  

The advantage of our novel method   is that it is based on low-level photometric and kinematical information, including only aperture photometry and aperture kinematics. These are typical standardized data provided by photometric large sky surveys, for example, the CSST, VR-LSST, and  EUCLID, and spectroscopic surveys, such as fiber-aperture velocity dispersion (e.g., from DESI and 4MOST). If successful, this method can provide a significant advantage with respect to standard analysis tools generally based on much  higher-level information in Jeans analysis, including accurate surface photometry analysis \citep[T+09 hereafter]{T09-CentralDM-2009MNRAS.396.1132Tortora09} and integral field spectroscopy \citep{2008Cappellari..JAM,2020Cappellari..JAM}. More importantly, it provides physically motivated inferences because they are based on realistic cosmological simulations. 

The other major novelty of the \mla\ project is that it is entirely data-driven. It provides, for the first time, a full application to real systems and a direct comparison of the predicted masses against classical dynamical methods on samples of early-type and late-type galaxies (\citealt{tortora2012spider,2023MaNGADynPop}) and dwarf systems (\citealt{SAMI-Fornax-Dwarfs-Survey-II}) (see \S\ref{sec:data} for details).

The outline of the paper is as follows. In \S\ref{sec:data} we describe the data and the main physical quantities therein. In \S\ref{sec:method} we describe how \mla\ works and how to evaluate the performance of   \mla's predictions. In \S\ref{sec:results} we show the result of self-prediction and apply the trained \mla\ to real data. In \S\ref{sec:discussion} we discuss the robustness of the \mla\ and the possible reason for the errors. In \S\ref{sec:conclusions} we draw conclusions and give future perspectives. 

Throughout this work, we adopt a flat Universe with a $\Lambda$CDM model: ($\Omega_\Lambda$, $\Omega_m$, h)\, =\, (0.6911, 0.3089, 0.6774) based on Planck2015 \citep{LCDM-cit_planck}. 

\section{Simulation and observation data}
\label{sec:data}
In this section, we introduce the simulation datasets we want to use for the training and testing of the {\sc Mela} and the different observational datasets we want to use as predictive samples. These latter are designated for testing only and are not  utilized for model training. In fact, they are used for assessing the performance of the algorithm previously trained using the entire simulation dataset. As anticipated, one of the novelties of this paper is a direct application to dynamical samples to check if {\sc Mela} can predict the mass content of galaxies consistently with standard dynamical analyses. For this reason, differently from our previous analysis, in vM+22, here we concentrate on the prediction of the central dark matter content of galaxies, namely the total mass \mtotre\ and the dark matter within effective radius \mdre. In a future work, we will test the same methods against analyses based on extended datasets able to constraint more total dark matter quantities \citep[e.g., based on planetary nebulae, see][]{Napolitano09-NGC4494-planetary-nebulae}

The simulation data we use for the training of the machine learning algorithms is based on the IllustrisTNG simulation (N+19), a state-of-the-art magneto-hydrodynamic simulation. 

As predictive samples, we  consider datasets covering different mass ranges (from dwarf to massive ETG systems) and different dynamical approaches. In particular, we have collected dynamical masses of massive ETGs from {\it Spheroids Panchromatic Investigation in Different Environmental Regions} project \citep[SPIDER;][T+12 hereafter]{tortora2012spider} and combined analysis of the Dynamics and stellar Population for the MaNGA survey \citep[MaNGA DynPop;][Z+23 hereafter]{2023MaNGADynPop}, while we also use dwarf spheroidals from the SAMI Fornax Dwarf survey \citep[][E+22 hereafter]{SAMI-Fornax-Dwarfs-Survey-II}.

\subsection{Features and targets: A necessary preamble}
\label{sec:preamble}
Before going into the details of the different datasets, we start by describing the physical quantities that we  use throughout the paper, as they are differently defined in simulations and observations. This is an important semantic preamble to set the following discussion and motivate some of the choices we need to make when combining simulations and observation in a single analysis. ``Observational realism'' is a complex chapter of this comparison \citep[see, e.g.,][]{2019MNRAS.490.5390Bottrell,2023A&A...677A.102F}, which also involves the impact of inappropriate definitions of the observational-like parameters derived from simulations (see, e.g., \citealt{2021MNRAS.508.3321Tang}). Addressing these issues is beyond the scope of this paper, except for some relevant aspects concerning the measurement errors (see, e.g., \citealt{Qiu2023_cosmo_cl_ML}, and \S\ref{sec:meas_err}). and will be fully addressed in future studies. However, here we need to discuss, and possibly quantify, all obvious mismatches of physical quantities in simulation and observation datasets and introduce some basic assumptions to align the two data types as much as possible. 

In Table~\ref{tab: features and targets} we show the most important physical properties that we want to use as input of the machine learning algorithm (features) and the quantities we want to predict (targets), with their broad meaning in simulations and observations, face-to-face. We also add some accessory features, that are indirectly used for the target definitions. In particular, we remark the stellar mass inside the half-light radius, \msre\, which we decided to exclude from the training/testing features because redundant with respect to the total stellar mass, \ms, at least in simulations, where, by definition, $\msre = \ms /2$. Here below are some notes to save for the rest of the paper. 

{\it 3D versus 2D features and targets}. Data dimensionality is a critical aspect of our study. On the one hand, simulations provide 3D quantities, inherently projection invariant. On the other hand, observational data are essentially 2D, derived from images; for example,  they represent projection variant views of astronomical objects, meaning that some of their attributes can change depending on different observation angles. For ``total'' quantities (luminosity, stellar masses), the use of 2D or 3D does not make a real difference, but for sizes and partial quantities (\re, and stellar and total masses inside it) it does. Generally speaking, for real galaxies the dynamical masses (i.e., the ones coming from Jeans modeling or equivalent) can be easily modeled as projected, 2D, or de-projected, 3D, although this conversion comes along with some geometrical assumption (from spherical symmetry, e.g., \citealt{Wolf10_2D_3D, 2022FrASS...8..197T&N22}, to axisymmetric, e.g., Z+23). 

{\it 3D half-light (\re$_\star$) versus half-mass ($r_{\rm 1/2}$) radii}. Half-light radii or effective radii are certainly the physical parameters that potentially carry the most complex systematics, due to their strong relation with galaxy masses and galaxy types \citep[see, e.g.,][]{2003MNRAS.343..978Shen03}. We stress in particular two of them: 1) the constant mass-to-light ratio, as in simulations the radius is computed on stellar mass particles; 2) 2D versus 3D definition (see also above). However: 1) M/L gradients might have a little impact on the mass-size relation of galaxies with an average ratio of the mass weighted, $R_m$, with respect to the light weighted, $R_l$, that is of the order of $R_m/R_l\sim$0.6-0.7, from low redshift \citep[see, e.g.,][]{Bernardi23_MLgrad_halfmassrad}, up to about redshift 1 \citep[e.g.,][]{Suess19_MLgrad_halfrad_high-z}, although with a large scatter (and assuming uniform initial mass function inside galaxies); 2) the ratio between the 2D and the 3D half-light radii can be quantified to be $\sim 3/4$ for pressure supported systems with a large variety of light distribution under spherical symmetry \citep[see][W+10 hereafter]{Wolf10_2D_3D}, while it is basically bias-less for disks where the 3D radius can be obtained from simple thin disk deprojection. Putting these two arguments together we finally conclude that, as the M/L gradients are computed in real galaxies using 2D quantities \citep[e.g.,][Eq. 4]{Bernardi23_MLgrad_halfmassrad}, and being the 2D radii more compact than the 3D radii, these gradients are an upper limit for the equivalent 3D ones. Hence, the $r_m/r_l$ (having used the $r$ for 3D quantities) has to be closer to unity than the one measured on 2D gradients, making it reasonable to use the 3D half-mass radius ($r_{\rm 1/2}$) in simulations as a good proxy of the half-light radius (\re$_\star$). In the following, we  use the symbol \re\ to indicate both ones, equivalently. 

{\it Predicting and comparing 3D targets}. A consequence of the conclusion above is that we can use the 3D features of simulations to predict the 3D targets, and so we can for real galaxies if we use analogous 3D features and targets. These latter can be compared with the equivalent obtained from dynamical models, only under the conditions that the conversion of the 2D to 3D mass properties of real galaxies is free from systematics. Alternatively, to make use of more observation-like features, we should train on 2D features, which are not yet a standard product for hydro-simulations. We will address this in the forthcoming paper on this project.

{\it Velocity dispersion}. The velocity dispersion in simulations is defined over the full set of particles {(i.e., DM, stellar, black hole, gas)}, which are distributed over a large range of radii. In observations, it is measured over a small aperture \citep{thomas-SDSS-III-velocity-dispersion,napolitano-lamost2020-central-velocity-dispersion} and possibly corrected with an empirical formula (e.g., \citealt{C06-cappellari2006}) to the effective radius, or directly derived at the \re\ via integral field spectroscopy (e.g., Z+23). The question arises of how comparable these two quantities are. The Observed velocity dispersion of galaxies shows a strong negative slope generally confined within the central regions, up to a few effective radii \citep[see, e.g.,][]{Gerhard-dynamical-properties-2001AJ....121.1936G, Coccato09,Napolitano09-NGC4494-planetary-nebulae,Napolitano2011,Pulsoni_ePNS}. On the other hand, the softening length of the TNG100 ($\sim0.7$ kpc/h) likely suppresses the typical central peak of the velocity dispersion profile of galaxies. The net effect is that in mid-resolution simulations the overall \sig, as in Table~\ref{tab: features and targets}, is basically sensitive to the large radii flatter part of the velocity dispersion profile that is smaller than the one measured in the central galaxies \citep[e.g.,][]{Pulsoni_ePNS}. This is expected to produce a systematic shift on the typical scaling relations involving \sig, like the mass-velocity dispersion relation \citep[e.g.,][]{FJ_96,Cappellari-2013-ATLAS3D-XX-2013MNRAS.432.1862C,napolitano-lamost2020-central-velocity-dispersion}. This is  shown in detail in \S\ref{sec:systematics}. For dwarf galaxies, though, the typical velocity dispersion profiles look flatter at all radii (see E+22, Fig.~4), hence this effect might be mitigated and the \sig\ from the simulations can be a realistic proxy of the real velocity dispersion (see also \S\ref{sec:systematics}).  

{\it Dark matter in galaxies}. This is naturally provided in the TNG100 catalog by the sum of all the DM particles inside a given 3D volume. In our case, this is the one enclosed in the effective radius, \mdre. For real galaxies, depending on the sample, it can be directly fitted assuming some form of dark matter halo (e.g., a \citealt{Navarro1997} or \citealt{NFW-Burkert1995} profile), or, if only the total mass of the galaxy is inferred, by the equation $\mdre=\mdyre-\msre$ -- see Table~\ref{tab: features and targets}. Either case, the underlying assumption is that the gas mass is negligible in the considered volume. This is somewhat reasonable for the regions inside \re\ we want to consider in this analysis. However, this might have a nonnegligible impact on the comparison with the simulations. In these latter cases, if we can reasonably assume that $\mdyre=\mtotre=\msre+\mdre+M_{\rm gas}(\re)$, the quantity $\mdyre-\msre$ equals, by definition, $\mdre+M_{\rm gas}(\re)=\tilde{M}_{\rm DM}(\re)$. Hence, this newly defined mass comes from the fact that in real galaxies the albeit limited, gas mass contributes as dark baryons to the \mdre. Thus, when explicitly checking the predictions of {\sc Mela} against the DM in real data, we should use $\tilde{M}_{\rm DM}(\re)$ as a target (see Appendix~\ref{app:mass_DM}). For convention, we  call this the ``augmented'' dark matter, meaning that it includes all the unaccounted mass contributions that exceed the stellar mass of a galaxy (aka ``missing mass'').  

\begin{figure*}
\hspace{-0.5cm}    
\includegraphics[width=1.04\columnwidth]{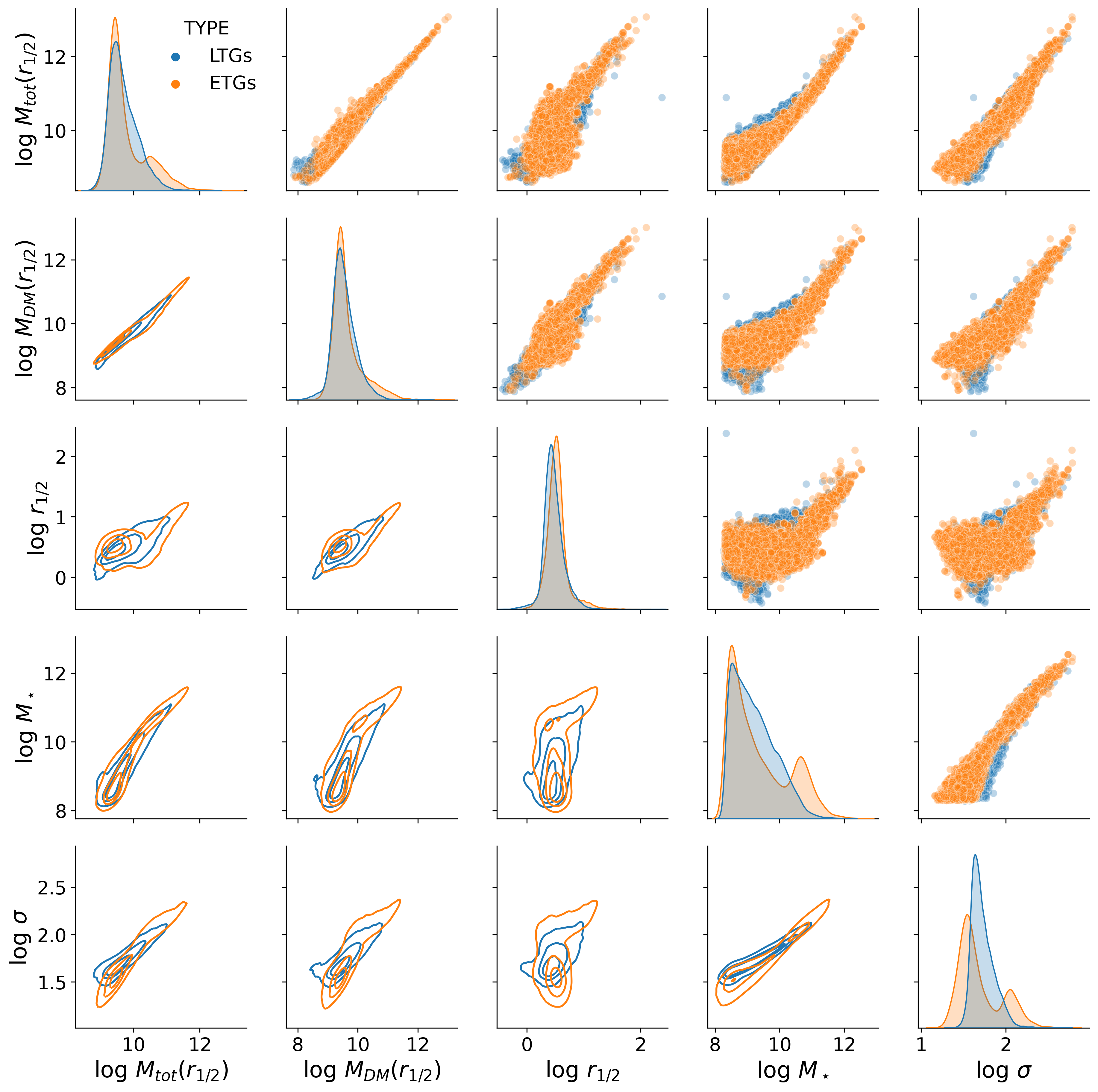}
\includegraphics[width=1.04\columnwidth]{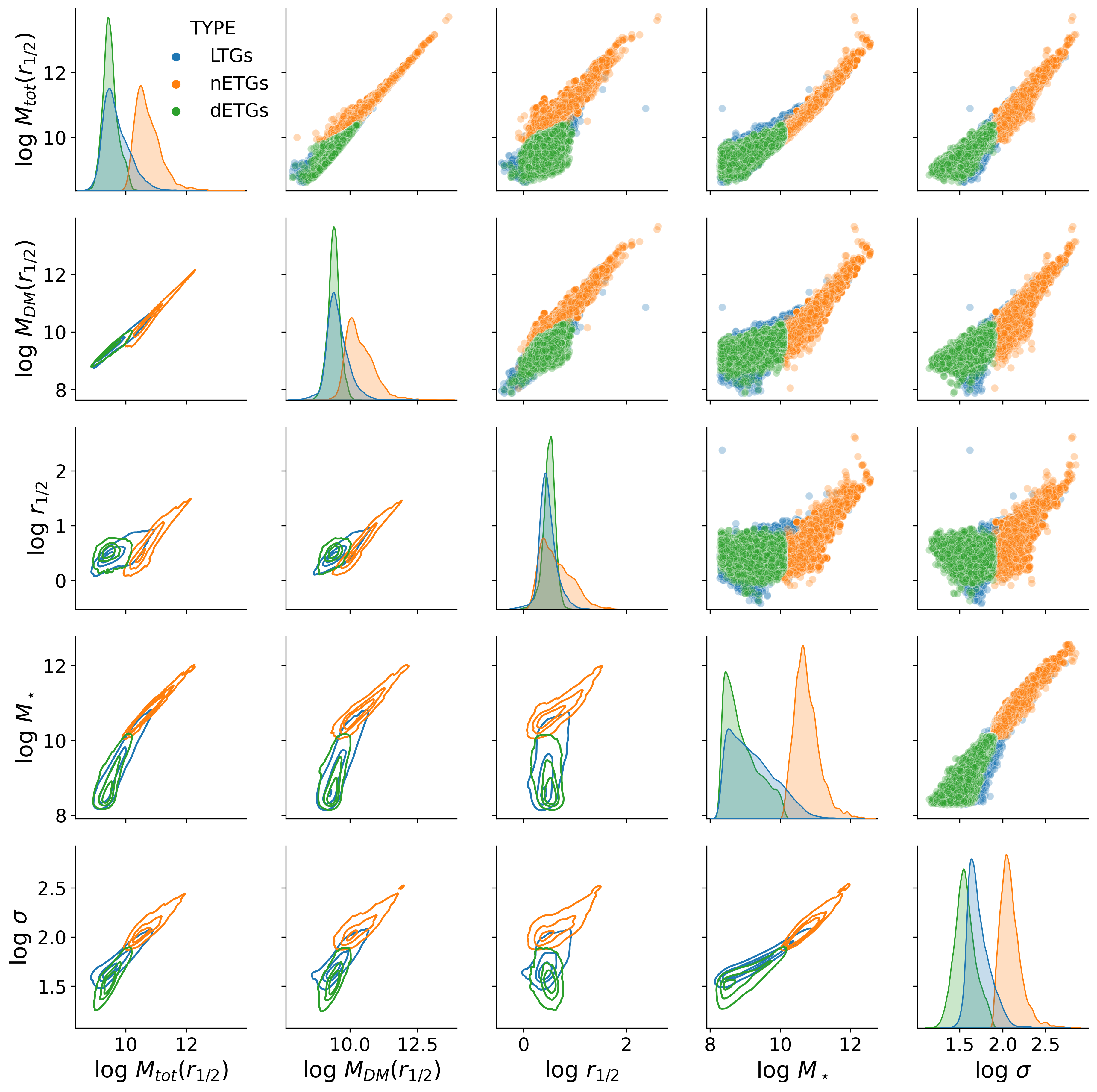}
\caption{Distribution of relevant features and targets as in Table~\ref{tab: features and targets}: total mass inside the stellar half-mass radius, augmented dark matter mass inside stellar half-mass radius, half-mass radius, stellar mass in half-mass radius, velocity dispersion, total and dark matter mass in half-light radius. Left: Galaxies are divided into ETGs and LTGs on the basis of their SFR. Right: ETGs are further divided into normal and dwarf ETGs based on the classification criteria outlined in Table~\ref{tab:classes}. The normalized distribution of the features and targets is shown along the diagonal. Units are as in Table~\ref{tab: features and targets}. This is the original data from TNG100 without considering mock measurement errors. To get a comparative picture,   a fixed value was set for the different types of galaxies. We randomly get a 20,000 galaxy subsample from the full dataset  and  from the three galaxy types.}
\label{fig:correlations}
\end{figure*}

\subsection{TNG100 simulation for training and testing}
\label{sec:TNG_sim}
IllustrisTNG (N+19) is a series of state-of-the-art magneto-hydrodynamic simulations using different box sizes: 50Mpc/h, 100Mpc/h, and 300Mpc/h. In this work, we use TNG100-1 (TNG100, for short), which is the highest resolution simulation with a volume of $106.5Mpc^3$ and $1024^3$ dark matter particles.  The mass resolution of dark matter particles is $7.5\times10^6M_{\odot}$, while the mean mass resolution of baryon particles is $1.4\times10^6M_{\odot}$. 
The Plummer equivalent gravitational softening  of the collisionless component in the comoving units at z=0 is $R_p=1.0~ \rm kpc/h$ (see N+19, Table~1) or $0.74 ~\rm kpc/h$ in the physical units at z=0. The cosmological parameters are based on Planck2015 (see \S\ref{sec:intro}), which is also the cosmology we use for  all the observed data throughout the paper. The hydrodynamical part of the simulation includes updated recipes for star formation and evolution, chemical enrichment, cooling, and feedbacks \citep{Weinberger2016/10.1093/mnras/stw2944,Pillepich2018MNRAS.475..648P,Nelson2018MNRAS.475..624N}. It also accounts for AGN feedback \citep{Weinberger2016/10.1093/mnras/stw2944} and galactic winds model \citep{Pillepich2018MNRAS.475..648P}, mimicking supernovae feedback.

The IllustrisTNG snapshots and Group catalogs in different redshift bins, from $z=0$ to $z=127$, are publicly available.\footnote{https://www.tng-project.org/data/downloads/TNG100-1/} In the following, to illustrate the main properties of the simulations, we will use a reference redshift window from 0 to 0.1. In particular, we only use the information of Subfind Groups, which contain groups of particles recognized as individual objects for which a series of physical quantities have been assigned as the integral of all particles belonging to the group. The main properties we are interested in are indicated as TNG in Table~\ref{tab: features and targets} and previously described in \S\ref{sec:preamble}. In the redshift range $z\in\left[0,0.1\right]$, they include 9 snapshots, corresponding to $z=$0.00, 0.01, 0.02, 0.03, 0.05, 0.06, 0.07, 0.08, 0.10. The Subfind Group Catalog represents the basic dataset to train our machine learning tools and subsequently test the performances, using the known target values as ground truth. If the galaxy-like systems belonging to the TNG simulations are a fair representation of the real galaxies, we can expect that the ML tool trained on the TNG galaxy catalog can predict the target quantities, specifically dark and total mass, not only on the simulated test sample, but also on the real galaxies. In the next section, we detail the selections needed to choose realistic galaxy systems from the Subfind Group Catalog.

\subsubsection{Selection and properties of the galaxy sample from TNG100}
\label{sec:gal_sel_TNG}
The galaxy dataset from the Subfind Group Catalog contains subhalos from friend-of-friend (FOF) and subfind algorithms, that represent galaxy-like groups, but not all of them are well-defined, realistic systems. To select physically meaningful galaxies, we use the following criteria:
\begin{itemize}
    \item SubhaloFlag=True. This selects subhalos with cosmological origin (i.e., they have not been produced by fragmentation of larger halos by baryonic processes). TNG provides a Flag for these spurious satellite systems.\footnote{https://www.tng-project.org/data/docs/background/\#subhaloflag} 
    \item The stellar half-mass (or effective) radius and dark matter half-mass radius are larger than 2 times of the Plummer radius $R_p=0.74\rm kpc$. This is to avoid that the internal properties of small galaxies are fully dominated by numerical softening. 
    \item Number of both stellar and dark matter particles is larger than $N=200$. This is a further criterion to have robust total quantities based on sufficiently large particle statistics in galaxies. This corresponds to a lower total stellar mass limit of $\ms=10^{8.3}M_\odot$, which also implies $\msre>10^{8.0}M_\odot$.
    \item We finally assume that every subhalo is a single galaxy.
\end{itemize}

\begin{figure*}
\hspace{-0.7cm}
\includegraphics[width=2.13\columnwidth]{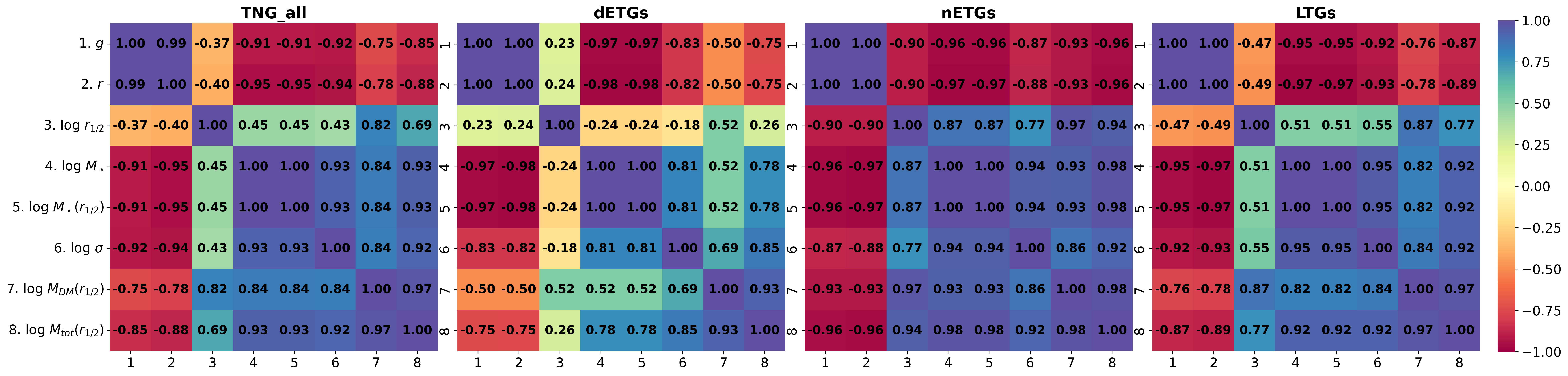}

\hspace{-0.7cm}
\includegraphics[width=2.13\columnwidth]{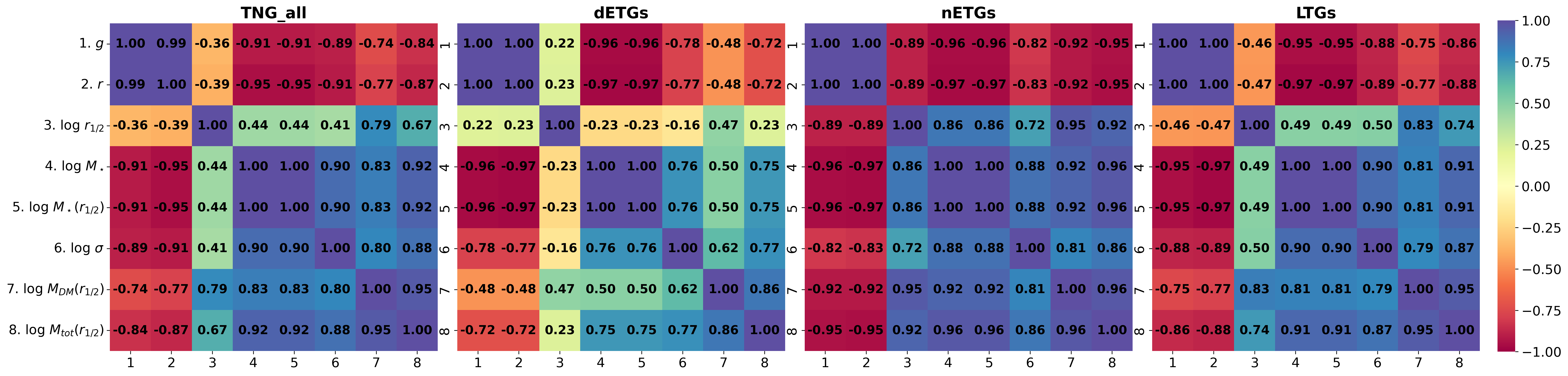}
\caption{Correlation heat map of the different TNG galaxy samples defined in \S\ref{sec:TNG_sim} when not considering (upper row) and considering  (bottom row) the mock measurement errors, as in \S\ref{sec:meas_err}. The correlation coefficients are calculated using the Pearson correlation coefficient (see Eq.~\ref{eq:Paerson1}).
}
\label{fig: heatmap-simuonly}
\end{figure*}

We divide this galaxy sample into passive/early-type galaxies (ETGs) and active/late-type galaxies (LTGs) using the specific star formation rate as the criterion. Following observational analyses \citep{ETG-LTG-sSFR-2023A&A...669A..11P}, we adopt $\log \rm sSFR/yr^{-1}<-11$ for ETGs and $\log \rm sSFR/yr^{-1}>-11$ for LTGs. This is possibly a more robust selection than one based on pure the color-stellar mass cut (see, e.g., \citealt{pulsoni20202_TNG_ETG}), which yet allows us a sharp separation of the red sequence galaxies from the blue cloud systems. In the absence of other relevant structural parameters in the TNG catalogs suitable for morphological (e.g., the $n$-index of the \citealt{Sersic1968adga.book.....S} profile) or kinematical selection criteria (e.g., galaxy spin, see \citealt{Rodriguez-Gomez_TNG100spin}), the sSFR criterion remains the best physical argument we can use to make the ETG/LTG separation. 

In Fig.~\ref{fig:correlations} we show the correlations/distributions of some of the relevant properties (see Table~\ref{tab: features and targets}) we will use in the rest of the paper as features and targets for these two classes. From the left panel, we clearly see the differences in the distributions of the ETGs and LTGs, which is also reflected in differences in the scaling relations, among the different quantities. We also remark the bimodal distribution of the ETGs, showing the presence of two populations of high-mass ETGs, at $\log \msre/M_\odot \gsim 9.8$, and ``dwarf'', low-mass sample, at $\log  \msre/M_\odot \lsim 9.8$, which is mirrored by the velocity dispersion distribution, which is also bimodal around $\log \sigma/\rm km s^{-1}\sim1.9$. 

``Normal'' ETGs (nETGs) and ``dwarf'' ETGs (dETGs) are known to have rather different scaling relations and possibly also different formation mechanisms \citep{Formation-dEs-2009MNRAS.396.2133K} which might reflect differences in their dark matter properties. Unsurprisingly, we also observe similar bimodalities in the total and DM masses. We then define nETGs as ETGs having $\log \msre/M_\odot > 9.8$ and $\log \sigma/\rm km s^{-1}>1.9$, and dETGs all the remaining ETGs. With these criteria, the LTGs, and n/dETGs are distributed as shown in the right panel of Fig.~\ref{fig:correlations}, where we can use the Pearson correlation coefficient $\rho$:
\begin{equation}
\rho=\frac{cov(y,x)}{\sigma(y)\sigma(x)}
\label{eq:Paerson1}
\end{equation}
to quantify the linear relationship between different pairs of variables (i.e., the features and targets in Table~\ref{tab: features and targets}). In Eq.~\ref{eq:Paerson1}, cov(y, x) represents the covariance of two variables, while the denominator $\sigma(y)$ and $\sigma(x)$ are the standard deviations of  y and x. These values normalize the covariance, ensuring that $\rho$ ranges between -1 (indicating a perfect negative linear relationship) and 1 (indicating a perfect positive linear relationship). From these correlation matrices, it is evident that the three classes have different distributions and scaling relations (see, e.g., mass-size relation, and the mass-$\sigma$ relation). This is more clearly illustrated in the correlation heatmap presented in Fig.~\ref{fig: heatmap-simuonly}, where there are groups of physical parameters (i.e., \ms, \msre, \sig, \mdre, \mtotre) that have a different level of correlation for the three distinct classes (three panels on the right). In particular, dwarfs show a looser correlations of the half-light radius, \re, with luminosity, mass, and $\sigma$, while the nETGs and LTG show tighter size-luminosity, mass-size relations, compatibly with observations \citep[see also,][]{2021MNRAS.508.3321Tang}. Also looking at the target quantities, \mdre\ and \mtotre, dETGs show shallower correlations with other parameters (e.g., luminosity, size, and stellar masses) with respect to nETGs and LTGs. When combined together, the full TNG galaxy sample (TNG\_all) shows a correlation matrix (first panel on the left) which is very similar to the one of LTGs, which is the numerically dominant population with intermediate properties. We will come back to these correlations later, when interpreting the results of the mass predictions.

Due to these different features in the correlations highlighted by the Figs.~\ref{fig:correlations} and~\ref{fig: heatmap-simuonly}, we will compare the mass predictions obtained by considering these three classes (i.e., nETGs, dETGs, LTGs) as separated and joined together, as done in vM+21. This will allow us to check whether {\sc Mela} can perform better on the individual classes rather than all galaxies together. 

In Table~\ref{tab:classes} we summarize the definition of the three galaxy classes from the TNG simulation. The number we   collected in the different redshift bins is used in the next analysis to predict the total mass and the dark mass for the observed samples.

\begin{table*}
\centering
\tiny
\begin{tabular}{llll}
\hline
\hline
\multicolumn{4}{c}{\bf TNG100 Simulations} \\ \multicolumn{4}{c}{SubhaloFlag=True; $N_\star>200$; $N_{\rm DM}>200$}\\
\hline
\bf Class & \bf Selection criteria & \bf Snapshots & \bf Number  \\
\hline
nETGs & $\re> 2R_{\rm p}$; $\log\ms/M_\odot$>8.3; $\log \rm sSFR\times yr<-11$; $\rm \log \sigma/kms^{-1} >1.9$ and $\log \ms/M_\odot>10.1$ & $z=0.00-0.10$ &  70,859\\
dETGs & $\re> 2R_{\rm p}$; $\log\ms/M_\odot$>8.3; $\log \rm sSFR\times yr<-11$; $\rm \log \sigma/kms^{-1} < 1.9$ and $\log \ms/M_\odot<10.1$ & $z=0.00-0.10$ &  21,416\\
LTGs & $\re> 2R_{\rm p}$; $\log\ms/M_\odot$>8.3; $\log \rm sSFR\times yr>-11$ & $z=0.00-0.10$ & 247,229\\
TNG\_all & nETGs+dETGs+LTGs & $z=0.00-0.10$ & 339,504\\
\hline
\multicolumn{4}{c}{\bf Observations}\\
\hline
\bf Class & \bf Selection criteria & \bf redshifts range & \bf Number  \\
\hline
SPIDER/nETGs & $\re> 2R_{\rm p}$; $\log\ms/M_\odot$>8.3; $\mtotre>\msre$; $\rm \log \sigma/kms^{-1} >1.9$ and $\log \ms/M_\odot>10.1$ & $z=0.05-0.095$ &  3,592\\
DynPoP/nETGs & as SPIDER/nETGs + Qual$\geq$1 and (DynPop\_E or DynPop\_S0) & $z=0.01-0.10$ &  1,271\\
DynPop/LTGs & $\re> 2R_{\rm p}$; $\log\ms/M_\odot$>8.3; $\mtotre>\msre$; Qual$\geq$1 and DynPop\_S & $z=0.01-0.10$ &  2,846\\
Fornax/dETGs & $\re> 2R_{\rm p}$; $\log\ms/M_\odot$>8.3; $\mtotre>\msre$; $\rm \log \sigma/kms^{-1} < 1.9$ and $\log \ms/M_\odot<10.1$  & $z=0.00$ & 9+6\\
\hline
\hline
\multicolumn{4}{l}{Where we have defined (see text):}\\
DynPop\_E & \multicolumn{3}{l}{($\rm P_{LTG}$<0.5) and (T-Type<0) and ($\rm P_{S0}$<0.5) and (VC=1) and (VF=0)}\\
DynPop\_S0 & \multicolumn{3}{l}{($\rm P_{LTG}$<0.5) and (T-Type<0) and ($\rm P_{S0}$>0.5) and (VC=2) and (VF=0)}\\
DynPop\_S & \multicolumn{3}{l}{($\rm P_{LTG}$<0.5) and (T-Type>0) and (VC=3) and (VF=0)}\\
\hline
\end{tabular}
\caption{Features and targets. The differences in physical quantities between the TNG100 simulation and observations are shown face-to-face. We show the features that are used in the training process and other features that are used in the definition of the targets.}
\label{tab:classes}
\end{table*}

\subsubsection{Mock measurement errors for TNG100 galaxies}
\label{sec:meas_err}
The physical quantities extracted from simulation usually are provided with no error (although they are themselves the product of a measurement process) and they ideally represent ``exact predictions'' of theoretical models. Statistically speaking they can be assumed to be the true value of physical quantities that, in observed galaxies, come along with measurement uncertainties. These measurement errors need to be taken into account when doing predictions, because their net effect is to dilute the correlations seen in Fig.~\ref{fig:correlations}, making also the predictions of the target broadened. This is a very basic form of ``observational realism'', discussed in \S\ref{sec:preamble}. 

To include the effect of observing measurements on the simulated quantities, we have assumed Gaussian errors with typical relative uncertainties of 10\% for the features (i.e., $g$ and $r$ magnitudes, \ms, \msre, \sig, \re) and 15\% for the targets (i.e., \mdre\ and \mtotre) to be consistent with typical uncertainties found for galaxy observations and dynamical samples in the galaxy centers (see, e.g., T+09, Z+23, \citealt{Cappellari-2013-ATLAS3D-XX-2013MNRAS.432.1862C}). Note that the adoption of errors on targets is not strictly necessary, under the assumption that the targets represent some ground truth properties of galaxies. However, our choice intends to conservatively account for the fact that targets are extracted from the simulation with a measurement process that brings some uncertainties. Finally, the mock measurements are obtained by randomly drawing the ``measured'' quantities from a Gaussian centered in their original (true) value and with standard deviation corresponding to the adopted relative errors. The new ``measured'' quantity is taken inside $3\sigma$ from the original ``true'' value. This step is needed to fully account for the intrinsic errors of the observed features in the training phase, which should reproduce the distribution of the observed features one wants to use to make the predictions of the chosen targets.

In Fig.~\ref{fig: heatmap-simuonly} we finally report the correlation matrices for the whole sample and the three galaxy classes after having added the measurement errors, face-to-face with the same matrices before the measurement errors. As seen the correlations have not been dramatically affected, although we can see a decrease of the correlation coefficients of the order of $\sim 5$\% or smaller for TNG\_all and $\sim 10$\% or smaller for the n/dETG and LTG classes. 

\subsection{Observational data: The predictive sample}

As mentioned in the introduction of this section, we use a variety of galaxy mass catalogs based on different dynamical methods as a predictive sample. These include SPIDER, MaNGA DynPop a SAMI Fornax Dwarf survey that we briefly describe below in this section.
Here we anticipate that according to the independent and identically distributed hypothesis in machine learning inferences, to have reliable ML predictions, we need to ensure that the feature and target distributions of the test sample are comparable with those of the training sample. To do that we apply to observational samples above the same selection criteria of the TNG classes, as summarized in Table~\ref{tab:classes}. Here we see that, for observations, we add the further condition that the dynamical mass inside the effective radius is larger than the stellar mass inside the effective radius (i.e., \mdyre$ > $\msre). This is because mass estimators, due to the statistical errors, might sometimes bring to such unphysical results. We also remark that the \re$\gsim1.5~\rm kpc$ criterion, imposed by the softening length of the simulations, is quite restrictive for observations as there are massive ultra-compact galaxies \citep[UCMGs, e.g.,][]{2016MNRAS.457.2845T,2018MNRAS.481.4728T,2020ApJ...893....4S,2021A&A...646A..28S} and dwarf galaxies \citep{2003AJ....125.2936G}, having sizes generally smaller than the threshold used above. If, on one hand, the UCMGs are rare in the local universe \citep{2014ApJ...780L..20T,2016MNRAS.457.2845T} and this selection does not impact the predictive sample, on the other hand, given the mass-size relation of dETGs (e.g., E+22) \re$>1.5~\rm kpc$ corresponds to stellar masses of the order of $\ms/M_\odot\sim10^9$ or higher, thus strongly impacting the accessible lower mass limit of observations. As we are interested in pushing the predictions of the \mla\ toward low stellar masses, then we decide to add a secondary dwarf predictive sample for which we use the \re$=1~\rm kpc$ as a lower size limit and use a separate analogous training sample to make predictions for it.

\begin{figure*}[ht]
\hspace{-0.7cm}
\includegraphics[width=2.13\columnwidth]{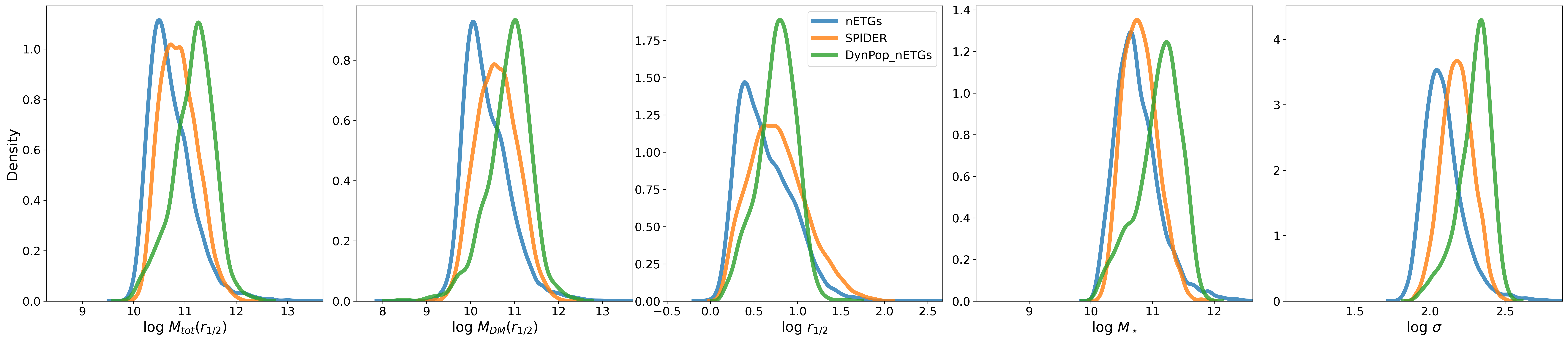}

\hspace{-0.7cm}
\includegraphics[width=2.13\columnwidth]{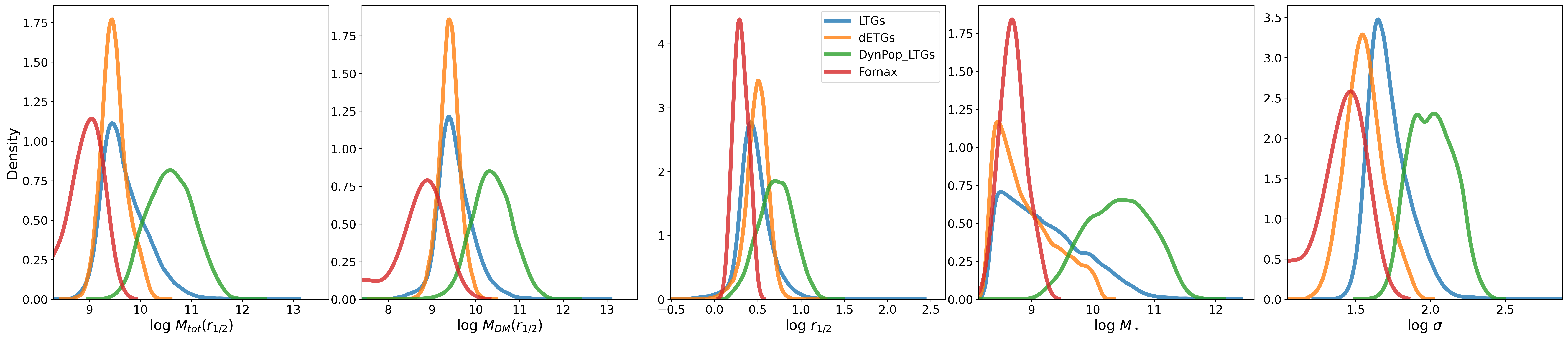}
\caption{Kernel density estimation (KDE) for each class of the dataset. Top row: KDE of the nETG dataset. Bottom row: KDE of the  LTG and dETG dataset. The number of each class of datasets is indicated in Table~\ref{tab:classes}. All the data points are within the x-axis limit. In the case of Fornax, an incompleteness of the smoothed estimate is evident due to the limited number of data points.}
\label{fig: kde-distribution}
\end{figure*}

\subsubsection{SPIDER ETGs}
\label{sec:spider}
The Spheroid's Panchromatic Investigation in Different Environmental Regimes (SPIDER) is a sample of 39,993 bright ETGs in the redshift range of 0.05 to 0.095, possessing SDSS optical photometry and optical spectroscopy. 5,080 galaxies also have YJHK photometric from DR2 of UKIDSS-LAS \citep[see][]{labarbera2010SPIDER-I/MNRAS.408.1313L}. ETGs are defined as bulge-dominated systems, with passive spectra in their centers. Following \citet{Bernardi2003a}, they select ETGs with {eClass}<0 and {fracDevr}>0.8, where the SDSS spectroscopic parameter {eClass} gives the spectral type of  a galaxy, while the SDSS photometric parameter {fracDevr} measures the fraction of galaxy light that is better fitted by a de Vaucouleurs (rather than an exponential) law. Here we are interested in a subsample of this catalog for which Jeans analysis has been used to derive the dynamical mass inside the \re\ from T+12. We note that the selections are very effective in removing late-type systems \citep[see][]{labarbera2010SPIDER-I/MNRAS.408.1313L}, but do not allow a clear separation of E and S0 galaxy types. However, this has no significant impact on the Jeans analysis results, as discussed in T+12. The final SPIDER sample contains 4,260 galaxies for which information about the quantities as in Table~\ref{tab: features and targets} are available from T+12, and that we briefly summarize here below:
\begin{itemize}
\item $g$- and $r$-band photometry from SDSS-DR6;
\item $R_e$: effective radius obtained from the S\'ersic fit to the SDSS imaging in the K-band. We convert it to 3D by $\re~=~1.35\times~R_{e}$;
\item \ms: total stellar mass \citet{SPIDER-V-stellar-mass-2011AJ....142..118S}, obtained by fitting synthetic stellar population models from \citet{BC03-2003MNRAS.344.1000B} from SDSS (optical) + UKIDS (NIR) using the software \textit{LePhare} \citep{Lephare-Ilbert2006}, assuming a extinction law \citep{Cardelli1989-extinction-law-ApJ...345..245C} and Chabrier IMF. 
\item \sig\ and \sigre: respectively, SDSS fiber velocity dispersion and aperture corrected velocity dispersion within 1 \re  following \citet{C06-cappellari2006};
\item \mdyre\ and \mdre\: respectively, dynamical mass from the Jeans equation and dynamical minus stellar mass in the 3D effective radius (see T+12).
\end{itemize}

In Fig.~\ref{fig: kde-distribution} we compare the distribution of the galaxy features and targets of the nETG sample against the same quantities from SPIDER ETGs selected according to the all set of criteria above. From the distribution, we can see that the SPIDER sample nicely overlaps with most of the distribution of the nETGs, except for the velocity dispersion. This is expected, as discussed in \S\ref{sec:preamble}. All other deviations from the ETG simulated sample of the observed sample should be tracked to the difference of the definitions highlighted in Table~\ref{tab: features and targets}. As anticipated in  sections~\ref{sec:preamble} and~\ref{sec:meas_err}, this matter is related to the ``observational realism'' of simulations. Here we want to check whether the application of uniform selection criteria, such as the one introduced above in this section, can provide reasonably realistic mass predictions, or whether we need physically motivated corrections to empirically align observations and simulations (see, e.g., \citealt{Pillepich2018MNRAS.475..648P}).

\subsubsection{MaNGA DynPop}
\label{sec:dynpop}
Mapping Nearby Galaxies at APO (MaNGA,\footnote{https://www.sdss4.org/surveys/manga/} \citealt{2015OverviewMaNGA}) is a spectroscopic program included in the Sloan Digital Sky Survey, (SDSS), released in the final Data Release 17 (DR17). Unlike previous SDSS surveys based on fiber spectroscopy, MaNGA obtained 3D spectroscopy of $\sim 10,000$ (10k) nearby galaxies, hence providing two-dimensional maps of stellar velocity and velocity dispersion, mean stellar age, and star formation history for an unprecedented sample at $z\sim0$. As no preliminary selections on size, inclination, morphology, or environment were applied, MaNGA is a volume limited sample fully representative of the local universe galaxy population, eventually including all varieties of galaxies as ETGs, LTGs, and also irregular systems. 

For our analysis, we are interested in the DynPop catalog (Z+23), combining the Jeans equation dynamical analysis with stellar population \citep{2023MNRAS.526.1022L} for the full 10k MaNGA sample.\footnote{The MaNGA DynPop catalogs are public on Github: https://manga\-dynpop.github.io/pages/data\_access/.} The stellar dynamics is performed using the Jeans anisotropic modeling (JAM) method \citep{2008Cappellari..JAM,2020Cappellari..JAM}, which was successfully adopted for extended analyses \citep[e.g.,][]{Cappellari-2013-ATLAS3D-XX-2013MNRAS.432.1862C}. The  JAM method contains a final combination of eight different set-ups. Namely: two orientations of the velocity ellipsoid (cylindrically-aligned $\rm JAM_{cyl}$ or spherically-aligned $\rm JAM_{sph}$) and four assumptions for the models' dark versus luminous matter distribution: 1) mass-follows-light \citep[e.g.,][]{2012Cappellari..Nature,2020MNRAS..Shetty}, 2) free NFW dark halo \citep{1996ApJ...NFW}, 3) cosmologically-constrained NFW halo\, 4) generalized NFW dark halo \citep{2001ApJ..gNFW} (i.e., using a free inner DM halo central slope). The catalog contains a series of parameters that we are interested in using in our analysis, and we briefly describe here below (see Z+23, their Appendix~B): 

\begin{itemize}
    \item \ms: total stellar masses from K-correction fit for Sersic fluxes which is from the NSA catalog \citep{Mstar-NSA-Blanton2007,Mstar-NSA-Blanton-2011} with Chabrie IMF. In the Dynpop catalog, they also provide the decomposition of the stellar mass from their total mass model.
    \item \re: 3D radius of the sphere which encloses half the total luminosity, based on JAM.
    \item \sig: effective velocity dispersion within elliptical half-light isophote.
    \item $M_{\rm dyn}(\re)$: dynamical masses derived via JAM.
    \item ``Qual'' flag: this is a quality flag that classifies galaxies according to the ``goodness'' of the dynamical model fit. Qual=-1 means irregular galaxies; Qual=0 no good fit for both velocity ($V$) and $V_{rms}=\sqrt{V^2+\sig^2}$ maps; Qual=1 means acceptable fit for the $V_{\rm rms}$ map. Qual=2 means have a good fit to the $V_{\rm rms}$ but a bad fit to the $v$. Qual=3 means that both $V_{\rm rms}$ and $V$ are well fitted (see Z+23, Sect 5.1).  
\end{itemize}

For our test, we choose as reference model the $\rm JAM_{sph}$ with the generalized NFW as a reference and use the related Qual flag to select the predictive sample.
We start by selecting a sample with quality flag { Qual$\geq$1} (i.e., good fit to either the velocity map, or the velocity dispersion map, or both) from the original sample of 10,296 galaxies. However, we will also report the {\sc Mela} predictions for all galaxies that are not clearly irregular (i.e., Qual$\geq0$) in Appendix~\ref{app:dynpop}. In the same Appendix, we also show the predictions by \mlaall\ as compared against all other JAM models. The Redshift range of the DynPop sample is 0.00-0.17, so for consistency with the training sample used for SPIDER, we further select only the $z<0.1$ systems to start with. 

Using Qual$\geq$1 and $z<0.1$, we are left with 5,737 galaxies. To separate ETGs from LTGs, we adopt the criteria proposed by \citet{MaNGA-DL-morph-2022} and consistently adopted by Z+23:
\begin{itemize}
    \item E: ($\rm P_{LTG}$<0.5) and (T-Type<0) and ($\rm P_{S0}$<0.5) and (VC=1) and (VF=0)
    \item S0: ($\rm P_{LTG}$<0.5) and (T-Type<0) and ($P_{S0}$>0.5) and (VC=2) and (VF=0)
    \item S: ($\rm P_{LTG}$>0.5) and (T-Type>0) and (VC=3) and (VF=0) 
\end{itemize}
where $\rm P_{LTG}$, $\rm P_{S0}$ and $\rm P_{LTG}$ are machine learning probabilities available as SDSS DR17 value added 
catalogs.\footnote{https://www.sdss4.org/dr17/data\_access/value-added-catalogs/?vac\_id=manga-morphology-deep-learning-dr17-catalog.} 

We finally fixed the E+S0 sample as the final DynPop ETG sample. 
According to Z+23, MANGA is not suitable to be effectively used to study dwarf galaxies, due to the low spectral resolution. Although some low-velocity dispersion systems are present in the sample, we decided to discard these ``dwarf'' systems. We finally show the features and target distribution in Fig.~\ref{fig: kde-distribution}, and find also in this case a good overlap with both the nETG and LTGs from TNG. We also see a nice overlap of the DynPop ETG sample with the SPIDER sample, especially looking at the central velocity dispersion distribution.

For the DynPop sample, Z+23 reports the use of Planck15 cosmology.

\subsubsection{SAMI-Fornax Dwarf Survey}
As a reference observational data on dwarf galaxies, we use the recent dynamical sample from the SAMI-Fornax Dwarf Survey (DSAMI, for short), reported in E+22. DSAMI is an integral field, high-resolution $R\sim5000$ survey of dwarf galaxies in the Fornax Cluster. The surveys provide spectroscopical data for the largest sample of low-mass ($10^7-10^8~M_\odot$) galaxies in a cluster to date. The full description of the sample and the spatially resolved stellar radial velocity and velocity dispersion
maps, together with their specific stellar angular momentum are given in \citet{SAMI-Fornax-I-Scott2020}, and
E+22 provides, in Table~1 and~5, the following parameters that we can use as predictive samples:
\begin{itemize}
    \item $M_\star$: the stellar mass within the effective radius, defined by the formula provided by \citet[][Eqn. 3]{Fornax-stellar-mass} with Chabrie IMF:
    \begin{equation}
        \log (M_\star/M_\odot)_e=1.15+0.70(g-i)-0.4M_{\rm r,e}+0.4(r-i)
    \end{equation}
    where $M_{\rm r,e}$ is the absolute magnitude inside an effective radius, assuming $\Omega_m=0.3$, $\Omega_\Lambda=0.7$, h=0.7, as cosmology;
    \item $R_e$: the effective radius obtained from $r$-band GALFIT model \citep{Fornax-size-mag-catalog-2018}. We convert it to 3D by $\re=1.33\times R_{e}$;
    \item $\sigma_e$: the velocity dispersion inside the effective radius. Due to the flat dispersion profile, they use the velocity dispersion with an aperture of 15'' diameter as a proxy of the \sigre;
    \item \mdyre\ and \mdre: respectively, the inferences of the total dynamical mass (see details below) and dark matter inside \re. The former is obtained via a simple mass estimator calibrated on the spherical Jeans equation from W+10 with $h=0.702$ from {WMAP5} cosmology:
\begin{equation}
\label{eq:mre-wolf}
    \mdyre\simeq 0.93\left(\frac{\sigma_e^2}{\rm km^2s^{-2}}\right)\left(\frac{R_{\rm e}}{\rm kpc}\right)10^6 \rm M_\odot.
\end{equation}    
where the \mdre\ is a 3D mass in the 3D half-light radius, although  \sigre\ and $R_{\rm e}$ are both projected, according to W+10.

\end{itemize}

According to W+10, Eq.~\ref{eq:mre-wolf} is a rather robust estimator that is little sensitive to orbital anisotropy and is valid if the projected velocity dispersion profile is fairly flat near the half-light radius. This is a good approximation for most of the observed dwarf kinematics in the E+22 sample, hence we expect it to provide fairly unbiased \mtotre\ estimates.

The total number of galaxies, which matches the training sample limits, in terms of stellar mass and effective radius, is 15 and the distribution is also shown in Fig.~\ref{fig: kde-distribution}.
  
To convert all features and targets on the same scale as the TNG data, we use a distance of 19.7 Mpc for the Fornax cluster, which corresponds to $z\sim$0.005 in Planck15 cosmology, and rescale all dynamical quantities from WMAP5 to this latter cosmology.

\section{The Mass Estimate machine Learning Algorithm}
\label{sec:method}
In this section we describe the principle of the Mass Estimate machine Learning Algorithm ({\sc Mela}) we want to develop in this paper. We first introduce the main architecture and training strategy and then the statistical indicators we will use to assess its performances.

\subsection{Random forests}
As a first model for {\sc Mela} we want to use random forests (RF). This is a powerful method for ensemble learning (the idea of combining the outputs of multiple models through some kind of voting or averaging). In particular, it is suitable for the specific goal of predicting the dark matter properties of galaxies, starting from a list of observations, as we have seen in vM+22, where RF has been the algorithm always picked by the Tree-based Pipeline Optimization Tool \citep[TPOT;][]{TPOT-OlsonGECCO2016}. RF makes use of decision trees and, more specifically, for regression problems, it is based on CART trees. Compared to a simple decision tree, RF results, based on averages from all the decision trees which is made, are more robust and less prone to overfit. We use the package {\it sklearn.ensemble.RandomForestRegressor} and keep the default structural parameters, with 100 trees, after having tested that the performances would not significantly change with the adoption of any variation around the default set-up. Finally, in order to make the results reproducible, we set the structure parameters random\_state=1.

\subsection{Training {\sc Mela}}
\label{sec:training}

In this section, we outline the various {\sc Mela} configurations we will utilize throughout the paper, each corresponding to different training samples. As previously mentioned, while the pipeline remains consistent and relies on the same features and targets, we plan to employ distinct training samples. These samples are anticipated to offer more specialized and consequently accurate predictions for individual classes. To differentiate between the various configurations, we will use the following \mla\ extensions:
\begin{itemize}
\item \mlaall. The {\sc Mela} trained using the full original TNG100 sample, with no classes. Training size: 339,504 galaxies.
\item \mlaetg. The {\sc Mela} trained using the ``normal'' ETG sample. Training size: 21,416 galaxies.
\item \mladw. The {\sc Mela} trained using the ``dwarf'' ETG sample. Training size: 70,859 galaxies.
\item \mlaltg. The {\sc Mela} trained using the LTG sample. Training size: 247,229 galaxies.
\end{itemize}
For the self-prediction (in \S\ref{sec:self_pred}), we allocate 80\% of the complete training dataset as a training set and the remaining 20\% as a test set to assess the performance of \mla. However, when applying the \mla\ to real observations, we use the entire dataset for training. 

Looking at the training sample sizes of the different \mla s above, we remark that they seem rather unbalanced, being the LTG sample the largest one in the TNG\_all sample. In principle, being the number counts of the individual classes a realistic representation of a complete sample of galaxies, their true distributions can be considered as a ``prior'' which realistically describes the observed samples.
However, one can expect that such unbalanced training samples can affect the relative performances of the different trained \mla s above. To check that, we will also adopt a more ``balanced'' training/testing approach to verify if this can impact the different \mla s predictions. 
In particular, in \S\ref{sec:self_pred}, we will use a sample consisting of 21,000 galaxies per class, which is randomly selected from the complete dataset of each respective class. This balanced sample is aligned with the less abundant test set among the three classes (i.e., the nETGs, with 21,416 entries). 

In this case, the class samples are divided into 16,800 galaxies for training and 4,200 for testing. We will compare both training approaches, referred to as ``full-counts'' and ``balanced-counts'' training samples, against the respective test samples.

\subsection{ML evaluator metrics}
To evaluate the performance of the {\sc Mela}, in terms of accuracy and precision, we use four different statistical estimators (see symbol definitions at the end of this list): 
\begin{enumerate}
\item The coefficient of determination $R^2$: 
\begin{equation}
R^2=1-\frac{\rm RSS}{\rm TSS}=1-\frac{\sum_{i=0}^N (y^i_{\rm pred}-y^i_{\rm true})^2}
{\sum_{i=0}^N (y^i_{\rm pred}-\hat{y}_{\rm true})^2}
\end{equation}

\item The Mean Absolute Error (MAE):
\begin{equation}
{\rm MAE}=\frac{1}{N}\sum_{i=0}^N|y^i_{\rm pred}-y^i_{\rm true}|
\end{equation}

\item The Mean Standard Error (MSE):
\begin{equation}
{\rm MSE}=\frac{1}{N}\sum_{i=0}^N(y^i_{\rm pred}-y^i_{\rm true})^2
\end{equation}

\item  The Pearson correlation coefficient $\rho$, already introduced in Eq.~\ref{eq:Paerson1}, but re-defined here as:
\begin{equation}
\rho=\frac{cov(y_{\rm pred},y_{\rm true})}{\sigma(y_{\rm pred})\sigma(y_{\rm true})}.
\end{equation}

\item The Median of Bias (MdBias), defined as:
\begin{equation}
{\rm MdBias}={\rm median(y_{pred}-y_{true})} 
\end{equation}

\item The outlier fraction: the fraction of prediction exceeding $\sim 2\sigma$ errors of the typical mass estimates from dynamical analyses. In particular, we have found that this corresponds to $\pm0.3$ dex for DynPop, being 0.15 dex about the average errors in that sample (see K+23). In case of log-normal errors, we expect a 5\% outliers, as an acceptable outlier fraction.
\end{enumerate}

In all equations above, $N$ represents the total number of data points, while the variable y refers to the output of the \mla, which is also known as the target (hence, $y^i$ represents the value of the $i$-th output), the subscripts ``true'' and ``pred'' indicate the true and predicted values respectively and $\hat{y}$ denotes the mean value of y. Furthermore, RSS indicates the residual sum of squares and TSS indicates the total sum of squares. $R^2$ is a statistical measure used to assess the goodness of fit of a regression. By definition, $R^2$ ranges from 0 to 1, with $R^2=1$ meaning the perfect fit. However, in practical cases  RSS can be larger than TSS, making $R^2<0$. In these cases, the ML fails to make the prediction. It is important to note that $R^2$ alone does not quantify the quality of a regression model; for example,  it does not consider the complexity of the model, the significance of individual predictors, or the presence of overfitting. Therefore, $R^2$ in used in conjunction with other metrics. MSE and MAE are both commonly used metrics for evaluating the performance of a regression model. They quantify the average magnitude of errors between the predicted values and the actual values. Both MSE and MAE are defined to be non-negative, with lower values indicating better performances of the ML tool. Finally, $\rho$ describes the linear correlation between the true value and the predicted value. It ranges from -1 (perfect anti-correlation) to 1 (perfect correlation). For the purpose of our analysis, when comparing predictions with ground truth, the closer $\rho$ and $R^2$ are to 1 and the closer MAE and MSE to 0, the better the performances of the \mla\ are.

\section{Results}
\label{sec:results}
\begin{figure*}
\centering
\includegraphics[width=1.8\columnwidth]{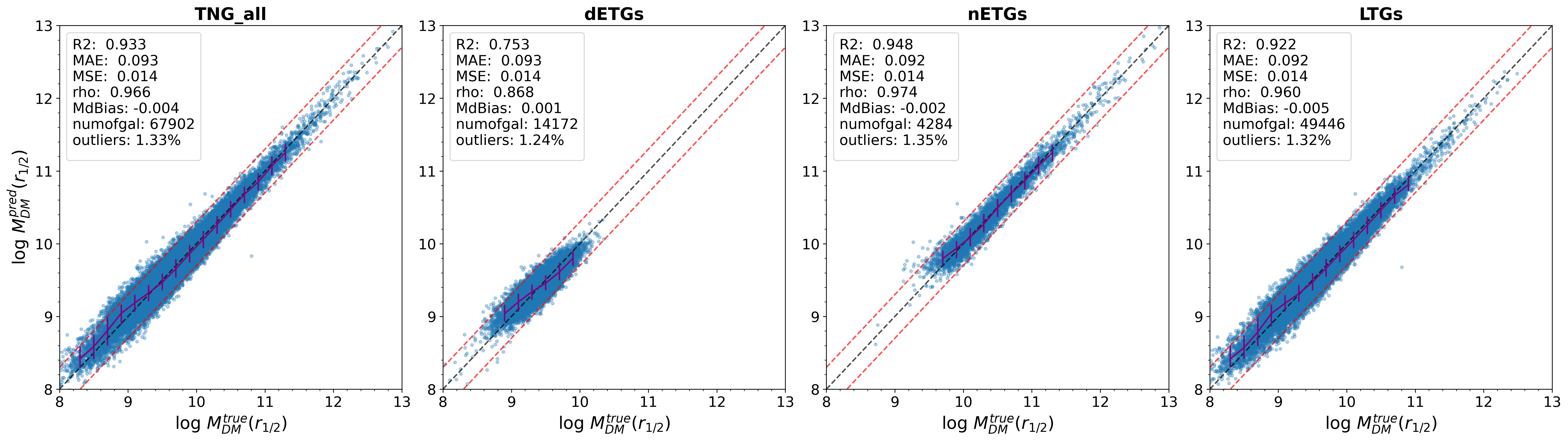}
\includegraphics[width=1.8\columnwidth]{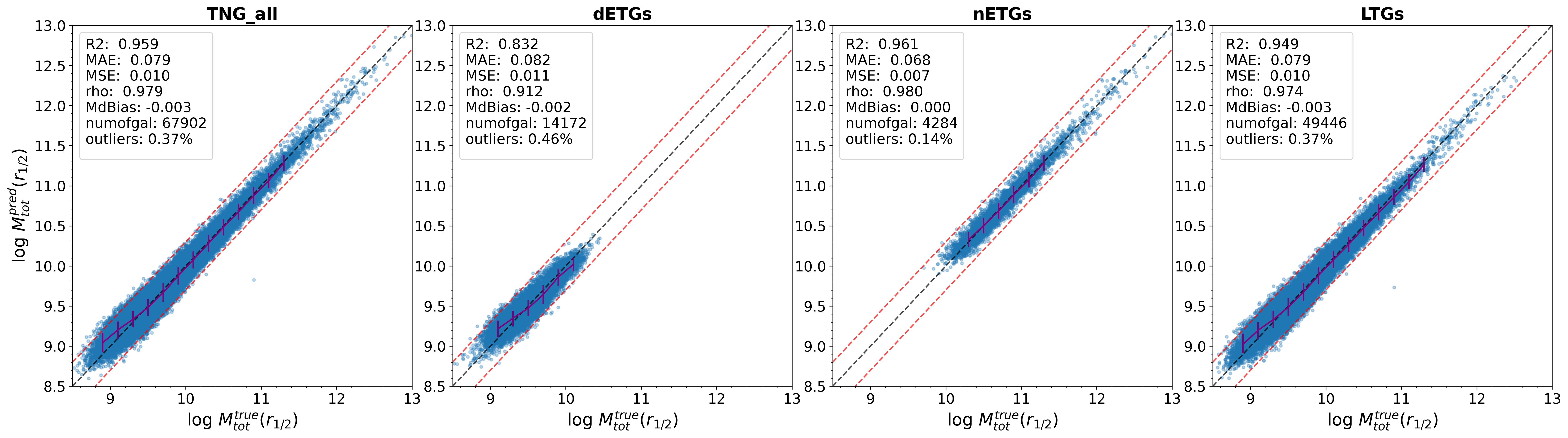}
\caption{Self-prediction test using full features as indicated in Table~\ref{tab: features and targets}, with the full-counts training sample incorporating added measurement errors, as described in \S\ref{sec:meas_err}. Top row: Target is \mdre. Bottom row: Target is \mtotre\ . The results without measurement errors are presented in  Appendix~\ref{app:errors}. The data is divided into 80\% for training and 20\% for testing. The x-axis represents the true values, while the y-axis represents the predicted values. ``numofgal'' is the number of the test set. The purple error bar represents the 16\%, 50\%, 84\% percentiles as a function of $M^{\rm true}(r_{\rm 1/2})$, with a bin size of 0.2 dex. The red dashed line is $\pm$ 0.30 dex (corresponding to $\sim2\sigma$ errors, see text). Outliers are defined as the fraction of data outside the red dashed line. In the case of accurate predictions, the data points are expected to lie along the dotted 1-to-1 line.}
\label{fig: self-prediction-simu}
\end{figure*}
\begin{figure*}
\centering
\includegraphics[width=1.8\columnwidth]{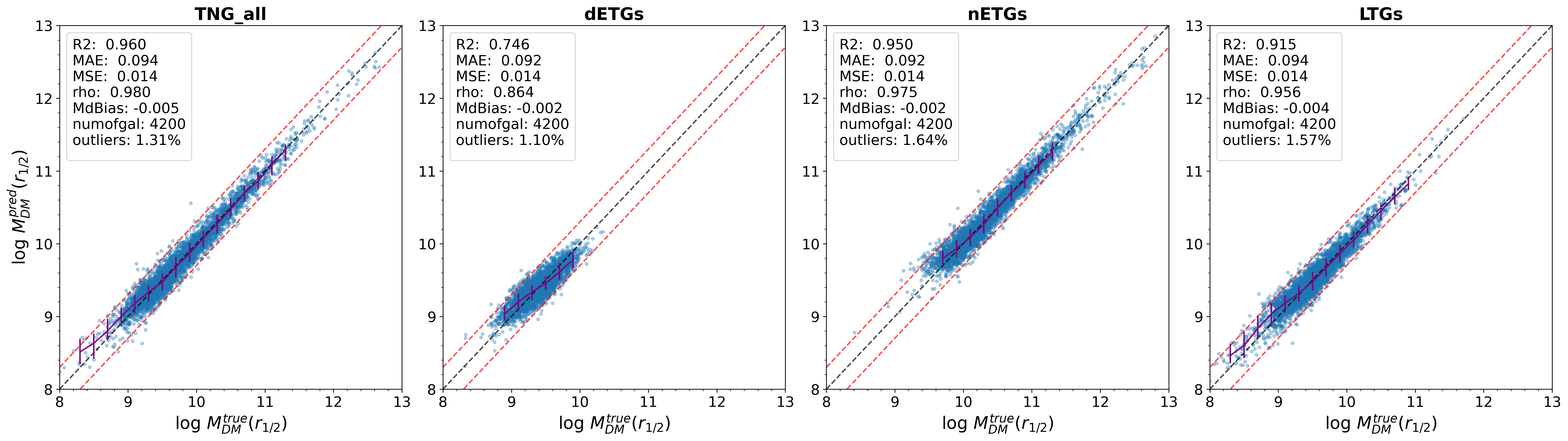}
\includegraphics[width=1.8\columnwidth]{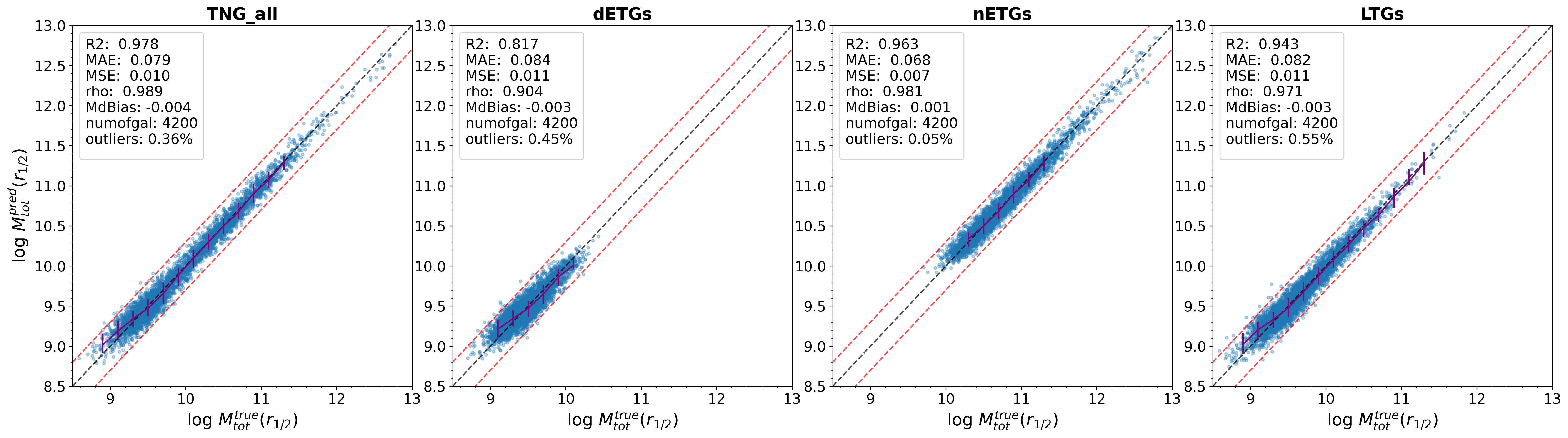}
\caption{Self-prediction test performed using the full set of features and a balanced-counts training sample, which includes measurement errors. The training-test sample has been adjusted to maintain an equal number of entries across all samples in Fig.~\ref{fig: self-prediction-simu} through random selection, aligning with the less populated class (nETGs). The training set consists of 80\% of the randomly selected subsample (16,800 entries), while the remaining 20\% (4,200 entries) is allocated for testing. Top row: Target is $M_{\rm DM}(r_{\rm 1/2})$. Bottom row: Target is \mtotre\ .}
\label{fig: self-prediction-simu-N4000}
\end{figure*}
\begin{figure*}[h]
\centering
\includegraphics[width=1.6\columnwidth]{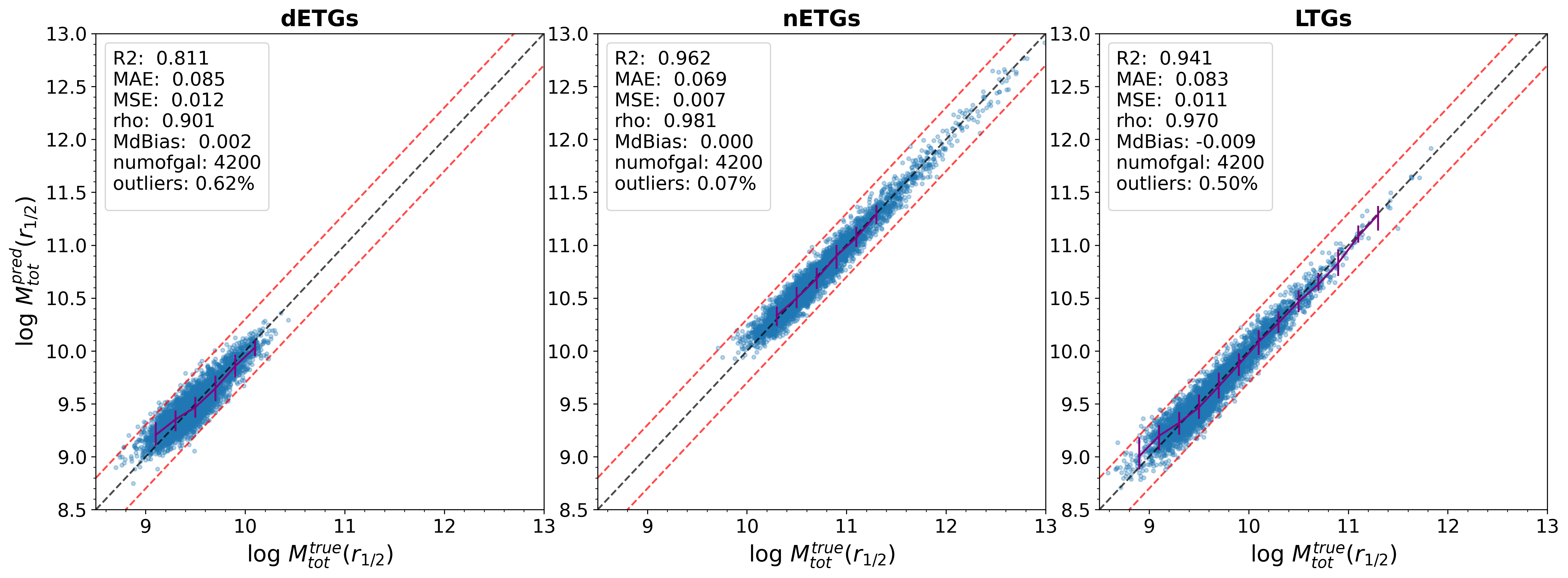}
\caption{Self-prediction test of the \mlaall\ is performed using the full set of the features, and the \mtotre\ as the target. As explained in \S\ref{sec:training}, for this test the training utilizes balanced-counts training samples. These training samples comprise $21,000\times 80\%=16,800$ galaxies for each class (i.e., dETGs, nETGs, LTGs), accompanied by 4,200 galaxies for testing. The predictions of the entire TNG sample are presented as the self-prediction test in Fig.~\ref{fig: self-prediction-simu-N4000}.}
\label{fig: self-prediction-simu-mela_all}
\end{figure*}

In this section, we first evaluate the performance of the {\sc Mela} to predict the total and the dark mass over a test sample derived by the TNG simulation. In our first analysis (see vM+22) we have demonstrated that ML can effectively predict the dark matter content of galaxies as a whole (i.e., without breaking these into different classes). Here, we want to check the performances of the {\sc Mela} on the different galaxy types. We start by using test samples containing simulated galaxies from one of the three TNG100 classes and checking the predictions using the corresponding trained \mla\ (see \S\ref{sec:training}) (i.e., the \mlaetg, the \mladw, and the \mlaltg) in turn, and compare these against the \mlaall\ trained over the full TNG sample. After having tested {\sc Mela} which trained on the all features, we also checked the performance of the algorithm with different combinations of a smaller number of them. As discussed in vM+22 and in \S\ref{sec:intro}, some of the features reported in Table~\ref{tab: features and targets} might be redundant and bring little contribution to the accuracy of the predictions. Hence, we will check whether we can find a minimal sample of features that can provide accurate enough predictions, for each of the three galaxy classes. 

\subsection{Training and testing on TNG100: Self-prediction}
\label{sec:self_pred}

In this section, we report the performances of {\sc Mela} on the three test samples from the different classes of simulated TNG100 galaxies, using the full set of features: $g$, $r$, \ms, \re, \sig. Since the training and test samples are derived from the same simulated dataset, we dub this test ``self-prediction''. For the sake of brevity, we illustrate here only the detailed results of the \mla\ trained with measurement errors included, while we will shortly report the results with no errors in Appendix~\ref{app:errors}.

\subsubsection{Self-prediction}

We start by showing the results by the ``full-counts'' training sample defined in \S\ref{sec:training}. In Fig.~\ref{fig: self-prediction-simu}, we show the predicted values of the two targets versus ground truth. Accuracy-wise, we can clearly see that nETGs and LTGs have $R^2$ and $\rho$ both larger than 0.9 for the two targets, although for \mtotre\ the indicators are systematically better than the \mdre. For the dETGs, the $R^2$ is $\sim 0.75$ and $\sim 0.83$, while $\rho\sim0.87$ and $\sim 0.91$, for \mdre\ and \mtotre, respectively (i.e.,  smaller than the other two classes). Looking at the same figure,  all predictions look quite nicely aligned to the 1-to-1 relation with a negligible number of outliers ($<2\% $) consistent with a log-normal scatter. We argue that the lower $R^2$ and $\rho$, for the dETGs, come from a lower correlation in the plots, due to the smaller mass range covered by the dETG sample. The scatter, on the other hand, as measured by the MAE and MSE, is rather similar for all the three classes (MAE$\sim0.07-0.09$, MSE$\sim0.007-0.014$) regardless of the target, suggesting very similar performances of \mla\ for all classes. We also note the emergence of a systematic deviation (within a few percent) at $\log$\mtotre$/M_\odot<9.3$, due to some incompleteness effect on objects close to the low-mass limit.

In vM+22 (see their Table~1), for TNG\_all with target \mdre\ they found $R^2\sim0.98$, $\rho\sim0.99$, MAE$\sim0.04$ and MSE$\sim0.004$ in the ``joint analysis" (i.e., using all features). Our accuracy and overall scatter is obviously larger because we are now considering the measurement errors, which were not taken into account in vM+22. The inclusion of errors ultimately returns a more realistic forecast of the accuracy and scatter we might expect in real applications. In Appendix~\ref{app:errors}, we will present our results without considering the measurement errors, to compare directly with what was done in vM+22, and show that these are in full agreement with the latter. 

Next, we move to the results of the ``balanced-counts'' training sample, as shown in Fig.~\ref{fig: self-prediction-simu-N4000}. Here we see that all self-predictions look almost unchanged, with all the statistical indicators remaining consistent with 1\%, as seen by comparing the $R^2$, MAE, MSE, and $\rho$ values in the insets with the ones in Fig.~\ref{fig: self-prediction-simu}. This shows significant stability of the \mla s with respect to the ``prior'' parameter distributions of the training samples. Most of all, the \mlaall\ performance remains insensitive to the relative balance of the three galaxy classes. This confirms the evidence that \mla\ can fully capture the diversity of the correlations of the three galaxy classes (nETG, dETGs, and LTGs) even if mixed together. To demonstrate that, in Fig.~\ref{fig: self-prediction-simu-mela_all} we show the predictions of the \mtotre\ for the same test sample of nETG, dETGs and LTGs, but using \mlaall\ for all classes. Compared to the same quantities predicted in Fig.~\ref{fig: self-prediction-simu-N4000} by \mlaetg, \mladw\ and \mlaltg, respectively, we find that the resulting $R^2$ are almost indistinguishable for the three classes.

This result seems rather surprising for two main reasons: 1) one could expect that the tilt of the different scaling relations in Fig.~\ref{fig:correlations} should give more sensitivity to \mla\ to the different galaxy classes, in particular for the dETGs, showing the more deviating correlations in Fig.~\ref{fig:correlations}; 2) one would also expect that the different distributions of the galaxy observed quantities (features), in Fig.~\ref{fig: kde-distribution}, should impact the prediction of \mla\ moving from one sample to the other. Instead, 
the result above seems to show that the \mlaall\ can use the combined information of all features from the different galaxy species, regardless of the specific correlations they have with the DM and the total mass, within the classes. Also, this result seems to show that \mlaall\ can correctly make predictions if the features and targets of a predictive sample are included in the dominion of the training sample, regardless of the detailed distribution of features and targets of the former of the two samples. 
We will return to this point in the discussion in \S\ref{sec:discussion}. 

As a final note of this section, one might guess that the equivalence of the performances of \mlaall\ and the customized \mla s can be a consequence of training over features (i.e., \sig\ and \ms) that are used to split the samples. First, this is true for the ETGs, for which we indeed use a condition on \sig\ and \ms\ to separate the nETGs from the dETGs (see Table~2), but not for LTGs, which are selected only according to the sSFR. Second, as discussed above, the \mla s eventually learns from correlations among features and targets, and thresholds in the features only define the range of the correlations to use in training, which cannot tightly correlate with the mass of the individual galaxies. To check that, we conduct two tests. First, we compare the results obtained from \mlaall\ and {\sc Mela}\_{\sc etg} (where \mla\ was trained on the whole ETG sample) applying on the whole ETG sample (i.e., applying no selection based on the features used for the predictions). We find that the $R^2$ values are nearly identical, meaning that \mla\ performs equally if any of the features used to split the sample is involved in the training. Second, we dig more into the details of the effect of the features in the splitting by predicting on the nETG, where we have used the $\sigma$ and \ms\ in their selection, but predicting using the {\sc Mela}\_{\sc etg} above, trained on the whole nETG+dETG sample. Here we find, again, almost no change in the results, this also shows that the information used to split the sample is not used by the \mla s to predict.

\subsubsection{Optimizing the features combination}
\label{sec:feat_imp}
\begin{table}
\small
\centering
\begin{tabular}{clc}
\toprule
$R^2$ & features & num \\ 
\hline
\multicolumn{3}{c}{\mlaetg}\\
\hline
0.8785 & \re & 1 \\ 
0.9407 & \re, $M_\star$ & 2 \\ 
0.9457 & \re, $M_\star$, $\sigma$ & 3 \\ 
0.9479 & g, \re, $M_\star$, $\sigma$ & 4 \\ 
0.9485 & g, r, \re, $M_\star$, $\sigma$ & 5 \\ 
\hline
\multicolumn{3}{c}{\mladw}\\
\hline
0.0983 & $\sigma$ & 1 \\ 
0.6807 & \re, $\sigma$ & 2 \\ 
0.7309 & \re, $M_\star$, $\sigma$ & 3 \\ 
0.7513 & g, \re, $M_\star$, $\sigma$ & 4 \\ 
0.7528 & g, r, \re, $M_\star$, $\sigma$ & 5 \\ 
\hline
\multicolumn{3}{c}{\mlaltg}\\
\hline
0.5698 & \re & 1 \\ 
0.9083 & \re, $M_\star$ & 2 \\ 
0.9167 & \re, $M_\star$, $\sigma$ & 3 \\ 
0.9211 & r, \re, $M_\star$, $\sigma$ & 4 \\ 
0.9217 & g, r, \re, $M_\star$, $\sigma$ & 5 \\ 
\hline
\multicolumn{3}{c}{\mlaall}\\
\hline
0.6113 & $M_\star$ & 1 \\ 
0.9097 & \re, $M_\star$ & 2 \\ 
0.9275 & \re, $M_\star$, $\sigma$ & 3 \\ 
0.9320 & g, \re, $M_\star$, $\sigma$ & 4 \\ 
0.9324 & g, r, \re, $M_\star$, $\sigma$ & 5 \\ 
\bottomrule
\end{tabular}
\caption{Accuracy as a function of the number of features for the \mdre, taking measurement errors into account. Each row displays the optimal feature combination for the corresponding feature count. The results without considering measurement errors can be found in Table~\ref{tab:feature_importance_app_1}.}
\label{tab: DM-number of features and accuracy}
\end{table}

\begin{table}
\small
\centering
\begin{tabular}{clc}
\toprule
$R^2$ & features & num \\ 
\hline
\multicolumn{3}{c}{\mlaetg}\\
\hline
0.8957 & $M_\star$ & 1 \\ 
0.9544 & \re, $M_\star$ & 2 \\ 
0.9584 & \re, $M_\star$, $\sigma$ & 3 \\ 
0.9604 & r, \re, $M_\star$, $\sigma$ & 4 \\ 
0.9606 & g, r, \re, $M_\star$, $\sigma$ & 5 \\  
\hline
\multicolumn{3}{c}{\mladw}\\
\hline
0.4242 & $\sigma$ & 1 \\ 
0.7139 & \re, $M_\star$ & 2 \\ 
0.8203 & \re, $M_\star$, $\sigma$ & 3 \\ 
0.8306 & g, \re, $M_\star$, $\sigma$ & 4 \\ 
0.8326 & g, r, \re, $M_\star$, $\sigma$ & 5 \\ 
\hline
\multicolumn{3}{c}{\mlaltg}\\
\hline
0.7803 & $M_\star$ & 1 \\ 
0.9405 & \re, $M_\star$ & 2 \\ 
0.9464 & \re, $M_\star$, $\sigma$ & 3 \\ 
0.9487 & r, \re, $M_\star$, $\sigma$ & 4 \\ 
0.9490 & g, r, \re, $M_\star$, $\sigma$ & 5 \\
\hline
\multicolumn{3}{c}{\mlaall}\\
\hline
0.8238 & $M_\star$ & 1 \\ 
0.9419 & \re, $M_\star$ & 2 \\ 
0.9560 & \re, $M_\star$, $\sigma$ & 3 \\ 
0.9581 & r, \re, $M_\star$, $\sigma$ & 4 \\ 
0.9585 & g, r, \re, $M_\star$, $\sigma$ & 5 \\
\bottomrule
\end{tabular}
\caption{Accuracy as a function of the number of features for the \mtotre\ , taking measurement errors into account. Each row displays the optimal feature combination for the corresponding feature count. The results without considering measurement errors can be found in Table~\ref{tab:feature_importance_app_2}.}
\label{tab: DYN-number of features and accuracy}
\end{table}

\begin{figure}
\centering
\includegraphics[width=1.0\columnwidth]{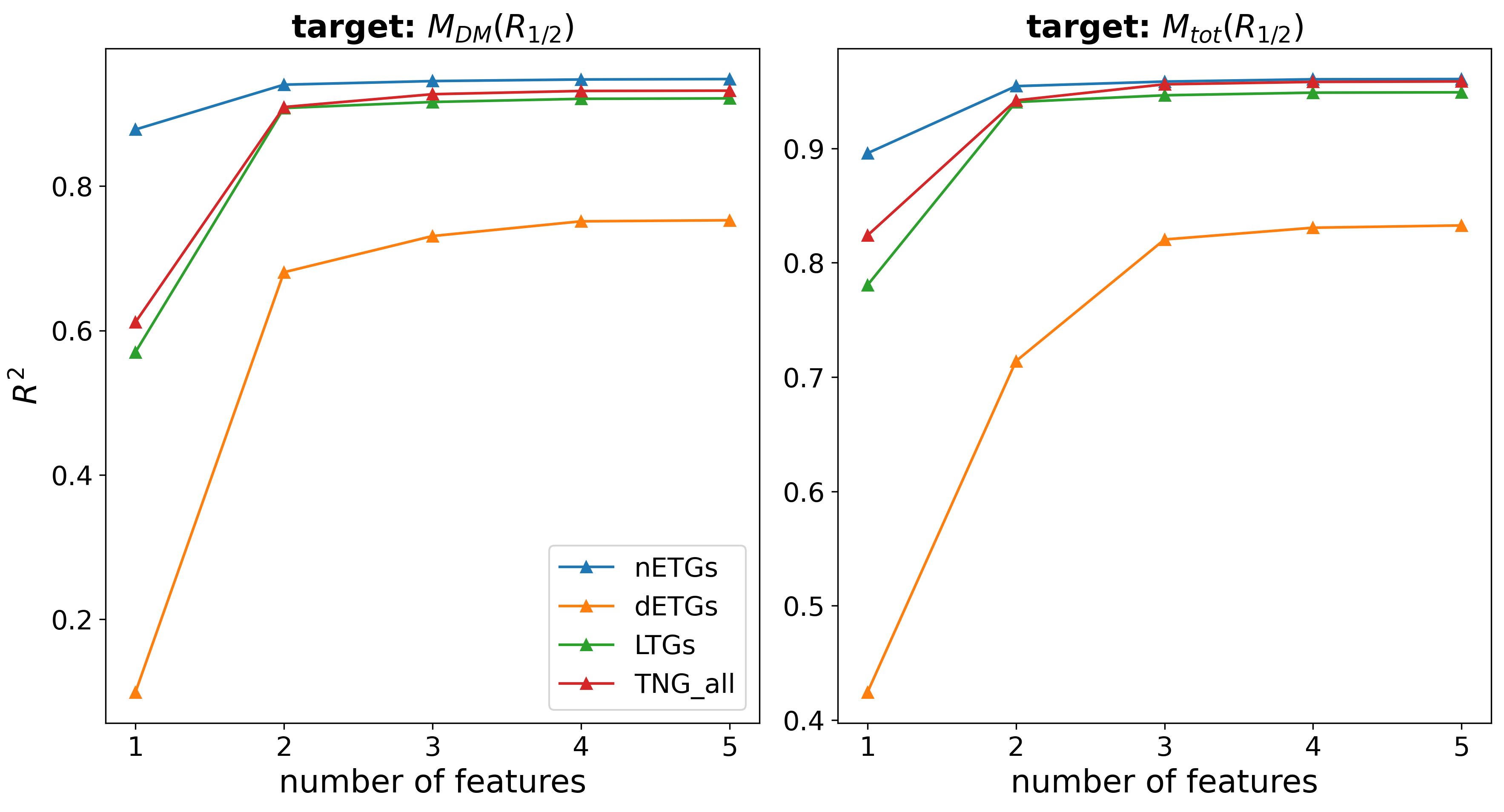}
\caption{Accuracy as a function of the number of features for  $M_{\rm DM}(\re)$ and \mtotre, taking measurement errors into account. This figure is based on Tables~\ref{tab: DM-number of features and accuracy} and ~\ref{tab: DYN-number of features and accuracy}. The results without considering measurement errors can be found in  Appendix~\ref{app:errors}.}
\label{fig: DM-feature-accuracy}
\end{figure}
One of the aims of this analysis is to find the optimal combination of the features needed to correctly predict the dark matter and/or the total mass of galaxies. In vM+22 we have grouped the canonical features one can collect from galaxy surveys into ``Photometric'' (including a series of broad optical and NIR bands), ``Structural'' (including the stellar mass and the \re), and ``Kinematical'' (including \sig\ and a global circular velocity parameter), for a total of 14 features. We have shown that Structural and Photometric features are particularly effective in the prediction of \mdre, being typically $R^2\sim0.88-0.94$, and that the best predictions are found when using all groups of features ($R^2\sim0.98-0.99$). In vM+22 we did not try to optimize the feature selection, although we noticed that some groups of features might be more relevant than others in the feature importance analysis. 

Here we want to check in detail the impact of the use of the individual features on the accuracy of the predictions. This is a heuristic ``feature importance'', which is more oriented to accuracy optimization by avoiding redundant features that might add noise rather than effective predictive power. This becomes particularly important for real applications, where the predictions might suffer more from the feature noise introduced by measurement errors (see above). In Tables~\ref{tab: DM-number of features and accuracy} and~\ref{tab: DYN-number of features and accuracy}, we report the $R^2$ estimator for \mdre\ and \mtotre, respectively, obtained by changing the number of features considered and selecting the first ranked feature combination giving the highest accuracy among all possible combinations allowed for that particular number of features. The content of both tables is graphically summarized in Fig.~\ref{fig: DM-feature-accuracy}, where we show the $R^2$ as a function of the number of features in Tables~\ref{tab: DM-number of features and accuracy} and~\ref{tab: DYN-number of features and accuracy}, regardless the features.  

The first thing to notice from Fig.~\ref{fig: DM-feature-accuracy} is that the accuracy ($R^2$) of both the targets, \mdre\ and \mtotre , reaches a ``plateau'' for all the galaxy groups with just 3 features, which eventually are the same for all classes and for both targets (i.e., \re\, \ms\, and \sig ), although the two highest ranked features can be different for the different classes and targets (see num=2 rows in Tables~\ref{tab: DM-number of features and accuracy} and~\ref{tab: DYN-number of features and accuracy}). According to the same tables, the stellar mass is the primary feature in almost all cases except the \mdre\ predictions of the dETGs, where \sig\ is the primary parameter. Interestingly, while \sig\ appears to be important for the dETGs for both targets, it seems to be less important for the nETGs, LTGs, and the full sample, TNG\_all, where it starts contributing to the predictions only after the stellar mass and the effective radius. The result of TNG\_all, in particular, seems to be consistent with vM+22, which also discussed the kinematics to have a feature importance in the DM predictions at the \re\, lower than the ``structural'' parameters (including \ms\ and \re ).

\begin{figure*}
\centering
\includegraphics[width=1.9\columnwidth]{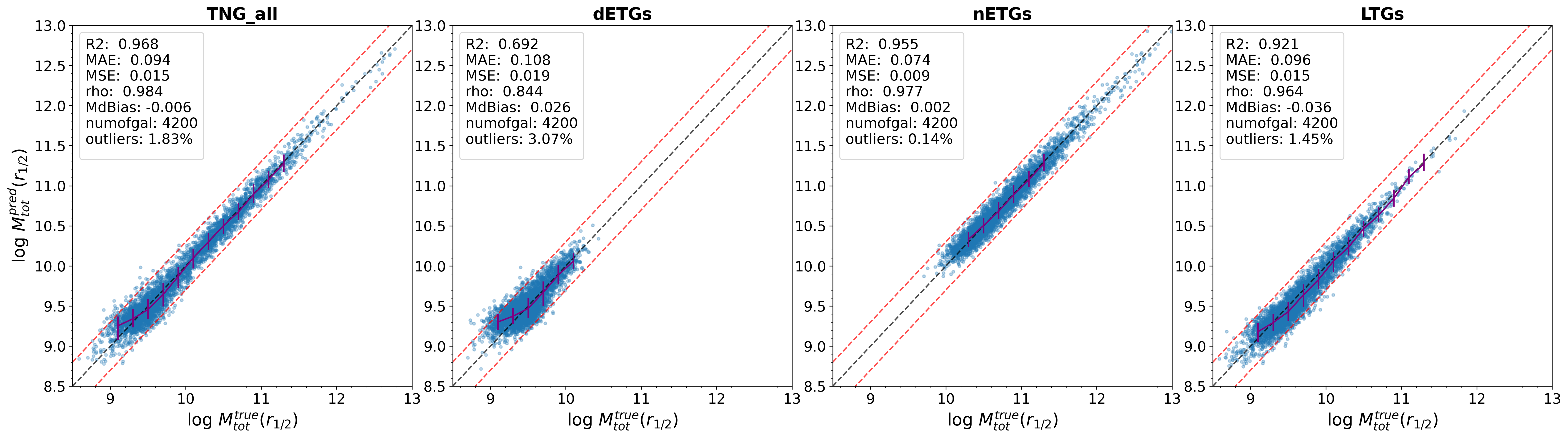}
\caption{Self-prediction test of the \mlaall\ with the target being \mtotre. This test uses the balanced-counts training sample and takes measurement errors into account. The test focuses on only two features: \re\ and \ms.}
\label{fig: self-prediction-simu-2features}
\end{figure*}

A second important result is that differently from the dETGs, which{, at least for \mtotre ,} clearly need three features to reach the ``accuracy plateau'', the nETGs and LTGs both seem to need only two features (i.e., the stellar mass and the \re) to reach the same plateau. This finding has interesting implications, which we need to explore further, in particular, on the real galaxies (see \S\ref{sec:pred_real}). 
For instance, this might be a reflection of the strong correlation of the \ms\ and the \mdre, which scores the highest in the correlation matrix in Fig.~\ref{fig: heatmap-simuonly}. However, this might be a partial explanation, as we notice that there is no direct connection between the feature ranking in Tables~\ref{tab: DM-number of features and accuracy} and~\ref{tab: DYN-number of features and accuracy}, with the correlation coefficients. For instance, for the TNG\_all sample in Fig.~\ref{fig: heatmap-simuonly}, the second highest correlation of the \mdre\ is \sig, but this latter, as commented before, is the third feature kicking in the feature ranking. 
Finally, we stress, that in selecting the features bringing the highest gain in accuracy in Tables~\ref{tab: DM-number of features and accuracy} and~\ref{tab: DYN-number of features and accuracy}, in some cases the difference among features is rather small, meaning that some features are just as good as others to make accurate predictions. This becomes clear when moving to the features that score fourth or higher. On the other hand, for the features ranked below the third, the feature ranking is rather robust (i.e., the first of the second-ranked features provides a larger gain in accuracy with respect to other features).

An obvious conclusion of this ``feature ranking'' test is that the \mla s do not need all the features in Table~\ref{tab: features and targets} to accurately predict the central total and dark matter content of galaxies. From Tables~\ref{tab: DM-number of features and accuracy} and~\ref{tab: DYN-number of features and accuracy} we can see that the combination of the 3 features [\re, \ms, \sig] is sufficient for all galaxy classes. 

Using only these 3 features, we also notice that \mlaall\ reaches the highest accuracy for both targets (i.e., $R^2\gsim0.93$ for the \mdre\ and $R^2\gsim0.96$ for the \mtotre). This copes with the result found in Fig.~\ref{fig: self-prediction-simu-mela_all} about the superiority of the performances of \mlaall\ with respect to customized \mla s. Based on these results, for the test on the real galaxies in \S\ref{sec:pred_real}, we decide: 1) to use \mlaall\ as a unique tool for all galaxy species, unless otherwise specified; 2) to use only [\re, \ms, \sig] as training/testing features.

Although these latter are standard physical products of imaging and spectroscopic surveys, it is still interesting to check the effective ability of the \mlaall\ to minimize the input information needed to make reliable predictions of the \mdre\ and \mtotre, with respect to the customized \mlaetg, \mlaltg, \mladw. In Fig.~\ref{fig: self-prediction-simu-2features} we show the predictions of \mlaall, trained on the ``balanced-counts'' sample using only \re\ and \ms\ as features. We can see that the prediction of dwarf galaxies has strongly degraded with respect to the same predictions using the 3 features in Fig.~\ref{fig: self-prediction-simu-mela_all} as the $R^2$ is decreased by $\gsim19\%$ (0.687 vs. 0.815), the MAE increased by $\gsim24\%$ (0.104 vs. 0.084), and the MSE increased up to $\sim73\%$ (0.019 vs. 0.011) for dETGs. On the other hand, as suggested by Table~\ref{tab: DYN-number of features and accuracy}, for the nETGs and LTGs the $R^2$ decreases by less than 3\% (0.956 vs. 0.963 for nETGs and 0.924 vs. 0.943 for LTGs) and the MAE and MSE increase by $<9$\% (0.074 vs. 0.068 for nETGs and 0.095 vs. 0.082 for LTGs) and $<16$\%  (0.009 vs. 0.007 for nETGs and 0.014 vs. 0.011 for LTGs), which are all much lower than the dETGs. Finally, the TNG\_all sample, despite keeping the least accuracy degradation ($<2$\% in R$^2$, i.e., 0.967 vs. 0.978), shows the largest increase in scatter (MAE: 0.094 vs. 0.078 and MSE: 0.015 vs. 0.010). This allows us to conclude that, when using \ms, \re\, and \sig\ as features, or even adding other features like the photometry, we can have a better accuracy using the \mlaall\ versus customized \mla s for each class, while the customized \mla s work equally accurately if the number of features is suboptimal (e.g., using only {\ms\ and \re},  at least for nETGs and LTGs). 

Finally, following the same logic of feature optimization, in \S\ref{sec:other_conf} we will explore other possible combinations of them,   excluding some that are more difficult to measure (e.g., \sig\ for dETGs) or more prone to systematics (e.g., stellar mass). To do that, we will test these combinations both on the TNG and the real datasets.

\subsection{Prediction on real data}
\label{sec:pred_real}
In this section, we can finally apply the \mla\ to the datasets introduced in \S\ref{sec:data}. This is the first attempt we are aware of, a ML tool trained on simulations is applied to perform mass predictions of galaxies. The fundamental premise here is that simulations are based on complex physical processes, which serve as a physically motivated ground truth. Alternatively, we could use real mass estimates and observables (see, e.g., \S\ref{sec:robustness}), but this would make \mla\ learn how to predict masses mimicking the process of dynamical modeling, including their assumptions and systematics. As discussed in \S\ref{sec:intro} this is not our goal, as we want to rather provide an orthogonal method to use for comparison with standard tools. 

As mentioned earlier, for the real galaxies, the obvious target to use is the \mtotre\ as this is a rather standard diagnostic for classical dynamical analysis of galaxies. We have also discussed in \S\ref{sec:preamble} that, at least for the dynamical analysis of ETGs, the usual assumption is the absence of gas, which, instead, cannot be excluded in the simulations. This brought us to define the ``augmented'' dark mass, $\tilde{M}_{\rm DM}(\re)$, in \S\ref{sec:preamble}, that will be tested in Appendix~\ref{app:mass_DM}. Here we anticipate that the line of arguments we want to use is that, if we can demonstrate that \mla\ is able to 1) correctly predict the total mass in galaxies, under the assumption that the dynamical analyses provide unbiased estimates of the galaxy total masses, and 2) also predict the augmented dark matter, then 3) we will deliberately conclude that also the DM estimates are correct, specifically in the context of the cosmological framework provided by the TNG100 simulation. 

As discussed in the \S\ref{sec:intro}, the combination of stellar masses, baryonic mass, and dark matter in a galaxy is the complex interplay of the cosmological parameters and the galaxy formation recipes, driving the star formation efficiency in galaxies. Hence, the fact that \mla\ can provide consistent predictions of the total mass of real galaxies is not a trivial result. For sure the predictions of both the \mtotre\ and \mdre\ are model-dependent by definition (i.e., those   expected exclusively in the cosmology + feedback model of the TNG100). Hence, if this cosmology/baryon physics mix is different than the one beyond the dynamical inferences of the real sample (e.g., the combination of the choice of the cosmological parameters and assumptions on the dark matter properties in the models), we should not expect \mla\ to return predictions consistently with the dynamical models. As far as the cosmological parameter choice concerned, {we have uniformed the units of the quantities (e.g., the distances and the stellar masses) that have a direct dependence on the cosmological parameters}, by aligning the real datasets to the TNG cosmology (i.e., Planck2015, see \S\ref{sec:intro}). Hence, we can expect that most of the deviation of the \mla\ predictions from the classical mass estimates can be tracked either to the assumptions behind the mass estimators from the real sample side, or the combination of the (cosmological + feedback) model + observational realism, from the simulation side. We will return to this point in detail in \S\ref{sec:discussion}.

\begin{figure*}[ht]
\centering
\includegraphics[width=1.9\columnwidth]{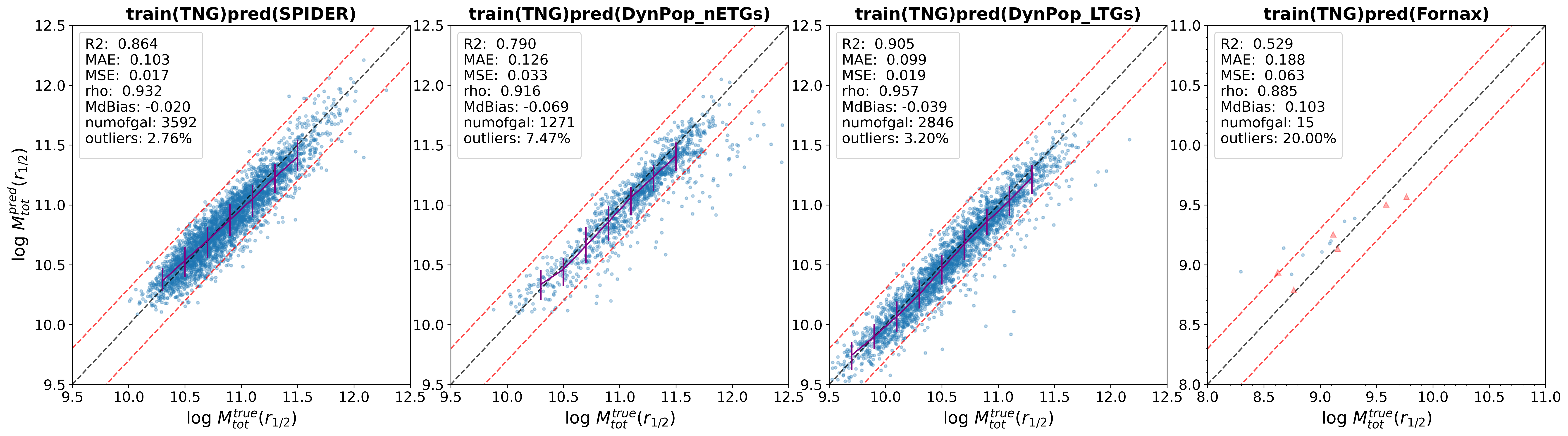}
\caption{\mlaall\ predictions of the central total mass, \mtotre, for the real galaxy dynamical samples. The optimal feature combination (i.e., \re, \ms\ and \sig) is used, as discussed in \S\ref{sec:self_pred} and Table~\ref{tab: DYN-number of features and accuracy}. Shown (from left to right) are predictions of the SPIDER sample;  the DynPop/nETG; the DynPop/LTG samples; and   the DSAMI sample. The dynamical model used as representative of the MaNGA Dynpop results is the $\rm JAM_{sph}$+generalized NFW profile (see \S\ref{sec:dynpop}). For the DSAMI sample, the red triangles represent the data points from the secondary test sample (1kpc<\re\ <2$R_{\rm p}$). The legend provides an overview of the statistical estimators for the different samples.}
\label{fig: train-simu-pred-real_1}
\end{figure*}

\begin{figure*}[ht]
\centering
\includegraphics[width=1.9\columnwidth]{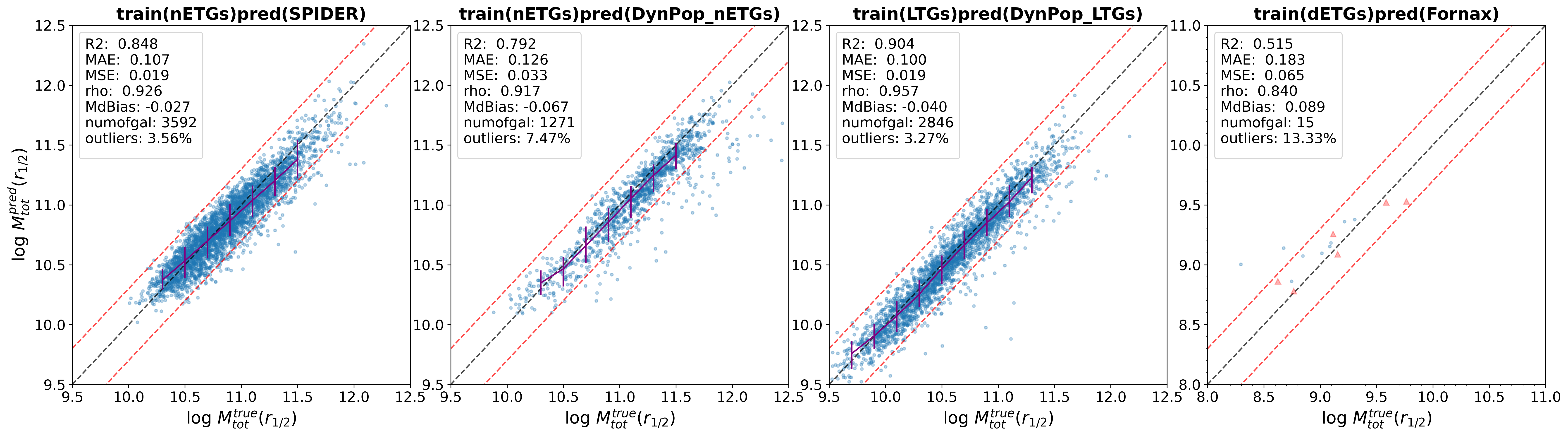}
\caption{As Fig.~\ref{fig: train-simu-pred-real_1}, but applying specialized \mla s on the different dynamical samples.}
\label{fig: train-simu-pred-real-customized}
\end{figure*}

After this premise, we can now show the results on the real data. We have anticipated in \S\ref{sec:feat_imp} that for the application to the observed datasets, we can use the \mlaall\ trained using the three most important features [\ms, \re, and \sig]. In particular, we use the ``full-counts'' training set. The results are shown in Fig.~\ref{fig: train-simu-pred-real_1}.

The immediate impression is a very good qualitative consistency of the \mlaall\ predictions with the dynamical estimates. The $R^2\gsim$0.8 in all cases except for the DSAMI sample, which also shows an MAE more than twice larger than the other samples, indicating an exceedingly high scatter. We recall here that the DSAMI sample is fully confined in the low end of the mass range covered by the training sample, in a region where the scatter of the scaling relations is systematically larger, at the level that the same scaling relations (see, e.g., the size-stellar mass relation in Fig.~\ref{fig:correlations}) are almost washed away for dETGs. Despite that, we see that the \mla\ predictions are still nicely aligned around the 1-to-1 relation. The $R^2$ values are in all cases smaller than the ones found in the self-predictions for nETGs, LTGs, and dETGs, if we use these latter as benchmarks. Looking at Fig.~\ref{fig: self-prediction-simu}, bottom row (but also Fig.~\ref{fig: self-prediction-simu-N4000} for the balanced training sample), the $R^2$ is smaller by 10\% and 18\%, respectively for the SPIDER and DynPop ETG samples with respect to the nETGs ($R^2\approx 0.96$), by 4\% for the DynPop LTG sample with respect to the self-predictions of the LTGs and by 35\% for the DSAMI sample with respect to the dETGs. In this latter case, we have included also the ``secondary'' dwarf predictive sample (red dots) to increase the statistics. In terms of scatter, both the MAE and the MSE can be between 32\% to 200\% larger than the self-prediction cases in Fig.~\ref{fig: self-prediction-simu}, although the maximum absolute values of both estimators (MAE$\lsim0.12$ and MAE$\lsim0.03$, if we exclude the DSAMI sample) is yet reasonably low. As comparison, in Fig.~\ref{fig: train-simu-pred-real_1}, we show the $\pm0.30$ dex limits, corresponding to the $\sim 2\sigma$ errors of the DynPop estimates, to check that the majority (typically $>95$\%) of the \mla\ predictions are enclosed within these limits (i.e., the outlier fraction is below 5\%). We also notice a small systematic effect, more evident in the \mla\ DynPop ETG estimates, that are 0.07 dex underestimated with respect to the corresponding classical estimates, with a tail of outliers below the $-0.30$ dex limit, which also exceeds the 5\% expected for an unbiased log-normal distribution. We track this misalignment between the \mla\ predictions and the classical Jeans analysis method of the DynPop sample, most of the more severe degradation (i.e., 18\%) of the $R^2$ with respect to the self-predictions, discussed above. Indeed, if we just artificially compensate for this small offset (i.e., by adding +0.07 dex to the \mla\ predictions) the $R^2$ would become 0.837 for DynPop ETGs, showing an improvement of 6\%. We will discuss more quantitatively the robustness of these results in \S\ref{sec:robustness} and \S\ref{sec:systematics}. Here we just anticipate that 1) the mentioned offsets seem statistically insignificant, as they are well within the scatter and 2) they partially come from the different definitions of the 3D structural quantities, which are difficult to align even between the observational samples.

The second, remarkable thing to notice, in Fig.~\ref{fig: train-simu-pred-real_1}, is that the good accuracy of the \mla\ predictions is insensitive to the methods adopted and, partially, to the sample adopted (i.e., DSAMI has a systematically lower accuracy, see above). In the figure, the classical dynamical methods adopted span from simple virial theorem (DSAMI), to the radial Jeans equation applied to fiber spectroscopy (SPIDER), to full 2D Jeans modeling of IFU rotation and velocity dispersion (DynPop). These methods are based on a variety of different data (deep imaging + multiobject spectroscopy, integral field spectroscopy) and contain a diversity of assumptions (e.g., geometry, orbital anisotropy, total mass model, if any), which potentially could bring a multiplicity of systematics. For these datasets, in the case of \mla\ estimates, we have adopted simple quantities characterizing the same systems. This is particularly impressive for the DynPop sample, which is, by far, the most complex predictive dataset. The dynamical analysis in K+23 is among the most sophisticated currently available on the market, and it is based on state-of-the-art observations, based on the largest IFU dataset ever observed, see \S\ref{sec:dynpop}. Despite that, \mla\ could use very basic information, like total stellar mass, size, and a single velocity dispersion values, to derive the mass of the DynPop sample with similar accuracy (see also \S\ref{sec:robustness}). As discussed earlier \ms\, \re\ and \sig\ are, nowadays, standard level 3 products of multiband photometric surveys (e.g., KiDS -- \citealt{2013ExA....35...25D}, DES -- \citealt{DESI-whitepaper-2013arXiv1308.0847L}) or by large sky spectroscopic surveys (e.g., SDSS --\citealt{thomas-SDSS-III-velocity-dispersion}, GAMA -- \citealt{Lange-2015-GAMA-mass-size-relation}, LAMOST -- \citealt{napolitano-lamost2020-central-velocity-dispersion}) for up to millions of galaxies, and will be obtained for hundreds of millions of galaxies by next-generations imaging surveys (e.g., LSST -- \citealt{LSST-2019}, EUCLID -- \citealt{Euclid-2011}, CSST -- \citealt{ZhanHu-CSST-2011SSPMA..41.1441Z}, ROMAN -- \citealt{Roman-2015arXiv150303757S}) or spectroscopic surveys (DESI -- \citealt{DESI-whitepaper-2013arXiv1308.0847L,DESI-Scienc-target-survey-design-2016arXiv161100036D}, 4MOST -- \citealt{4MOST-project-overview-2019Msngr.175....3D}). This gives an idea of the great potential of \mla\ applications. This potential is not diminished by the only apparent shortcoming of this approach: the cosmology + feedback dependency. We will come back to this in detail in \S\ref{sec:systematics}.

We conclude this section by showing the \mtotre\ predictions for all the observational samples, based on the same three features but using the customized \mla s. These are shown in Fig.~\ref{fig: train-simu-pred-real-customized}. By comparing this latter with Fig.~\ref{fig: train-simu-pred-real_1}, we see that all data samples show almost no changes in all statistical indicators. This confirms the conclusions based on the self-prediction tests based on TNG100 mock galaxies in \S\ref{sec:feat_imp}, that \mlaall\ is as good as, or even better than, the customized \mla s.

\subsection{Other relevant feature combinations on simulations and real galaxies}
\label{sec:other_conf}
In this section, we finally test different feature combinations to check whether 1) they provide similar accuracy with respect to the optimal feature combination seen in the \S\ref{sec:pred_real}; 2) there are combinations of features easier to measure that can still provide sufficient accuracy on the targets and that can be considered for applications on real data. 

\subsubsection{Using two features: \re\ and \ms}
\label{sec:2features}
\begin{figure*}[ht]
\centering
\includegraphics[width=1.9\columnwidth]{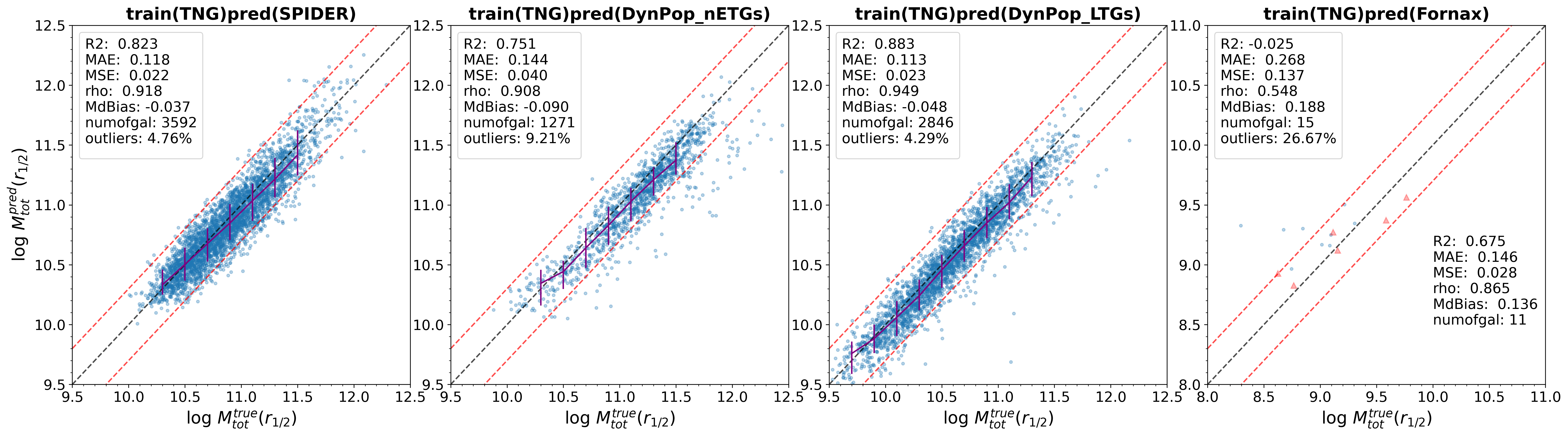}
\caption{\mlaall\ predictions of the central total mass, \mtotre, as in Fig.~\ref{fig: train-simu-pred-real_1}, but using only two features: \re\ and \ms. In the case of DSAMI, there are three obvious outliers above +0.30 dex. In the bottom right corner of the plot are shown the statistical estimators obtained excluding these outliers.}
\label{fig: train-simu-pred-real_ReMRe}
\end{figure*}

The first obvious test is the ``minimal feature'' scenario anticipated in \S\ref{sec:feat_imp} (i.e., the use of two features, \re\ and \ms). 
The results for \mlaall\ and ``full-counts'' training are shown in Fig.~\ref{fig: train-simu-pred-real_ReMRe}. Overall we see again a remarkable agreement for SPIDER and DynPop (both nETGs and LTGs), with $R^2$, MAE, MSE, and $\rho$ only slightly degraded with respect to the case of the three features shown in Fig.~\ref{fig: train-simu-pred-real_1}. This is in line with the accuracy degradation reported in Table~\ref{tab: DYN-number of features and accuracy}. However, we also notice that the very massive end of the SPIDER sample ($\log\ms/M_\odot>11.5$) shows a large scatter and a positive offset than the much tighter estimates in Fig.~\ref{fig: train-simu-pred-real_1}. On the other hand, the predictions for DSAMI are much worse than the 3 feature case, which is also expected, as discussed in \S\ref{sec:feat_imp} and seen in Table~\ref{tab: DYN-number of features and accuracy}. We see though that most of the poor $R^2$ might come from a small but not insignificant fraction of notable exceptions (2/15, the two blue points on the left), while the majority of the prediction looks well predicted. This is yet promising and needs to be tested possibly on larger samples. Indeed, the possibility of predicting correctly the total mass of dwarf galaxies without kinematical measurements is interesting as these latter are notoriously difficult outside the local universe \citep{Battaglia-Nature-dwarf-local-group-2022NatAs...6..659B}. Finally, we have checked that for the \mladw\ the results are not significantly better ($R^2=-0.100$) than the \mlaall, although the self-prediction test shows otherwise (see, e.g., Table~\ref{tab: DYN-number of features and accuracy}), again, very likely for the poor statistics.

\subsubsection{Excluding the stellar mass}
\label{sec:no_mstar}
Moving forward, with tests beyond the ``feature ranking'' analysis, we want to check how good \mla\ can predict if one of the highest ranked  features is missing (i.e., the \ms\ and the \re\ in turn), still keeping all other features from our original catalog.

We start by excluding the \ms, assuming that one does not possess enough data (either multiband photometry or spectroscopy) to have a robust estimate of the stellar mass. In this case, the features we can use are the two photometric bands ($g$ and $r$), the size (\re ), and the velocity dispersion (\sig ). We stress here that, as the TNG100 simulations are based on the Chabrier IMF, the mass prediction we derive even excluding \ms\ from the features, remains bound to the same IMF. This is to avoid explaining any difference we might find using total luminosity instead of the total mass, with an IMF variation.

\begin{figure}
\includegraphics[width=\columnwidth]{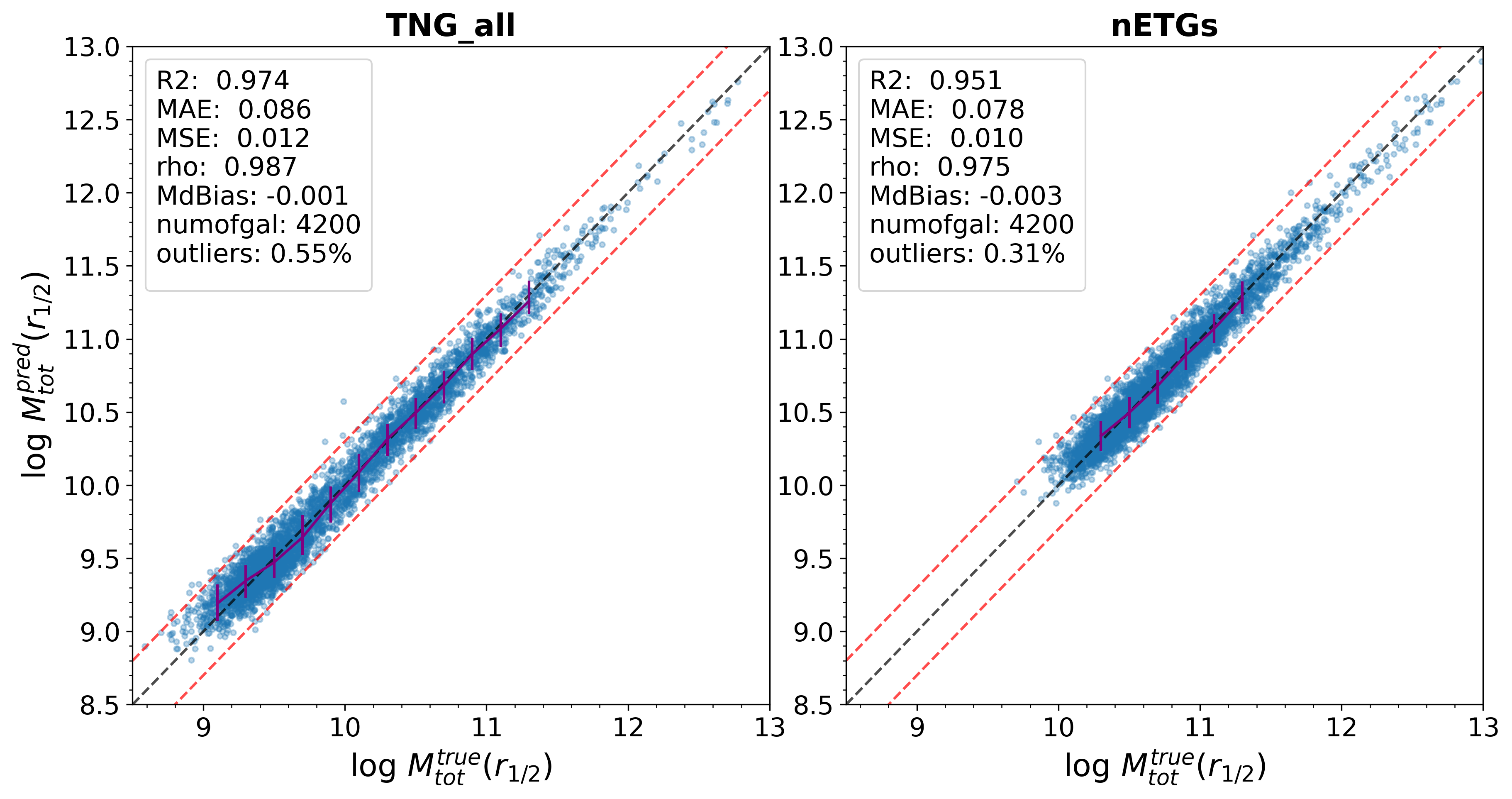}
\includegraphics[width=1.01 \columnwidth]{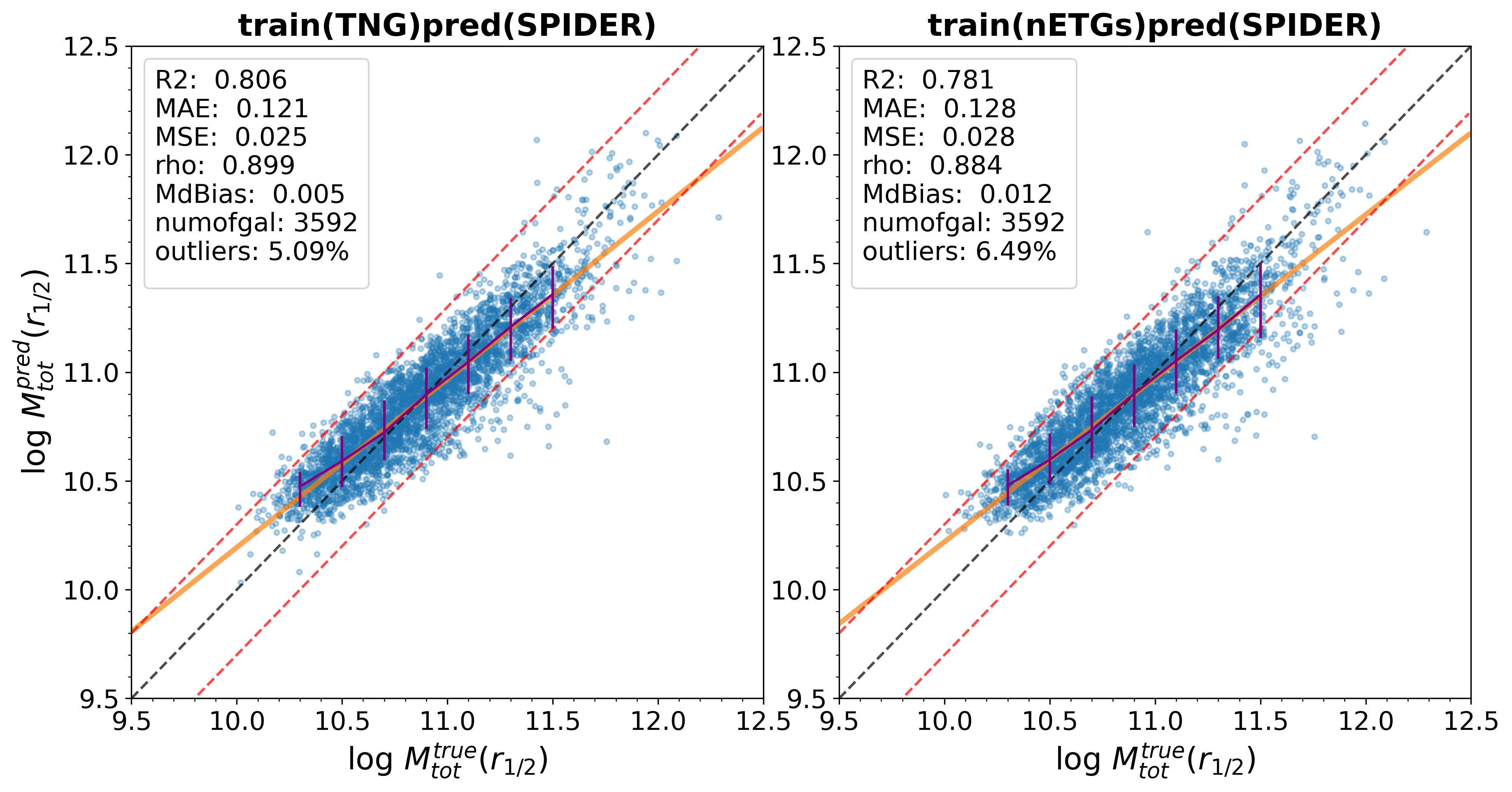}
\caption{\mlaall\ predictions of the central total mass, \mtotre, using four features:  $g$, $r$, \re, and \sig. Top row: Self-prediction test with balanced-counts training sample. Bottom row: Application of \mlaall\ and \mlaetg\ on the SPIDER dataset. Only the SPIDER sample was used as a real data test as it is the only dataset providing the broadband luminosities. }
\label{fig: spider-no-mstar-with-gr}
\end{figure}

In Fig.~\ref{fig: spider-no-mstar-with-gr} we show both the self-prediction results, i.e., the prediction obtained using a random TNG100 test sample (upper row), using the \mlaall\ (left) and \mlaetg\ (right), and the corresponding SPIDER predictions (bottom row), being this latter the only observed sample which provides all the features (included $g$ and $r$ band photometry) in a consistent way with the TNG100 (as in Table~\ref{tab: features and targets}). For the self=prediction test, we find a very good accuracy, which is not far from what found for the results with all features, as in Figs.~\ref{fig: self-prediction-simu} and~\ref{fig: self-prediction-simu-N4000}. This confirms the trend seen in Fig.~\ref{fig: DM-feature-accuracy} and Table~\ref{tab: DYN-number of features and accuracy}, where the accuracy flattens to $R^2>0.95$ if the number of features is larger than four. We notice that the result here is even better than in Table~\ref{tab: DYN-number of features and accuracy}. This is because here we are using ``balanced-counts'' while in Table~\ref{tab: DYN-number of features and accuracy} we are using the ``full counts'' (The ``full counts'' results are $R^2$=0.955, 0.949 for \mlaall\ and \mlaetg). The fact that these good accuracies are found even in case the stellar mass is missing, though, seems in contrast with Table~\ref{tab: DYN-number of features and accuracy}, showing that the \ms\ is the most relevant feature. In fact, there is no contradiction, as the total luminosity carries almost the same information as the total mass, if one can change one into another via a constant stellar mass-to-light ratio, $M/L$. This is clearly seen in the correlation matrices of all galaxy species in Fig.~\ref{fig: heatmap-simuonly}, where the largest correlation coefficients of the \ms\ are the ones with the $g$ and $r$ magnitudes.     
 
In Fig.~\ref{fig: spider-no-mstar-with-gr} lower-row, the predictions of the SPIDER sample, we find an acceptable $R^2$, MAE, and MSE, which are comparable with the ones found for the 2-feature test (\S\ref{sec:2features}). Most of the degradation of the estimator comes from the larger scatter and tendency of \mla\ to underestimate the masses at the high-mass end ($\log \mtotre/M_\odot >11.5$).
This tilt is not statistically significant. However, due to the tight correlation between the luminosity-stellar mass-total mass, we can argue that the mismatch might come from a steeper slope, $\alpha$, of the $M/L\propto L^\alpha$ relation (where $L$ can be the luminosity of $g$ or $r$-band in this case of the TNG100 nETGs with respect to the SPIDER ETGs. Given the adoption of the same IMF in the two samples, this has to be tracked to the stellar population parameters, star-formation history, and, ultimately, feedback in simulation. The reason this does not affect the stellar mass is that, generally, this latter is calibrated in simulation in order to reproduce the observed stellar mass function, while the consistency of the luminosity function with real observations, even introducing realistic calibrations, is yet disputable \citep{TNG_LF_2022}.

As an extreme case of minimal information to use for predictions, we can limit to photometry-only observations and exclude the kinematical information. For instance, we can check if using $g$ magnitude and $r$ magnitude \mla\ can still provide reasonable estimates. We apply the trained \mla\ on the SPIDER sample. The prediction shows a tilt similar to the one in Fig.~\ref{fig: spider-no-mstar-with-gr} lower-row, but with a worse accuracy ($R^2$= 0.740 vs. 0.806), meaning that adding \sig\ to photometry does improve the \mla\ performance, although not dramatically (in line with Table~\ref{tab: DYN-number of features and accuracy}).

\begin{figure}
\includegraphics[width=\columnwidth]{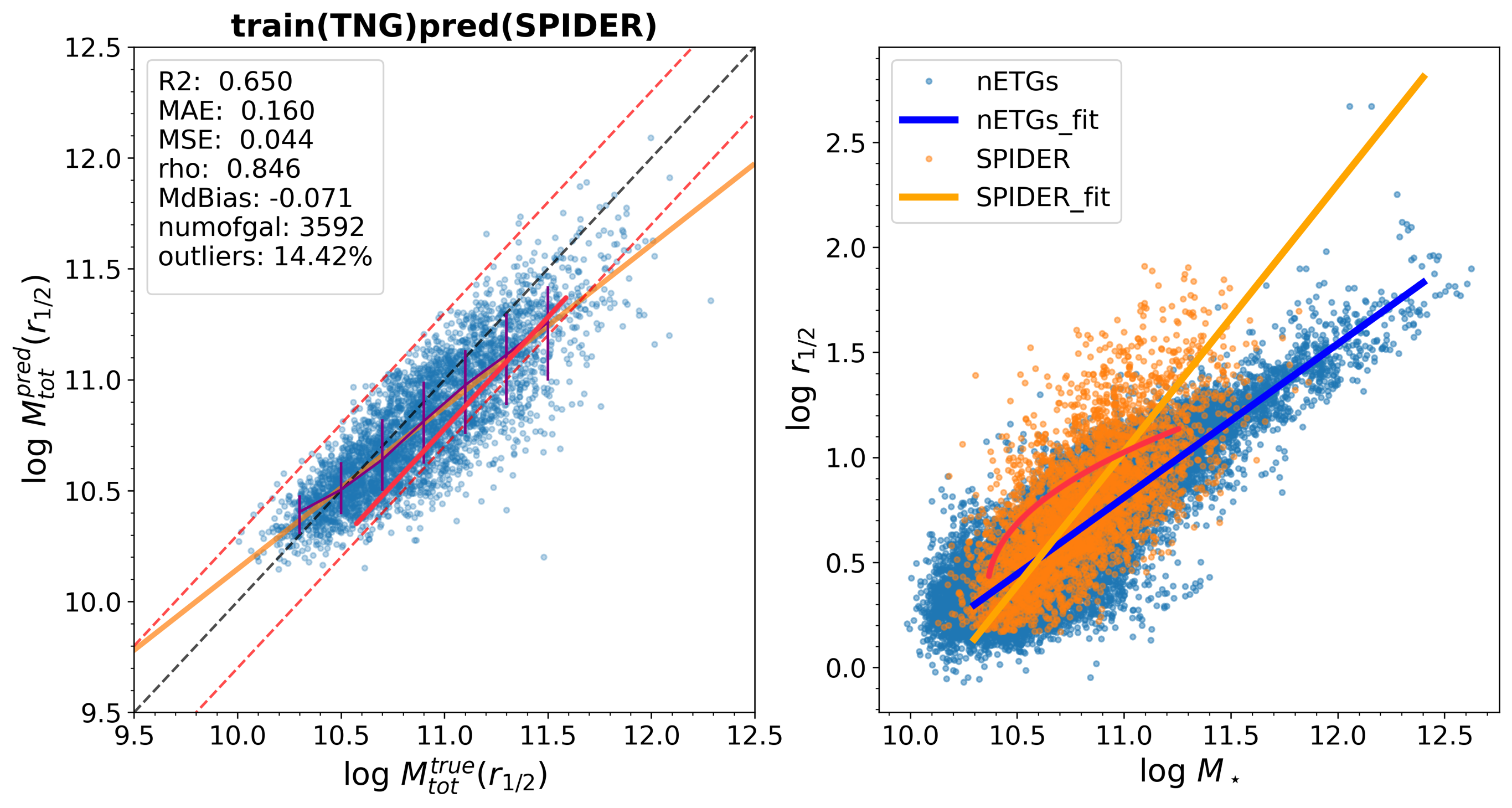}
\caption{\mlaall\ prediction and the scaling relation analysis of SPIDER.
Left panel: Result of applying \mlaall\ on SPIDER with four features (g, r, \ms, \sig) and missing the \re. The green circle is the median value for a given x-axis bin, where the bin size is 0.1 dex. The orange line represents the linear fit of the blue data points. Right panel: \ms-\re\ relation of nETGs and SPIDER. A linear fit line is displayed for both data points. The best linear fit of the SPIDER sample is determined by  averaging the linear fit using $\log\ms $ and $\log\re$ as independent variables. This approach compensates for the more unbalanced distribution observed towards $\log \ms /M_\odot\lsim11.2$. Completeness effects on the \ms-\re\ relation are not   considered here, as the focus is on understanding the origin of the tilt in the predictions (left panel).}
\label{fig: spider-no-Re-with-gr}
\end{figure}

\subsubsection{Excluding the effective radius}
\label{sec:no_reff}
We conclude the round of feature combinations, with a note about excluding the \re . We do not extensively consider this case as this has no practical applications, because we cannot predict the total and dark matter quantities within the \re , if this latter is unknown. Anyways, we have tested this situation as this can provide further insight into the way \mla\ handles missing features, specifically the second most important feature after \ms, as seen in Tables~\ref{tab: DM-number of features and accuracy} and~\ref{tab: DYN-number of features and accuracy}. Starting with self-predictions of the TNG100 galaxies, using $g$, $r$, \ms , and \sig\ as input, the predictions of the \mtotre\ show accuracies and precisions of the order of the ones found for the missing \ms. In particular, we find $R^2=0.951, 0.949$ for self-prediction using \mlaall\ and \mlaetg , once again in line with what is expected for 4 features from Fig.~\ref{fig: DM-feature-accuracy} and Table~\ref{tab: DYN-number of features and accuracy}. Going to the prediction of the SPIDER sample (we remind this is the only sample for which there are explicit cataloged $g$ and $r$ magnitudes), besides a much lower $R^2=0.65$ and very large MAE and MSE ($\sim$0.16 and $\sim$0.04 respectively), once again, we register a tilt between the \mlaall\ predictions and the dynamical values from T+12, although this time it is statistically significant, which is shown in Fig.~\ref{fig: spider-no-Re-with-gr} (left panel). Here, the bulk of the predicted distribution, between $\log\mtotre/M_\odot=10$ and 11, stays within the $\pm0.30$ dex from the 1-to-1 relation, while strong deviations are seen above $\log \mtotre /M_\odot=11$. This is mirrored by the behavior of the two datasets in the $\ms -\re$ relation, shown in the same Fig.~\ref{fig: spider-no-Re-with-gr} (right panel), where we see a steeper slope of the linear fit to the data for the SPIDER sample (orange line) with respect to the nETGs (blue line) of TNG100, that starts to strongly deviate at $\log\mtotre/M_\odot\sim11$. Overall, we notice that as long as the SPIDER galaxies remain consistent with the $\ms -\re$ of TNG100 galaxies (e.g., below the red curve as a qualitative example), also the predictions remain aligned along the 1-to-1 line (see the correspondent red line in the left panel). We believe these deviating galaxies explain the tilt in the predictions because \mlaall, in the absence of any knowledge about \re s, guesses their values from the TNG100 $\ms -\re$ relation, which, being shallower in slope, returns an underestimated \re. Due to the tight correlation between the \re\ and \mtotre\ (see, e.g., the correlation matrix in Fig.~\ref{fig: heatmap-simuonly}), this causes an underestimated \mtotre , particularly at $\log\mtotre/M_\odot>11$, and the subsequent tilt. 

\subsection{Final remarks}
\label{sec:fin_rem}
We want to conclude this section with a recap of the results we have collected using different test samples, that, we believe, have some profound implications.

1) The first clear evidence (\S\ref{sec:self_pred}) is that \mla\ can correctly predict the total mass of a galaxy (inside \re) if this belongs to the parent population of the training sample (self-prediction). This is true either if we separate the different galaxy types, or if we keep all species together (see also vM+22). Hence, \mla\ seems to be insensitive to the differences among the scaling relations of the different galaxy types, and able to predict independently the multiple correlations among the features and targets (i.e., stellar properties and DM/total mass). 

2) The second evidence (\S\ref{sec:pred_real}) is that \mla\ can accurately predict the \mtotre\ of the real sample, with a scatter which is consistent with the typical dynamical analysis errors, if a sufficient number of features is available. The minimal combination to keep the maximum accuracy includes the stellar mass, the effective radius, and the velocity dispersion (\S\ref{sec:feat_imp}). This might not be surprising as these are the structural parameters that enter into the Virial theorem, which ultimately should govern the physics behind the equilibrium of gravitational systems. 

3) The third evidence is that \mla\ cannot accurately predict the total masses of the real sample, if either the stellar mass or the effective radius are missed (i.e., using only two features in the training). These are the most important features providing high accuracy to the \mtotre\ predictions (see \S\ref{sec:feat_imp}) and missing either of them causes not only a larger scatter, but also a tilt in the predictions with respect to the real sample. Since the tilt is not present in the self-predictions test, this cannot be tracked to the inability of \mla\ to guess what is the true mass if any of these features are missing, but rather it has to be tracked to the differences (i.e., a tilt in the scaling relations between the simulation and the real data). We have given proof of that in Fig.~\ref{fig: spider-no-Re-with-gr}. The question arises of why these tilted scaling relations do not affect the \mtotre\ predictions whatsoever feature combination one uses, including the 3-feature (\ms, \re, \sig) case.

4) To answer this question, we argue that  the ML has indeed learned the physics behind the equilibrium of the galaxies in the TNG100. This is basically the Jeans equation, which can be written to explicit the total mass of stellar systems as (\citealt{Binney_Tremaine}):
\begin{equation}
    M_{\rm tot}(r)=\frac{\sigma^2 r}{G}\left(2\beta -\frac{d\ln\rho_*}{d\ln r}-\frac{d\ln\sig^2}{d\ln r}\right)
    \label{eq:rad_jeans_eq}
\end{equation}
where $G$ is the gravitational constant, $\rho_*$ is the 3D light density profile, $\sigma$ is the radial component of the velocity dispersion,\footnote{This is not equivalent to the observed velocity dispersion, which is the result of projection effects and the anisotropy parameter, but it can be reasonably be assumed proportional to the measured quantities.} and $\beta$ is the anisotropy parameters. By incorporating all the bracket content in a single ``virial factor'', $k$, which eventually depends on the slope of the light profile, $n$, and orbital anisotropy, as well as the slope of the velocity dispersion profile, $\gamma$, and also using the \citet{FJ_96}, $\ms\propto \sig^\delta$ (where, canonically, $\delta\sim4$), then Eq.~\ref{eq:rad_jeans_eq} writes (e.g., at the 3D effective radius, \re):

\begin{equation}
    \mtotre=k(n,\gamma,\beta) \frac{\sigma^2 \re}{G}= k'(n,\gamma,\beta) \frac{
    \ms^{2/\delta} \re}{G}.
    \label{eq:new_virial_eq}
\end{equation}
where, in the second equation, the new constant $k'$ incorporates the units of the conversions from the velocity dispersion to the stellar mass from the FJ. 
\begin{figure}
\centering
\includegraphics[width=0.9\columnwidth]{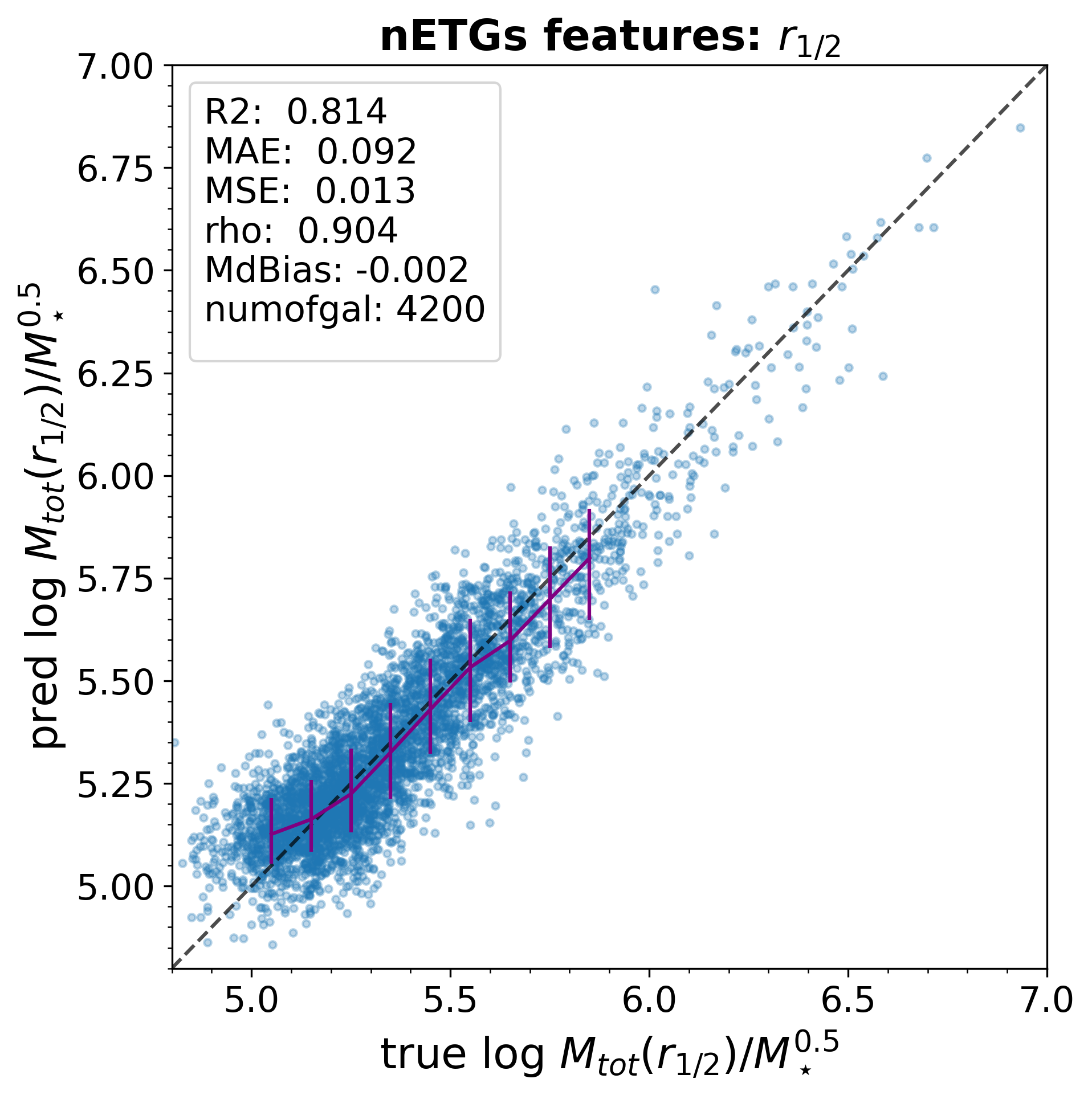}
\caption{\mla\ predictions of $\mtotre/\ms^{2/\delta}$, using only one feature \re. The data used is the  balanced-counts nETGs sample, where $\delta=4$.}
\label{fig: Eqn9}
\end{figure}
These are the equations that \mla\ eventually learns from the training sample and that are behind all results seen in this section. The proof of this assertion is beyond the purpose of this paper but it can be traced back to the ability of the ML to find universal relations in subhalo properties (see, e.g.,  \citealt{2022subhalo-galaxy-MLP}). However, we note that the equation on the right seems consistent with \ms\ and \re\ being the most important features for nETGs and LTGs found in \S\ref{sec:feat_imp} and also explains why the \sig\ is the least important feature, in another word because it is needed only to set the slope of the FJ, $\delta$, which is defined with a little scatter (Fig.~\ref{fig:correlations}, bottom right). If this interpretation is correct then \mla\ should be able to correctly predict the quantity $\mtotre/\ms^{2/\delta}$ with $\delta=4$ in Eq.~\ref{eq:new_virial_eq}, using \re\ only as a feature for the nETG sample. This is because we have fully incorporated the effect of \sig\ by fixing $\delta=4$, as argued above. Indeed, in Fig.~\ref{fig: Eqn9} we see that \mla\ predicts the $\mtotre/\ms^{2/4}$ with a rather high accuracy, as expected, showing that it might have learned Eq.~\ref{eq:new_virial_eq}, where \ms\ is a fundamental quantity. On the other hand, for dwarf galaxies, the main driver of the feature importance is the poorer correlation between \ms\ and \mtotre\ (and \mdre\, see Figs.~\ref{fig:correlations} and~\ref{fig: heatmap-simuonly}) than the \sig-\mtotre, meaning that \mla\ becomes more accurate using the equation on the left in Eqs.~\ref{eq:new_virial_eq}. This explains why \sig\ and \re\ or \ms\ and \re\ are, in turns, the highest ranked features in Tables~\ref{tab: DM-number of features and accuracy} and~\ref{tab: DYN-number of features and accuracy}, for dETGs and nETGs, respectively. Finally, the reason this equation works also for the LTGs, which generally are rotation-dominated, can reside in the fact that the total mass of the galaxies is strongly correlated to the mass of their bulges, which are governed by Eq.~\ref{eq:rad_jeans_eq}.

\begin{figure*}
\centering
\includegraphics[width=1.8\columnwidth]{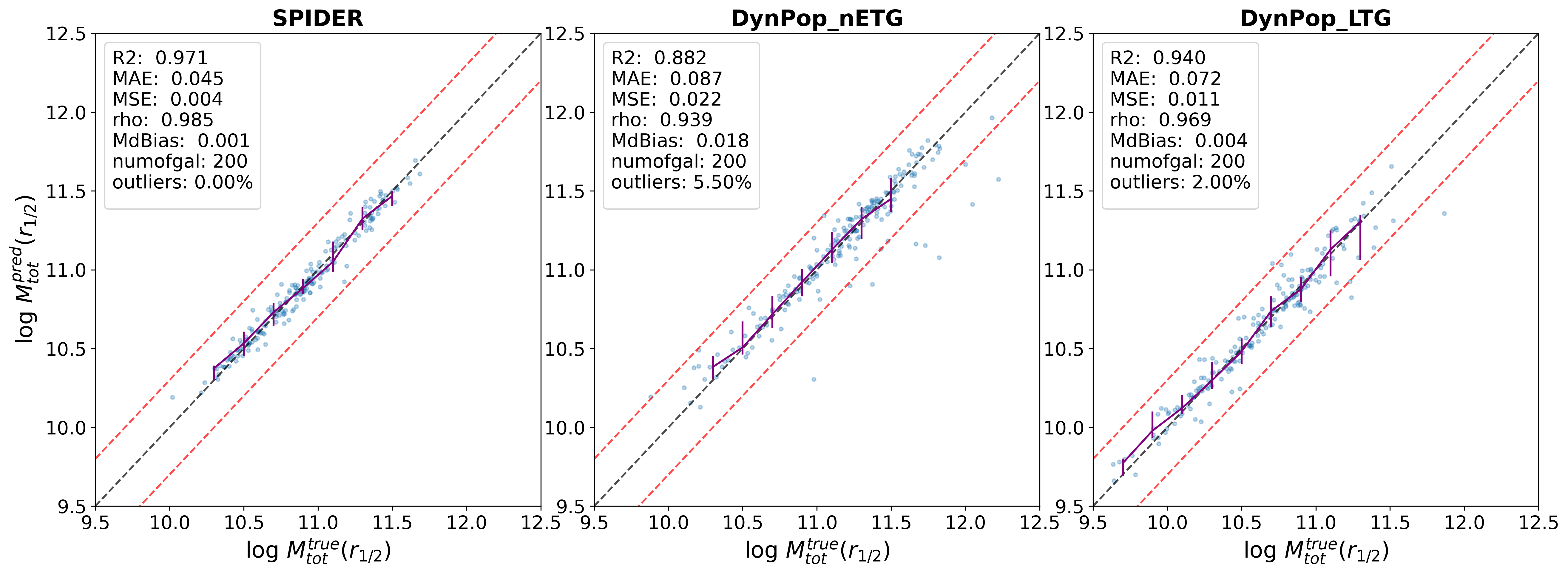}
\caption{Self-prediction test for observation using three features: \ms, \re,\ and \sig\ with the balanced-counts training sample. The training--test sample has been re-scaled to ensure an equal number of entries across all samples through random selection, aligning with the less populated class (DynPop-nETGs). The training set consists of 80\% of the randomly selected subsample (800 entries), while the remaining 20\% (200 entries) is allocated for testing. The Fornax sample is too small, with fewer than 20 entries, making it impossible to conduct the self-prediction test.}
\label{fig: self-pred-real}
\end{figure*}

A first corollary of this assertion is that every time one of the three features is missing, the ML tool tries to replace it with other scaling relations it has also learned from the training sample. If these are consistent with the observations, then the predictions remain correct. If the scaling relations are tilted, then the predictions are tilted (Figs.~\ref{fig: spider-no-mstar-with-gr} and~\ref{fig: spider-no-Re-with-gr}). A second corollary of our assertion is that if there is an offset between the scaling relations of the observations and simulations, then one can expect an offset in the predictions. We will discuss this in more detail in \S\ref{sec:systematics}, dedicated to systematics. In particular, we will demonstrate that this is a necessary but not sufficient condition for accurate predictions.

Finally, a last note about Eq.~\ref{eq:new_virial_eq}. The dependence of the $k$ factor on the S\'ersic suggests that the $n$-index is responsible for part of the scatter in the predictions. Hence, the accuracy and precision of the predictions would strongly benefit from the inclusion of this parameter in the training. We are currently working on this implementation in future analyses (De Araujo Ferrera et al., in preparation), however, we stress here that being the $n$-index generally known with little accuracy, this might also introduce some extra-scatter itself. Hence, here we acknowledge that, despite not knowing  $n$, \mla\ can still return a reasonable scatter in its estimates. 

\section{Discussion}
\label{sec:discussion}
In this section, we want to revisit all the results discussed in the previous section and quantify the robustness and the source of the systematics of our machine learning mass estimator algorithm, where possible. In particular, we will discuss the impact of the results of this paper in a wider context of the use of galaxies to infer cosmology, using machine learning tools trained on hydro-dynamical cosmological simulations (see CAMELS, VN+23).

\subsection{Robustness of the results}
\label{sec:robustness}
In \S\ref{sec:results}, we have seen that typical accuracies obtained for the best combinations of features in the \mtotre\ predictions of real data (e.g., the 3-feature cases) give typical $R^2$ which are up to 18\% worse that the self-predicting case for ETGs (e.g., DynPop) and up to 35\% for dETGs (i.e., DSAMI), and even larger MAE and MSE degradation. However, the statistical indicators can be somehow too optimistic in the test on simulated data. To check this, we have let \mlaall\ make predictions on both the training and test sets of TNG100. We have found that all the evaluation metrics are nearly identical,  confirming that there is no overfitting in \mlaall. Furthermore, to better assess whether the statistical indicators on real samples are still sufficient to claim a good accuracy and precision of \mla\ predictions, we have benchmarked them against the best accuracy the \mla\ can provide if effectively trained on real galaxies as ground truth (and not the TNG simulation as done in \S\ref{sec:training}). This means we can simply use the self-predictions of the real datasets, to estimate the best indicators one can expect to achieve, in the ideal conditions of knowing the ground truth (accuracy) and taking into account all the uncertainties on the measurements (precision). This is done by training and testing using the same features from the observed datasets but using the classical mass estimates as a target. Due to the smaller training size, this ``self-prediction test'' on the real data can be performed only for the massive ETG samples and LTG samples which have up to thousands of galaxies, allowing training sets of several hundred entries, but not for the dwarf samples. The scope of this test is to find out the best $R^2$ and MAE and MSE of the real sample as compared to the TNG100 self-predictions, which might be too idealized, even if accounting for realistic statistical errors. This is particularly important for the scatter, as measured by the MAE and MSE, that shows up much larger when moving from the TNG100 to the real sample, as discussed above. In Fig.~\ref{fig: self-pred-real}, we can see the results of such a test from SPIDER and DynPop. 
All indicators of the real data self-prediction result in scores closer to the maximum scores found in the self-predictions with a $R^2\sim$ 3\% smaller and MSE not more than 60\% larger except for SPIDER.\footnote{This is possible because of the simpler dynamical method adopted by T+12.} Hence, based on the absolute values of the statistical indicators and the vicinity to the benchmarks found for the self-predictions of the real samples, we conclude that the overall accuracy and precision of the \mla\ predictions of the real sample based on the TNG100 training are rather satisfactory.

\subsection{Source of systematics}
\label{sec:systematics}
In \S\ref{sec:preamble} we have listed a series of potential factors that can produce a misalignment between the simulation data and the observational data, which might introduce some systematic errors in the ML prediction based on simulations. We have also discussed that, by construction, the physical properties of galaxies in a single simulation are model (i.e., cosmology, feedback, DM flavor) dependent, this means that training on a single simulation might give a wrong answer about the true mass content of a galaxy. In \S\ref{sec:results} we have presented the first attempt at central mass content predictions of different observational datasets from \mla , a machine learning-based mass estimator trained to use a set of very basic observables from multiband imaging and spectroscopic surveys. 
It's noteworthy that we have achieved an excellent agreement with classical mass(e.g., in Fig.~\ref{fig: train-simu-pred-real_1}), since there is  a well-known misalignment between real galaxy data and simulations \citep{Cosmological-simulation-galaxy-formation-2020NatRP...2...42V}, which is also reported in e.g., Fig.~\ref{fig: kde-distribution} or Fig.~\ref{fig: spider-no-Re-with-gr}. However, here below we want to address, in a more quantitative, albeit not exhaustive way, some obvious sources of systematics and provide possible pathways for future improvements of the \mla\ results.     

\begin{figure}
\centering
\hspace{-0.4cm}
\includegraphics[width=0.9\columnwidth]{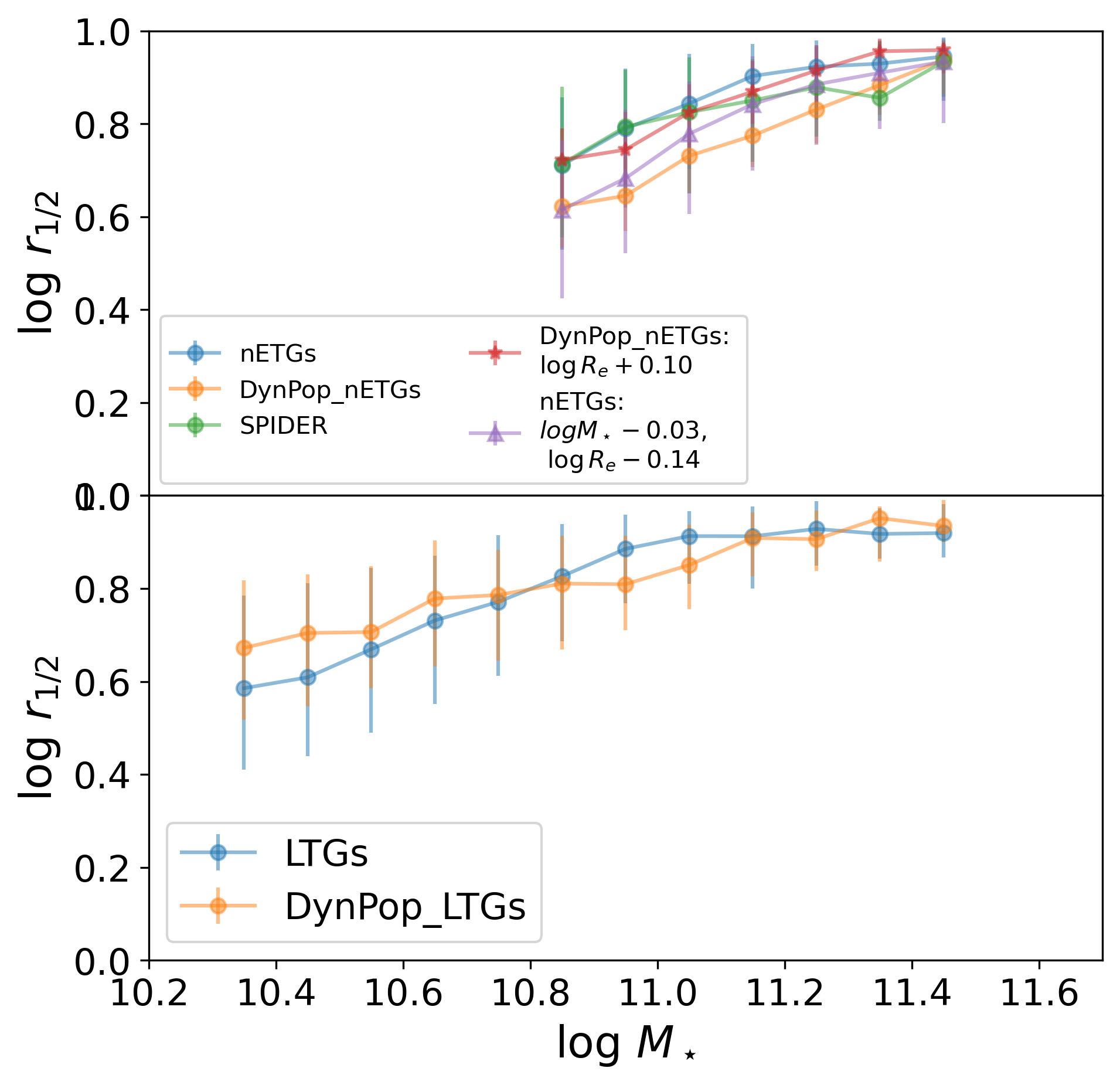}
\caption{Median mass-size relation of the TNG100 simulated galaxies and the observed datasets. Top: ETG galaxies from TNG100 (nETGs), SPIDER, and DynPop\_nETGs (see inset for legend). Bottom: LTG galaxies from TNG100 and DynPop\_LTGs, as in the legend. Error bars represent the 0.16, 0.84 percentiles of the data within different bins. We use $\log\ms/M_\odot=10.8$, as a reasonable completeness limit based on the DynPop\_nETGs and $\log\ms/M_\odot=10.3$ for the LTGs samples. }
\label{fig:mass-size}
\end{figure}

\subsubsection{Simulations versus observations}
\label{sec:syst_1}
Following the discussion at point 4) of \S\ref{sec:fin_rem} we can now better understand the impact of the systematics of the feature definitions in the \mla\ predictions. In \S\ref{sec:preamble} we have mentioned that among the quantities involved in Eq.~\ref{eq:new_virial_eq}, the most prone to mismatch in definitions is the effective radius. 
As mentioned in \S\ref{sec:preamble}, in this paper we make use of the 3D radius, consistently with the simulation. However, the definition of such a radius is different in SPIDER and DynPop. The former starts from the 2D effective radius from S\'ersic fit to the $r-$band imaging from SDSS and then converts this to 3D using a simple formula (i.e., $R_{\rm 3D}=1.35\times R_{\rm 2D}$ from W+10). In DynPop, the effective radius is determined via multi-Gaussian expansion (see Z+23, their Appendix~B)  of the SDSS $r$-band imaging, and it is defined as the radius of the sphere enclosing half of the total light of the galaxy. Although the original data of both SPIDER and DynPop are the same (SDSS imaging), the techniques are different and the assumptions are also different, hence it is possible that the (small) difference in the prediction of the data (e.g., seen in Figs.~\ref{fig: train-simu-pred-real_1},~\ref{fig: train-simu-pred-real-customized}, and~\ref{fig: train-simu-pred-real_ReMRe}) come from observed quantities' definition. If so, according to the second corollary at the end of \S\ref{sec:fin_rem}, we might expect an offset between the mass-size relations of the SPIDER and DynPop, with SPIDER being better aligned to the TNG galaxies \ms-\re\ relation. A proper analysis of the  \ms$-$\re\ relation should take into account the completeness of the samples \citep{Roy-2018-KiDS-stellar-mass}, which is beyond the purpose of this discussion that we want to keep yet qualitative at this stage. 

\begin{figure}
\hspace{-0.4cm}
\includegraphics[width=1.05\columnwidth]{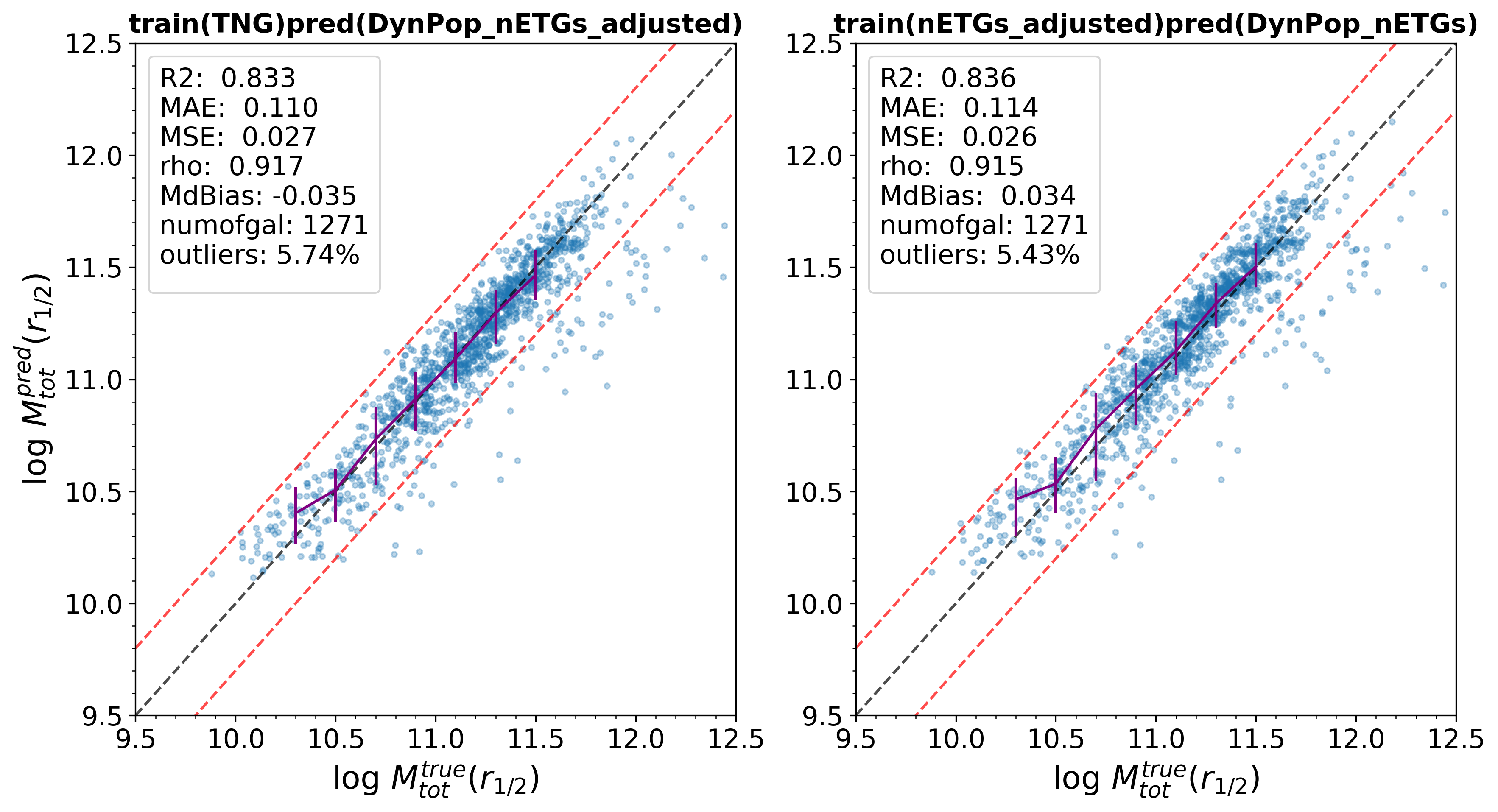}
\caption{Prediction results of the DynPop\_nETGs sample with three features: \ms, \re,\ and \sig, as indicated in Fig.~\ref{fig: train-simu-pred-real_1} and Fig.~\ref{fig: train-simu-pred-real-customized}. Left panel: Prediction   made by correcting the \re\ of the Dynpop\_nETGs by increasing 0.10 dex with \mlaall,\ as discussed in \S\ref{sec:syst_1}. This correction is applied due to a potential offset between observation and simulation, as shown in Fig.~\ref{fig:mass-size} top row. Right panel: Prediction  made by correcting both \re\ and \ms\ of the nETGs (i.e., the training sample of \mlaetg). Specifically, the \re\ is decreased by 0.14 dex, and the \ms\ is decreased by 0.30 dex. The adjusted mass-size relation is shown in Fig.~\ref{fig:mass-size}. This correction is applied because of the anticipated offset arising from different cosmology settings between simulation and observation, as discussed in \S\ref{sec:cosmology}. In comparison to Fig.~\ref{fig: train-simu-pred-real_1} and Fig.~\ref{fig: train-simu-pred-real-customized}, the prediction is observed to improve after these adjustments.}
\label{fig:dynpop_re+0.05}
\end{figure}

In Fig.~\ref{fig:mass-size}, we show the median mass-size relation of the TNG100, SPIDER, and DynPop galaxies, divided into ETGs and LTGs as a comparison, excluding the stellar mass ranges where there is a clear incompleteness, for example for the DynPop sample (ETG: $\log\ms/M_\odot\lsim10.8$; LTG: $\log\ms/M_\odot\lsim10.3$, see, e.g., Fig.~\ref{fig: train-simu-pred-real_1}). From the figure, we can see that there is a substantial agreement for the DynPop LTG and TNG100 LTG samples, which is somehow compatible with almost no-offstet in the LTG predictions, especially in Figs.~\ref{fig: train-simu-pred-real-customized} and~\ref{fig: train-simu-pred-real_ReMRe}. We also see an agreement between the SPIDER sample and the nETGs from TNG100, which corresponds also to quite unbiased predictions in the same figures. On the other hand, the DynPop ETG sample looks systematically offset with respect to the nETGs in the \re\ direction by $\sim0.10$ dex, except for the most massive data points where all ETG relations seem to converge. Assuming that the DynPop offset is entirely due to a \re\ bias, we can check whether by naively correcting the $\log$\re\ of the DynPop sample by +0.10 dex, we can also correct the 0.07 dex bias of the predictions discussed in \S\ref{sec:pred_real}. After adding a constant value 0.10 dex to the \re\ , we can see that in Fig.~\ref{fig:mass-size} nETGs (green circle) and DynPop nETG (red star) overlap nicely, which means we could correct the systematics error. We repeat the predictions using \re, \ms, and \sig\ as input and show the results in Fig.~\ref{fig:dynpop_re+0.05} left panel, using \mlaall. We observe indeed a much better agreement of the machine learning predictions with the K+23 ETG estimates, with the offset completely solved. We believe this is a confirmation of the second corollary in \S\ref{sec:fin_rem}, although, in the specific case, a constant correction to the \ms-\re\ is just a rough approximation. 

We conclude this section by noticing that the velocity dispersion also has a different definition, especially because in simulations there is no attempt to subtract the rotation component. However, as discussed in \citet{T09-CentralDM-2009MNRAS.396.1132Tortora09}, the rotation is expected to have an impact that cannot exceed the 10\%, also for $v/\sigma\sim 1$ systems. Furthermore, due to the least importance in the ranking of the features (\S\ref{sec:feat_imp}), this is found to have a minor impact on the case of small systematics. For instance, we have seen in \S\ref{sec:no_mstar}, that excluding the stellar mass, we can still find relatively good results, and having found the insignificant tilt mainly coming from the mass-to-light ratio and not from the $\sigma$. Indeed, luminosity is the feature substituting the stellar mass as a significant feature used to make predictions. 

\subsubsection{Feedback}
\label{sec:feedback}
In the previous subsection, we have discussed a first series of systematics causing a mismatch between the predictions of different datasets, residing on the definition of the measured quantities. However, assuming no definition mismatch, the scaling relations are expected to differ for the physics, especially the one of the baryons, behind them \citep{Wechsler-connectioin-galaxy-DM-halo-2018ARA&A..56..435W}. In this case, the offset and tilt of the scaling relations discussed at point 4) of the \S\ref{sec:fin_rem}, can be the consequence of the feedback model of the assumed simulation. If, on one side, this represents a problem if one wants to use ML tools trained on simulations to make total or dark mass predictions, on the other side, this also holds a great potential to reverse engineering the approach to use some ``unbiased'' dark or total mass measurements from galaxies as features and thus put some constraints on the feedback model. In this latter case the only unknown in Eq.~\ref{eq:new_virial_eq} are the slopes and normalizations of the scaling relations like \ms$-$\re\ and \ms$-$\sig\ (\S\ref{sec:syst_1}), and $M/L \sim L^\alpha$ (\S\ref{sec:no_mstar}), that eventually depend on the feedback. This is the basic philosophy behind the recent experiments trying to use machine learning tools trained on multi-cosmology simulations, eventually including also a variation of the cosmological parameters (see also \S\ref{sec:cosmology}), to constrain the combination of the feedbacks model and the cosmology using galaxies (e.g., \citealt{2022_one_galaxy_camels}), or galaxy clusters (e.g., \citealt{Qiu2023_cosmo_cl_ML}). 

To illustrate how a different series of predictions from simulations can produce a better agreement with observations, we use again the case of the DynPop ETGs, and put ourselves in the hypothesis that the K+23 are the ground truth and the TNG100 nETGs have the wrong feedback. Then we can imagine that to match the observation we need predictions that have a smaller \ms\ and a more compact radius in order to align the \ms-\re\ relation of the simulation with the DynPop in Fig.~\ref{fig:mass-size} (note that in the previous section we have increased the \re\ of DynPop to make the same correction). This qualitatively would correspond to the impact of a lower AGN feedback that produces more stellar mass and possibly more concentrated in the central regions which (see, e.g., \citealt{2023MNRAS.522.3912Ceverino}). In Fig.~\ref{fig:dynpop_re+0.05} right panel, we have arbitrarly reduced the $\log \ms$ of the nETGs training sample by 0.03 dex and the $\log \re$ by 0.14 dex in order to match the \ms\ -\re\ in Fig.\ref{fig:mass-size}. After this adjustment, we can see the adjusted nETGs are aligned well with the DynPop nETGs. Then we used \mlaetg\ to train and predict again the DynPop nETGs \mtotre. As it can be seen in Fig.~\ref{fig:dynpop_re+0.05} right panel, the prediction is much improved with respect to the one in Fig.~\ref{fig: train-simu-pred-real_1}, showing again the connection between the prediction and the scaling relations, but, most of all, the sensitivity of the method to the physics reproduced in the hydro-simulation used as training sample. 

\subsubsection{Cosmology}
\label{sec:cosmology}
In \S\ref{sec:results} we have commented that the results presented in this paper are model-dependent as all the mass predictions are based on TNG100 simulations which are characterized by a precise set of cosmological parameters (Planck15), plus a given feedback model (see \S\ref{sec:TNG_sim}). 

As far as mass estimation is concerned, this is not a limitation as, broadly speaking, this is a more general and physically motivated approach than typical mass modeling analyses where one uses assumptions on the dark matter density profile \citep{1996ApJ...NFW,2001ApJ..gNFW,NFW-Burkert1995,Einasto-DM-halo-1965TrAlm...5...87E} that are generally based on DM only simulations. The advantage of this latter approaches is that one can predict the halo properties (e.g., concentration, virial mass) including either a certain variance on the cosmological parameters \citep[see, e.g.,][]{2014MNRAS.441.3359D} or even different DM ``flavors'' \citep[e.g.,][]{2021JCAP...08..062N,2020JCAP...02..024B}. More rarely, dynamical studies try to account for the effect of the baryonic physics \citep{2010MNRAS.405.2351N,Napolitano2011}. 

The use of direct prediction from hydro-simulations, as proposed here, has the advantage of including all the physics of the interplay between DM and baryons. One can even generalize the results by considering simulations, with different cosmology, DM flavor, and feedback recipes. For example, CAMELS simulations\footnote{https://camels.readthedocs.io/en/latest/} have produced more than 5,000 simulations where they have changed two cosmological parameters (namely the mass cosmic density, $\Omega_m$, the amplitude of the fluctuation within 8Mpc, $\sigma_8$) and four astrophysical parameters regulating the supernova (ASN1, ASN2) and the AGN feedback (AGN1, AGN2). The combination of these parameters can produce a variety of scaling relations including baryonic and dark matter parameters. In this case, the caution to use is to adopt only those simulations showing baryonic scaling relations that do not systematically deviate from the ones of the real data, allowing some freedom to the ``unresolved'' biases between observations and simulations discussed in \S\ref{sec:syst_1}. In this latter case, one can train on a wider range of simulations and check the variance of the mass estimates from different cosmologies and feedback, then fully generalize the final results. Machine learning is an ideal environment to perform this, for its natural capability to handle very large and highly complex datasets, with tiny computational time. We will address this kind of application in a future analysis. 

Another obvious application is the one proposed by the CAMELS team \citep{2022_one_galaxy_camels,2023cosmo_multiple_gal}, also known as whether cosmology with one or more galaxies, but it is not clear whether it would work. Answering this question in a quantitative way is beyond the purpose of this paper. However, we believe we have proved that this is possible using real data, as the ML tools are sensitive to small variations of the scaling relations, regardless these are introduced by feedback or cosmology. Of course, the sensitivity to the cosmology can be inferior to other more effective cosmological probes (see, e.g., galaxy clusters, \citealt{Qiu2023_cosmo_cl_ML}). However, independent tests using more classical statistical methods have also shown that this is possible using galaxy observations \citep{busillo2023casco}.

\subsubsection{Centrals versus satellites}
\label{sec:cen_sat}
A final possible source of systematics is the existence of differences in the predictions between central and satellite galaxies. This might be due to intrinsic different DM properties, due to the different assembly history, or the difficulty of extracting the physical quantities of satellites embedded into larger parent halos. To check for the presence of biases, we have performed a test by training \mla\ on central and satellite galaxies and comparing the self-predictions of the two classes against the cross-predictions (i.e., training on centrals and predicting the targets of the satellites and vice versa). For nETGs we have seen this to produce insignificant variation on the final $R^2$, MSE, and MAE, and just the outlier fraction increasing from $<1$\%  to $\gsim 1\%$. Hence, we can exclude that this might have an impact on the final results of our analysis. 

\section{Conclusions}
\label{sec:conclusions}
The next-generation photometric and spectroscopic surveys will collect high-quality multiband imaging and spectroscopy for   billions of galaxies. For instance, we will be able to  measure the  sizes, stellar masses, stellar population properties, and internal kinematics for hundreds of millions of systems. To have access to their dark matter content, though, it is necessary to use dynamical methods, which have limitations due to modeling assumptions (e.g., geometry, orbital anisotropy) or complexity (e.g., jeans modeling vs. orbital superposition) that might require complementary data (e.g., 2D kinematics) to be compellingly fulfilled. This makes the direct estimates of the DM in such large galaxy samples prohibitive. On the other hand, indirect methods can be used based on semi-empirical scaling relations (e.g., from abundance matching; \citealt{2020MNRAS.496L.101M}), but these also imply two types of biases: 1) the assumed cosmology, as these scaling relations are built on the expectation of the halo mass function which is cosmology dependent, and 2) the halo model (see, \citealt{2023OJAp....6E..39A}; \citealt{2000MNRAS.318.1144P}; \citealt{2005ApJ...633..791Z}), which is generally challenging to constrain. In either case (i.e., direct or indirect), the DM content of galaxies of the upcoming galaxy compilations will remain highly unconstrained or biased. 

Nevertheless, such an unprecedented dataset holds an enormous potential to deeply answer fundamental questions of galaxy formation and evolution, especially related to the star-halo connection \citep{Wechsler-connectioin-galaxy-DM-halo-2018ARA&A..56..435W} and mass assembly across time \citep{2017MNRAS.470.3720T}. Ultimately, these collections of data, in combination with their analogs from cosmological simulations, can allow us to break the contribution of the dark matter and the physics of baryons in galaxies and their dependence on cosmological parameters \citep{CAMLES-public-data-release-2023ApJS..265...54V,busillo2023casco}. Hence, it is of pivotal importance to find effective ways to connect data and simulations in high-dimensional parameter spaces \citep[e.g.,][]{2022subhalo-galaxy-MLP}, and finally fully exploit the capability of ML to constrain cosmology and galaxy formation in a single framework \citep[see also][]{Qiu2023_cosmo_cl_ML}.  

In this work, which follows the first test made in vM+22, we   developed a novel mass estimator based on random forests, called \mla, which can learn the relation between luminous properties (namely the $g$- and $r$-band luminosity, the effective radius, the velocity dispersion, the stellar mass) and the mass of galaxies in hydrodynamical simulations and can predict the total and dark matter content of galaxies. For the first time, after having tested  \mla\ on a sample of simulated galaxies \citep[see][]{von2022inferring}, we  applied a ML-based estimator to different samples of real systems, including an ETG sample from SPIDER, an ETG/LTG sample from DynPop/MANGA, and a dwarf sample from the  SAMI project. In order to implement some observational realism, we  added  Gaussian noise to the original simulation data used for the training of \mla\ to reproduce the observational errors, while we  used the 3D mass inference from the dynamical models to consistently align the quantity to be predicted in real galaxies (targets). 

We summarize here  the major results of our analysis, based on  tests on catalogs of simulated galaxies from TNG100 (self-predictions) and the different real dynamical samples (real-predictions):
\begin{enumerate}
\item In the self-prediction test, \mla\  achieved $R^2=0.933$ for \mdre\ and $R^2=0.959$ for \mtotre\ using all galaxies without separating them into classes. The predictions for \mtotre\ are more accurate than \mdre, which mainly reflects the low scatter of the  scaling relations of the former. Normal ETGs (nETGs) always have the highest accuracy when separating galaxies into classes, followed by the LTGs and dwarf ETGs (dETGs). In particular, the outlier fraction is less than 0.5\% for dETGs and LTGs; and less than 0.2\% for nETGs.
\item We  also   used the TNG100 mock catalog sample to investigate different combinations of features optimizing the $R^2$. We   found that for all classes of galaxies, using three features (i.e., \re, \ms,\ and \sig) we reach a plateau in the accuracy, $R^2\sim 0.93$ for \mdre\ and $R^2\sim 0.96$ for \mtotre, that is a little improved using further features. We   also found that these accuracies are almost equivalent if we use a customized \mla\ trained on the different classes or a generalized tool, \mlaall, trained on all classes mixed. 
\item We finally applied both \mlaall\ and the customized \mla\ to the different classes of real galaxies with dynamical masses from different methods. We find that all the statistical indicators show almost no difference between \mlaall\ and \mla. 
The \mlaall\ performs well in the nETGs (SPIDER: $R^2$=0.864, DynPop/nETGs: $R^2$=0.790) and the LTGs (DynPop/LTGs: $R^2$=0.905); it performs a little worse in the dETGs (Fornax: $R^2$=0.529), possibly for the larger scatter of the dwarf galaxies' size-mass relation. Outlier fractions are generally lower than 5\%, consistent with log-normal errors of the \mla\ predictions.
\item We finally checked the robustness of the results by applying a self-prediction test on the observations. We find that if \mla\ is trained using the real galaxy catalog of features and targets, \mla\ correctly predicts the targets, and it does so with an accuracy slightly better than if trained on the TNG100 training sample. We also showed the impact of the bias between simulations and observations under the form of offset and tilt of the scaling relations.
For instance, we tracked the statistically insignificant offset found on the DynPop/nETGs ($<0.07$dex) to the offset in the \re-\ms\ relation of this sample with respect to the TNG100 and the SPIDER sample (which is nicely aligned with simulations), perhaps due to the definition of DynPop \re. 
We also discussed the impact of the feedback model on the scaling relations, which is incorporated in the data, but that can be varied in the simulations, thus producing other sets of predictions that can closely match independent dynamical analyses. 
\item From this point of view, taking the dynamical estimates at face value, and considering the surprising consistency of the \mla\ predictions with the dynamical masses, we  conclude that the TNG100 feedback model is good enough to predict \mtotre\ in real galaxies, having fixed the cosmological parameters to Planck15. We have made the \mla\ and \mlaall\ publicly available at this URL\footnote{https://github.com/wsr1998/MELA\_galaxy\_total\_mass\_estimator \label{foot:release}} to be used to predict total and dark matter for other datasets having features consistent with those used in this paper (possibly \re, \ms,\ and \sig). Together with the code, we have also made available the catalog of the features and targets used for the three data samples analyzed in this paper: SPIDER, MaNGA/DynPop, and DSAMI.
\item We tested that \mla\ can equally well predict the augmented DM (or missing mass), $M_{\rm DM}(\re)=\mtotre-\msre$, which is the only directly inferred with dynamics. This led us to conclude that it can also correctly predict the pure dark matter component, \mdre, in real galaxies (in simulations, it is trivial as this is a known quantity). In this latter case, the \mdre\ is possibly more tightly dependent on the feedback model. Overall, the obvious advantage of \mla\ with respect to standard dynamical models is that it can estimate the dark mass of galaxies.
\item We  finally discussed what equilibrium equations \mla\ might have learned during the training, although these are not used in simulations, but should be encoded in the physics of the collapse of baryons and dark matter. This led us to argue that ML can extract the physics from data. In a forthcoming paper, we will use symbolic regression to constrain a ML-based physical formula from \mla\ .   
\end{enumerate}

In terms of perspectives for future applications, we expect to extend the training of \mla\ to other cosmologies and feedback models, for example, using publicly available CAMELS simulations (\citealt{CAMLES-public-data-release-2023ApJS..265...54V}). We foresee the ability of the tool to provide mass predictions in multiple cosmology--feedback combinations. We also plan to train other \mla-like algorithms over catalogs of TNG100 simulated galaxies for which we have extracted the S\'ersic parameters of the light and the DM density profiles, to predict the galaxy total slope and other DM halo density properties (De Araujo Ferrera et al. in preparation). 

We believe this work has provided evidence that ML techniques are indeed mature enough to move to applications to real data, like those we expect to collect from next-generation surveys (EUCLID, CSST, VR-LSST, DESI, 4MOST). We showed we can obtain robust predictions, provided that observations and simulations use homogeneous quantities for features and targets. For this ``observational realism'' step, in this paper, we have to make a significant effort to match data and simulations, and the results are very convincing. We believe that we are now motivated to take even another step, and use some unbiased dark matter estimate as a further feature (e.g., via strong gravitational lensing), and train a \mla-like tool, having cosmology and feedback parameters as a target, to finally try to address the cosmology with the multiple galaxies problem (\citealt{2023cosmo_multiple_gal}), but using real data.


\begin{acknowledgements}
      We thank the anonymous referee for the stimulating report which helped to strengthen some of our results and improve the clarity of the paper. We thank Lanlan Qiu for insightful discussions. NRN acknowledges that part of this work was supported by the National Science Foundation of China, the Research Fund for Excellent International Scholars (grant n. 12150710511), and from the research grant from China Manned Space Project n. CMS-CSST-2021-A01. C.T. acknowledges the INAF grant 2022 LEMON. WL acknowledges the support from NSFC grant (No. 12073089).
\end{acknowledgements}


\bibliographystyle{aa}
\bibliography{aa}

@ARTICLE{DM_distribution_CNN_2023MNRAS.525.6015D,
       author = {{de los Rios}, Mart{\'\i}n and {Peta{\v{c}}}, Mihael and {Zaldivar}, Bryan and {Bonaventura}, Nina R. and {Calore}, Francesca and {Iocco}, Fabio},
        title = "{Determining the dark matter distribution in simulated galaxies with deep learning}",
      journal = {\mnras},
         year = 2023,
        month = nov,
       volume = {525},
       number = {4},
        pages = {6015-6035},
          doi = {10.1093/mnras/stad2614},
archivePrefix = {arXiv},
       eprint = {2111.08725},
 primaryClass = {astro-ph.GA},
       adsurl = {https://ui.adsabs.harvard.edu/abs/2023MNRAS.525.6015D}
}

@ARTICLE{jiani2023arXiv231110351C,
       author = {{Chu}, Jiani and {Tang}, Hongming and {Xu}, Dandan and {Lu}, Shengdong and {Long}, Richard},
        title = "{Galaxy stellar and total mass estimation using machine learning}",
      journal = {\mnras},
         year = 2024,
        month = mar,
       volume = {528},
       number = {4},
        pages = {6354-6369},
          doi = {10.1093/mnras/stae406},
archivePrefix = {arXiv},
       eprint = {2311.10351},
 primaryClass = {astro-ph.GA},
       adsurl = {https://ui.adsabs.harvard.edu/abs/2024MNRAS.528.6354C}
}

@ARTICLE{GalaxyNet_2021MNRAS.507.2115M,
       author = {{Moster}, Benjamin P. and {Naab}, Thorsten and {Lindstr{\"o}m}, Magnus and {O'Leary}, Joseph A.},
        title = "{GalaxyNet: connecting galaxies and dark matter haloes with deep neural networks and reinforcement learning in large volumes}",
      journal = {\mnras},
         year = 2021,
        month = oct,
       volume = {507},
       number = {2},
        pages = {2115-2136},
          doi = {10.1093/mnras/stab1449},
archivePrefix = {arXiv},
       eprint = {2005.12276},
 primaryClass = {astro-ph.GA},
       adsurl = {https://ui.adsabs.harvard.edu/abs/2021MNRAS.507.2115M}
}

@ARTICLE{2013ExA....35...25D,
       author = {{de Jong}, Jelte T.~A. and {Verdoes Kleijn}, Gijs A. and {Kuijken}, Konrad H. and {Valentijn}, Edwin A.},
        title = "{The Kilo-Degree Survey}",
      journal = {Experimental Astronomy},
         year = 2013,
        month = jan,
       volume = {35},
       number = {1-2},
        pages = {25-44},
          doi = {10.1007/s10686-012-9306-1},
archivePrefix = {arXiv},
       eprint = {1206.1254},
 primaryClass = {astro-ph.CO},
       adsurl = {https://ui.adsabs.harvard.edu/abs/2013ExA....35...25D}
}

@ARTICLE{2022A&A...666A..85L,
       author = {{Li}, Rui and {Napolitano}, Nicola R. and {Feng}, Haicheng and {Li}, Ran and {Amaro}, Valeria and {Xie}, Linghua and {Tortora}, Crescenzo and {Bilicki}, Maciej and {Brescia}, Massimo and {Cavuoti}, Stefano and {Radovich}, Mario},
        title = "{Galaxy morphoto-Z with neural Networks (GaZNets). I. Optimized accuracy and outlier fraction from imaging and photometry}",
      journal = {\aap},
         year = 2022,
        month = oct,
       volume = {666},
          eid = {A85},
        pages = {A85},
          doi = {10.1051/0004-6361/202244081},
archivePrefix = {arXiv},
       eprint = {2205.10720},
 primaryClass = {astro-ph.GA},
       adsurl = {https://ui.adsabs.harvard.edu/abs/2022A&A...666A..85L}
}

@ARTICLE{2021A&A...645A..87B,
       author = {{Baqui}, P.~O. and {Marra}, V. and {Casarini}, L. and {Angulo}, R. and {D{\'\i}az-Garc{\'\i}a}, L.~A. and {Hern{\'a}ndez-Monteagudo}, C. and {Lopes}, P.~A.~A. and {L{\'o}pez-Sanjuan}, C. and {Muniesa}, D. and {Placco}, V.~M. and {Quartin}, M. and {Queiroz}, C. and {Sobral}, D. and {Solano}, E. and {Tempel}, E. and {Varela}, J. and {V{\'\i}lchez}, J.~M. and {Abramo}, R. and {Alcaniz}, J. and {Benitez}, N. and {Bonoli}, S. and {Carneiro}, S. and {Cenarro}, A.~J. and {Crist{\'o}bal-Hornillos}, D. and {de Amorim}, A.~L. and {de Oliveira}, C.~M. and {Dupke}, R. and {Ederoclite}, A. and {Gonz{\'a}lez Delgado}, R.~M. and {Mar{\'\i}n-Franch}, A. and {Moles}, M. and {V{\'a}zquez Rami{\'o}}, H. and {Sodr{\'e}}, L. and {Taylor}, K.},
        title = "{The miniJPAS survey: star-galaxy classification using machine learning}",
      journal = {\aap},
         year = 2021,
        month = jan,
       volume = {645},
          eid = {A87},
        pages = {A87},
          doi = {10.1051/0004-6361/202038986},
archivePrefix = {arXiv},
       eprint = {2007.07622},
 primaryClass = {astro-ph.IM},
       adsurl = {https://ui.adsabs.harvard.edu/abs/2021A&A...645A..87B}
}

@ARTICLE{2024MNRAS.527.3347V,
       author = {{von Marttens}, R. and {Marra}, V. and {Quartin}, M. and {Casarini}, L. and {Baqui}, P.~O. and {Alvarez-Candal}, A. and {Galindo-Guil}, F.~J. and {Fern{\'a}ndez-Ontiveros}, J.~A. and {del Pino}, Andr{\'e}s and {D{\'\i}az-Garc{\'\i}a}, L.~A. and {L{\'o}pez-Sanjuan}, C. and {Alcaniz}, J. and {Angulo}, R. and {Cenarro}, A.~J. and {Crist{\'o}bal-Hornillos}, D. and {Dupke}, R. and {Ederoclite}, A. and {Hern{\'a}ndez-Monteagudo}, C. and {Mar{\'\i}n-Franch}, A. and {Moles}, M. and {Sodr{\'e}}, L. and {Varela}, J. and {V{\'a}zquez Rami{\'o}}, H.},
        title = "{J-PLUS: galaxy-star-quasar classification for DR3}",
      journal = {\mnras},
         year = 2024,
        month = jan,
       volume = {527},
       number = {2},
        pages = {3347-3365},
          doi = {10.1093/mnras/stad3373},
archivePrefix = {arXiv},
       eprint = {2212.05868},
 primaryClass = {astro-ph.GA},
       adsurl = {https://ui.adsabs.harvard.edu/abs/2024MNRAS.527.3347V}
}

@ARTICLE{2018MNRAS.481.4728T,
       author = {{Tortora}, C. and {Napolitano}, N.~R. and {Spavone}, M. and {La Barbera}, F. and {D'Ago}, G. and {Spiniello}, C. and {Kuijken}, K.~H. and {Roy}, N. and {Raj}, M.~A. and {Cavuoti}, S. and {Brescia}, M. and {Longo}, G. and {Pota}, V. and {Petrillo}, C.~E. and {Radovich}, M. and {Getman}, F. and {Koopmans}, L.~V.~E. and {Trujillo}, I. and {Verdoes Kleijn}, G. and {Capaccioli}, M. and {Grado}, A. and {Covone}, G. and {Scognamiglio}, D. and {Blake}, C. and {Glazebrook}, K. and {Joudaki}, S. and {Lidman}, C. and {Wolf}, C.},
        title = "{The first sample of spectroscopically confirmed ultra-compact massive galaxies in the Kilo Degree Survey}",
      journal = {\mnras},
         year = 2018,
        month = dec,
       volume = {481},
       number = {4},
        pages = {4728-4752},
          doi = {10.1093/mnras/sty2564},
archivePrefix = {arXiv},
       eprint = {1806.01307},
 primaryClass = {astro-ph.GA},
       adsurl = {https://ui.adsabs.harvard.edu/abs/2018MNRAS.481.4728T}
}

@ARTICLE{Rubin1970ApJ...159..379R,
       author = {{Rubin}, Vera C. and {Ford}, W. Kent, Jr.},
        title = "{Rotation of the Andromeda Nebula from a Spectroscopic Survey of Emission Regions}",
      journal = {\apj},
         year = 1970,
        month = feb,
       volume = {159},
        pages = {379},
          doi = {10.1086/150317},
       adsurl = {https://ui.adsabs.harvard.edu/abs/1970ApJ...159..379R}
}

@ARTICLE{2020MNRAS.496L.101M,
       author = {{Macci{\`o}}, Andrea V. and {Courteau}, St{\'e}phane and {Ouellette}, Nathalie N. -Q. and {Dutton}, Aaron A.},
        title = "{Abundance matching tested on small scales with galaxy dynamics}",
      journal = {\mnras},
         year = 2020,
        month = jul,
       volume = {496},
       number = {1},
        pages = {L101-L105},
          doi = {10.1093/mnrasl/slaa094},
archivePrefix = {arXiv},
       eprint = {2006.00818},
 primaryClass = {astro-ph.GA},
       adsurl = {https://ui.adsabs.harvard.edu/abs/2020MNRAS.496L.101M}
}

@ARTICLE{2017MNRAS.470.3720T,
       author = {{Tojeiro}, Rita and {Eardley}, Elizabeth and {Peacock}, John A. and {Norberg}, Peder and {Alpaslan}, Mehmet and {Driver}, Simon P. and {Henriques}, Bruno and {Hopkins}, Andrew M. and {Kafle}, Prajwal R. and {Robotham}, Aaron S.~G. and {Thomas}, Peter and {Tonini}, Chiara and {Wild}, Vivienne},
        title = "{Galaxy and Mass Assembly (GAMA): halo formation times and halo assembly bias on the cosmic web}",
      journal = {\mnras},
         year = 2017,
        month = sep,
       volume = {470},
       number = {3},
        pages = {3720-3741},
          doi = {10.1093/mnras/stx1466},
archivePrefix = {arXiv},
       eprint = {1612.08595},
 primaryClass = {astro-ph.CO},
       adsurl = {https://ui.adsabs.harvard.edu/abs/2017MNRAS.470.3720T}
}

@ARTICLE{2005ApJ...633..791Z,
       author = {{Zheng}, Zheng and {Berlind}, Andreas A. and {Weinberg}, David H. and {Benson}, Andrew J. and {Baugh}, Carlton M. and {Cole}, Shaun and {Dav{\'e}}, Romeel and {Frenk}, Carlos S. and {Katz}, Neal and {Lacey}, Cedric G.},
        title = "{Theoretical Models of the Halo Occupation Distribution: Separating Central and Satellite Galaxies}",
      journal = {\apj},
         year = 2005,
        month = nov,
       volume = {633},
       number = {2},
        pages = {791-809},
          doi = {10.1086/466510},
archivePrefix = {arXiv},
       eprint = {astro-ph/0408564},
 primaryClass = {astro-ph},
       adsurl = {https://ui.adsabs.harvard.edu/abs/2005ApJ...633..791Z}
}

@ARTICLE{2000MNRAS.318.1144P,
       author = {{Peacock}, J.~A. and {Smith}, R.~E.},
        title = "{Halo occupation numbers and galaxy bias}",
      journal = {\mnras},
         year = 2000,
        month = nov,
       volume = {318},
       number = {4},
        pages = {1144-1156},
          doi = {10.1046/j.1365-8711.2000.03779.x},
archivePrefix = {arXiv},
       eprint = {astro-ph/0005010},
 primaryClass = {astro-ph},
       adsurl = {https://ui.adsabs.harvard.edu/abs/2000MNRAS.318.1144P}
}

@ARTICLE{2023OJAp....6E..39A,
       author = {{Asgari}, Marika and {Mead}, Alexander J. and {Heymans}, Catherine},
        title = "{The halo model for cosmology: a pedagogical review}",
      journal = {The Open Journal of Astrophysics},
         year = 2023,
        month = nov,
       volume = {6},
          eid = {39},
        pages = {39},
          doi = {10.21105/astro.2303.08752},
archivePrefix = {arXiv},
       eprint = {2303.08752},
 primaryClass = {astro-ph.CO},
       adsurl = {https://ui.adsabs.harvard.edu/abs/2023OJAp....6E..39A}
}

@ARTICLE{2010MNRAS.405.2351N,
       author = {{Napolitano}, Nicola R. and {Romanowsky}, Aaron J. and {Tortora}, Crescenzo},
        title = "{The central dark matter content of early-type galaxies: scaling relations and connections with star formation histories}",
      journal = {\mnras},
         year = 2010,
        month = jul,
       volume = {405},
       number = {4},
        pages = {2351-2371},
          doi = {10.1111/j.1365-2966.2010.16710.x},
archivePrefix = {arXiv},
       eprint = {1003.1716},
 primaryClass = {astro-ph.CO},
       adsurl = {https://ui.adsabs.harvard.edu/abs/2010MNRAS.405.2351N}
}

@ARTICLE{busillo2023casco,
       author = {{Busillo}, V. and {Tortora}, C. and {Napolitano}, N.~R. and {Koopmans}, L.~V.~E. and {Covone}, G. and {Gentile}, F. and {Hunt.}, L.~K.},
        title = "{CASCO: Cosmological and AStrophysical parameters from Cosmological simulations and Observations - I. Constraining physical processes in local star-forming galaxies}",
      journal = {\mnras},
         year = 2023,
        month = nov,
       volume = {525},
       number = {4},
        pages = {6191-6213},
          doi = {10.1093/mnras/stad2691},
archivePrefix = {arXiv},
       eprint = {2308.14822},
 primaryClass = {astro-ph.GA},
       adsurl = {https://ui.adsabs.harvard.edu/abs/2023MNRAS.525.6191B}
}

@ARTICLE{Einasto-DM-halo-1965TrAlm...5...87E,
       author = {{Einasto}, J.},
        title = "{On the Construction of a Composite Model for the Galaxy and on the Determination of the System of Galactic Parameters}",
      journal = {Trudy Astrofizicheskogo Instituta Alma-Ata},
         year = 1965,
        month = jan,
       volume = {5},
        pages = {87-100},
       adsurl = {https://ui.adsabs.harvard.edu/abs/1965TrAlm...5...87E}
}

@ARTICLE{2014MNRAS.441.3359D,
       author = {{Dutton}, Aaron A. and {Macci{\`o}}, Andrea V.},
        title = "{Cold dark matter haloes in the Planck era: evolution of structural parameters for Einasto and NFW profiles}",
      journal = {\mnras},
         year = 2014,
        month = jul,
       volume = {441},
       number = {4},
        pages = {3359-3374},
          doi = {10.1093/mnras/stu742},
archivePrefix = {arXiv},
       eprint = {1402.7073},
 primaryClass = {astro-ph.CO},
       adsurl = {https://ui.adsabs.harvard.edu/abs/2014MNRAS.441.3359D}
}

@ARTICLE{2003AJ....125.2936G,
       author = {{Graham}, Alister W. and {Guzm{\'a}n}, Rafael},
        title = "{HST Photometry of Dwarf Elliptical Galaxies in Coma, and an Explanation for the Alleged Structural Dichotomy between Dwarf and Bright Elliptical Galaxies}",
      journal = {\aj},
         year = 2003,
        month = jun,
       volume = {125},
       number = {6},
        pages = {2936-2950},
          doi = {10.1086/374992},
archivePrefix = {arXiv},
       eprint = {astro-ph/0303391},
 primaryClass = {astro-ph},
       adsurl = {https://ui.adsabs.harvard.edu/abs/2003AJ....125.2936G}
}

@ARTICLE{2021A&A...646A..28S,
       author = {{Spiniello}, C. and {Tortora}, C. and {D'Ago}, G. and {Coccato}, L. and {La Barbera}, F. and {Ferr{\'e}-Mateu}, A. and {Napolitano}, N.~R. and {Spavone}, M. and {Scognamiglio}, D. and {Arnaboldi}, M. and {Gallazzi}, A. and {Hunt}, L. and {Moehler}, S. and {Radovich}, M. and {Zibetti}, S.},
        title = "{INSPIRE: INvestigating Stellar Population In RElics. I. Survey presentation and pilot study}",
      journal = {\aap},
         year = 2021,
        month = feb,
       volume = {646},
          eid = {A28},
        pages = {A28},
          doi = {10.1051/0004-6361/202038936},
archivePrefix = {arXiv},
       eprint = {2011.05347},
 primaryClass = {astro-ph.GA},
       adsurl = {https://ui.adsabs.harvard.edu/abs/2021A&A...646A..28S}
}

@ARTICLE{2020ApJ...893....4S,
       author = {{Scognamiglio}, Diana and {Tortora}, Crescenzo and {Spavone}, Marilena and {Spiniello}, Chiara and {Napolitano}, Nicola R. and {D'Ago}, Giuseppe and {La Barbera}, Francesco and {Getman}, Fedor and {Roy}, Nivya and {Raj}, Maria Angela and {Radovich}, Mario and {Brescia}, Massimo and {Cavuoti}, Stefano and {Koopmans}, L{\'e}on V.~E. and {Kuijken}, Konrad H. and {Longo}, Giuseppe and {Petrillo}, Carlo E.},
        title = "{Building the Largest Spectroscopic Sample of Ultracompact Massive Galaxies with the Kilo Degree Survey}",
      journal = {\apj},
         year = 2020,
        month = apr,
       volume = {893},
       number = {1},
          eid = {4},
        pages = {4},
          doi = {10.3847/1538-4357/ab7db3},
archivePrefix = {arXiv},
       eprint = {2002.12922},
 primaryClass = {astro-ph.GA},
       adsurl = {https://ui.adsabs.harvard.edu/abs/2020ApJ...893....4S}
}

@ARTICLE{2022MNRAS.510..500G,
       author = {{Gentile}, Fabrizio and {Tortora}, Crescenzo and {Covone}, Giovanni and {Koopmans}, L{\'e}on V.~E. and {Spiniello}, Chiara and {Fan}, Zuhui and {Li}, Rui and {Liu}, Dezi and {Napolitano}, Nicola R. and {Vaccari}, Mattia and {Fu}, Liping},
        title = "{Lenses In VoicE (LIVE): searching for strong gravitational lenses in the VOICE@VST survey using convolutional neural networks}",
      journal = {\mnras},
         year = 2022,
        month = feb,
       volume = {510},
       number = {1},
        pages = {500-514},
          doi = {10.1093/mnras/stab3386},
archivePrefix = {arXiv},
       eprint = {2105.05602},
 primaryClass = {astro-ph.GA},
       adsurl = {https://ui.adsabs.harvard.edu/abs/2022MNRAS.510..500G}
}

@ARTICLE{2016MNRAS.457.2845T,
       author = {{Tortora}, C. and {La Barbera}, F. and {Napolitano}, N.~R. and {Roy}, N. and {Radovich}, M. and {Cavuoti}, S. and {Brescia}, M. and {Longo}, G. and {Getman}, F. and {Capaccioli}, M. and {Grado}, A. and {Kuijken}, K.~H. and {de Jong}, J.~T.~A. and {McFarland}, J.~P. and {Puddu}, E.},
        title = "{Towards a census of supercompact massive galaxies in the Kilo Degree Survey}",
      journal = {\mnras},
         year = 2016,
        month = apr,
       volume = {457},
       number = {3},
        pages = {2845-2854},
          doi = {10.1093/mnras/stw184},
archivePrefix = {arXiv},
       eprint = {1507.00731},
 primaryClass = {astro-ph.GA},
       adsurl = {https://ui.adsabs.harvard.edu/abs/2016MNRAS.457.2845T}
}

@ARTICLE{2014ApJ...780L..20T,
       author = {{Trujillo}, Ignacio and {Ferr{\'e}-Mateu}, Anna and {Balcells}, Marc and {Vazdekis}, Alexandre and {S{\'a}nchez-Bl{\'a}zquez}, Patricia},
        title = "{NGC 1277: A Massive Compact Relic Galaxy in the Nearby Universe}",
      journal = {\apjl},
         year = 2014,
        month = jan,
       volume = {780},
       number = {2},
          eid = {L20},
        pages = {L20},
          doi = {10.1088/2041-8205/780/2/L20},
archivePrefix = {arXiv},
       eprint = {1310.6367},
 primaryClass = {astro-ph.CO},
       adsurl = {https://ui.adsabs.harvard.edu/abs/2014ApJ...780L..20T}
}

@ARTICLE{2023cosmo_multiple_gal,
       author = {{Chawak}, Chaitanya and {Villaescusa-Navarro}, Francisco and {Echeverri Rojas}, Nicolas and {Ni}, Yueying and {Hahn}, ChangHoon and {Angles-Alcazar}, Daniel},
        title = "{Cosmology with multiple galaxies}",
      journal = {arXiv e-prints},
         year = 2023,
        month = sep,
          eid = {arXiv:2309.12048},
        pages = {arXiv:2309.12048},
archivePrefix = {arXiv},
       eprint = {2309.12048},
 primaryClass = {astro-ph.CO},
       adsurl = {https://ui.adsabs.harvard.edu/abs/2023arXiv230912048C}
}

@ARTICLE{Cosmological-simulation-galaxy-formation-2020NatRP...2...42V,
       author = {{Vogelsberger}, Mark and {Marinacci}, Federico and {Torrey}, Paul and {Puchwein}, Ewald},
        title = "{Cosmological simulations of galaxy formation}",
      journal = {Nature Reviews Physics},
         year = 2020,
        month = jan,
       volume = {2},
       number = {1},
        pages = {42-66},
          doi = {10.1038/s42254-019-0127-2},
archivePrefix = {arXiv},
       eprint = {1909.07976},
 primaryClass = {astro-ph.GA},
       adsurl = {https://ui.adsabs.harvard.edu/abs/2020NatRP...2...42V}
}

@ARTICLE{Formation-dEs-2009MNRAS.396.2133K,
       author = {{Koleva}, Mina and {de Rijcke}, Sven and {Prugniel}, Philippe and {Zeilinger}, Werner W. and {Michielsen}, Dolf},
        title = "{Formation and evolution of dwarf elliptical galaxies - II. Spatially resolved star formation histories}",
      journal = {\mnras},
         year = 2009,
        month = jul,
       volume = {396},
       number = {4},
        pages = {2133-2151},
          doi = {10.1111/j.1365-2966.2009.14820.x},
archivePrefix = {arXiv},
       eprint = {0903.4393},
 primaryClass = {astro-ph.CO},
       adsurl = {https://ui.adsabs.harvard.edu/abs/2009MNRAS.396.2133K}
}

@ARTICLE{ETG-LTG-sSFR-2023A&A...669A..11P,
       author = {{Paspaliaris}, E. -D. and {Xilouris}, E.~M. and {Nersesian}, A. and {Bianchi}, S. and {Georgantopoulos}, I. and {Masoura}, V.~A. and {Magdis}, G.~E. and {Plionis}, M.},
        title = "{Star-forming early-type galaxies and quiescent late-type galaxies in the local Universe}",
      journal = {\aap},
         year = 2023,
        month = jan,
       volume = {669},
          eid = {A11},
        pages = {A11},
          doi = {10.1051/0004-6361/202244796},
archivePrefix = {arXiv},
       eprint = {2209.13437},
 primaryClass = {astro-ph.GA},
       adsurl = {https://ui.adsabs.harvard.edu/abs/2023A&A...669A..11P}
}

@ARTICLE{Gerhard-dynamical-properties-2001AJ....121.1936G,
       author = {{Gerhard}, Ortwin and {Kronawitter}, Andi and {Saglia}, R.~P. and {Bender}, Ralf},
        title = "{Dynamical Family Properties and Dark Halo Scaling Relations of Giant Elliptical Galaxies}",
      journal = {\aj},
         year = 2001,
        month = apr,
       volume = {121},
       number = {4},
        pages = {1936-1951},
          doi = {10.1086/319940},
archivePrefix = {arXiv},
       eprint = {astro-ph/0012381},
 primaryClass = {astro-ph},
       adsurl = {https://ui.adsabs.harvard.edu/abs/2001AJ....121.1936G}
}

@ARTICLE{SPARC-Mass-model-rotation-curve-Lelli-2016AJ....152..157L,
       author = {{Lelli}, Federico and {McGaugh}, Stacy S. and {Schombert}, James M.},
        title = "{SPARC: Mass Models for 175 Disk Galaxies with Spitzer Photometry and Accurate Rotation Curves}",
      journal = {\aj},
         year = 2016,
        month = dec,
       volume = {152},
       number = {6},
          eid = {157},
        pages = {157},
          doi = {10.3847/0004-6256/152/6/157},
archivePrefix = {arXiv},
       eprint = {1606.09251},
 primaryClass = {astro-ph.GA},
       adsurl = {https://ui.adsabs.harvard.edu/abs/2016AJ....152..157L}
}

@ARTICLE{virial-theorem-FP-1997A&A...320..415B,
       author = {{Busarello}, G. and {Capaccioli}, M. and {Capozziello}, S. and {Longo}, G. and {Puddu}, E.},
        title = "{The relation between the virial theorem and the fundamental plane of elliptical galaxies.}",
      journal = {\aap},
         year = 1997,
        month = apr,
       volume = {320},
        pages = {415-420},
       adsurl = {https://ui.adsabs.harvard.edu/abs/1997A&A...320..415B}
}

@ARTICLE{2023MNRAS.526.1022L,
       author = {{Lu}, Shengdong and {Zhu}, Kai and {Cappellari}, Michele and {Li}, Ran and {Mao}, Shude and {Xu}, Dandan},
        title = "{MaNGA DynPop - II. Global stellar population, gradients, and star-formation histories from integral-field spectroscopy of 10K galaxies: link with galaxy rotation, shape, and total-density gradients}",
      journal = {\mnras},
         year = 2023,
        month = nov,
       volume = {526},
       number = {1},
        pages = {1022-1045},
          doi = {10.1093/mnras/stad2732},
archivePrefix = {arXiv},
       eprint = {2304.11712},
 primaryClass = {astro-ph.GA},
       adsurl = {https://ui.adsabs.harvard.edu/abs/2023MNRAS.526.1022L}
}

@inproceedings{TPOT-OlsonGECCO2016,
    author = {Olson, Randal S. and Bartley, Nathan and Urbanowicz, Ryan J. and Moore, Jason H.},
    title = {Evaluation of a Tree-based Pipeline Optimization Tool for Automating Data Science},
    booktitle = {Proceedings of the Genetic and Evolutionary Computation Conference 2016},
    series = {GECCO '16},
    year = {2016},
    isbn = {978-1-4503-4206-3},
    location = {Denver, Colorado, USA},
    pages = {485--492},
    numpages = {8},
    url = {http://doi.acm.org/10.1145/2908812.2908918},
    doi = {10.1145/2908812.2908918},
    acmid = {2908918},
    publisher = {ACM},
    address = {New York, NY, USA},
}

@ARTICLE{Auger+10_SLACSX,
   author = {{Auger}, M.~W. and {Treu}, T. and {Bolton}, A.~S. and {Gavazzi}, R. and
    {Koopmans}, L.~V.~E. and {Marshall}, P.~J. and {Moustakas}, L.~A. and
    {Burles}, S.},
    title = "{The Sloan Lens ACS Survey. X. Stellar, Dynamical, and Total Mass Correlations of Massive Early-type Galaxies}",
  journal = {\apj},
archivePrefix = "arXiv",
   eprint = {1007.2880},
 primaryClass = "astro-ph.CO",
     year = 2010,
    month = nov,
   volume = 724,
    pages = {511-525},
      doi = {10.1088/0004-637X/724/1/511},
   adsurl = {http://adsabs.harvard.edu/abs/2010ApJ...724..511A}
}

@ARTICLE{Tortora+10lensing,
   author = {{Tortora}, C. and {Napolitano}, N.~R. and {Romanowsky}, A.~J. and
    {Jetzer}, P.},
    title = "{Central Dark Matter Trends in Early-type Galaxies from Strong Lensing, Dynamics, and Stellar Populations}",
  journal = {\apjl},
archivePrefix = "arXiv",
   eprint = {1007.3988},
 primaryClass = "astro-ph.CO",
     year = 2010,
    month = sep,
   volume = 721,
    pages = {L1-L5},
    doi = {10.1088/2041-8205/721/1/L1},
   adsurl = {http://adsabs.harvard.edu/abs/2010ApJ...721L...1T}
}

@ARTICLE{TK04,
   author = {{Treu}, T. and {Koopmans}, L.~V.~E.},
    title = "{Massive Dark Matter Halos and Evolution of Early-Type Galaxies to z \~{} 1}",
  journal = {\apj},
   eprint = {arXiv:astro-ph/0401373},
     year = 2004,
    month = aug,
   volume = 611,
    pages = {739-760},
      doi = {10.1086/422245},
   adsurl = {http://adsabs.harvard.edu/abs/2004ApJ...611..739T}
}

@ARTICLE{Koopmans+06_SLACSIII,
   author = {{Koopmans}, L.~V.~E. and {Treu}, T. and {Bolton}, A.~S. and
    {Burles}, S. and {Moustakas}, L.~A.},
    title = "{The Sloan Lens ACS Survey. III. The Structure and Formation of Early-Type Galaxies and Their Evolution since z \~{} 1}",
  journal = {\apj},
   eprint = {arXiv:astro-ph/0601628},
     year = 2006,
    month = oct,
   volume = 649,
    pages = {599-615},
      doi = {10.1086/505696},
   adsurl = {http://adsabs.harvard.edu/abs/2006ApJ...649..599K}
}

@ARTICLE{Sonnenfeld+13_SL2S_IV,
   author = {{Sonnenfeld}, A. and {Treu}, T. and {Gavazzi}, R. and {Suyu}, S.~H. and
    {Marshall}, P.~J. and {Auger}, M.~W. and {Nipoti}, C.},
    title = "{The SL2S Galaxy-scale Lens Sample. IV. The Dependence of the Total Mass Density Profile of Early-type Galaxies on Redshift, Stellar Mass, and Size}",
  journal = {\apj},
archivePrefix = "arXiv",
   eprint = {1307.4759},
 primaryClass = "astro-ph.CO",
     year = 2013,
    month = nov,
   volume = 777,
      eid = {98},
    pages = {98},
      doi = {10.1088/0004-637X/777/2/98},
   adsurl = {http://adsabs.harvard.edu/abs/2013ApJ...777...98S}
}

@ARTICLE{Battaglia-Nature-dwarf-local-group-2022NatAs...6..659B,
       author = {{Battaglia}, Giuseppina and {Nipoti}, Carlo},
        title = "{Stellar dynamics and dark matter in Local Group dwarf galaxies}",
      journal = {Nature Astronomy},
         year = 2022,
        month = may,
       volume = {6},
        pages = {659-672},
          doi = {10.1038/s41550-022-01638-7},
archivePrefix = {arXiv},
       eprint = {2205.07821},
 primaryClass = {astro-ph.GA},
       adsurl = {https://ui.adsabs.harvard.edu/abs/2022NatAs...6..659B}
}

@ARTICLE{SAMI-multi-object-integral-field-spectrograph-2012MNRAS.421..872C,
       author = {{Croom}, Scott M. and {Lawrence}, Jon S. and {Bland-Hawthorn}, Joss and {Bryant}, Julia J. and {Fogarty}, Lisa and {Richards}, Samuel and {Goodwin}, Michael and {Farrell}, Tony and {Miziarski}, Stan and {Heald}, Ron and {Jones}, D. Heath and {Lee}, Steve and {Colless}, Matthew and {Brough}, Sarah and {Hopkins}, Andrew M. and {Bauer}, Amanda E. and {Birchall}, Michael N. and {Ellis}, Simon and {Horton}, Anthony and {Leon-Saval}, Sergio and {Lewis}, Geraint and {L{\'o}pez-S{\'a}nchez}, {\'A}. R. and {Min}, Seong-Sik and {Trinh}, Christopher and {Trowland}, Holly},
        title = "{The Sydney-AAO Multi-object Integral field spectrograph}",
      journal = {\mnras},
         year = 2012,
        month = mar,
       volume = {421},
       number = {1},
        pages = {872-893},
          doi = {10.1111/j.1365-2966.2011.20365.x},
archivePrefix = {arXiv},
       eprint = {1112.3367},
 primaryClass = {astro-ph.CO},
       adsurl = {https://ui.adsabs.harvard.edu/abs/2012MNRAS.421..872C}
}

@ARTICLE{Blanton-SDSS-multiobject-spectro-2003AJ....125.2276B,
       author = {{Blanton}, Michael R. and {Lin}, Huan and {Lupton}, Robert H. and {Maley}, F. Miller and {Young}, Neal and {Zehavi}, Idit and {Loveday}, Jon},
        title = "{An Efficient Targeting Strategy for Multiobject Spectrograph Surveys: the Sloan Digital Sky Survey ``Tiling'' Algorithm}",
      journal = {\aj},
         year = 2003,
        month = apr,
       volume = {125},
       number = {4},
        pages = {2276-2286},
          doi = {10.1086/344761},
archivePrefix = {arXiv},
       eprint = {astro-ph/0105535},
 primaryClass = {astro-ph},
       adsurl = {https://ui.adsabs.harvard.edu/abs/2003AJ....125.2276B}
}

@ARTICLE{CAMLES-public-data-release-2023ApJS..265...54V,
       author = {{Villaescusa-Navarro}, Francisco and {Genel}, Shy and {Angl{\'e}s-Alc{\'a}zar}, Daniel and {Perez}, Lucia A. and {Villanueva-Domingo}, Pablo and {Wadekar}, Digvijay and {Shao}, Helen and {Mohammad}, Faizan G. and {Hassan}, Sultan and {Moser}, Emily and {Lau}, Erwin T. and {Machado Poletti Valle}, Luis Fernando and {Nicola}, Andrina and {Thiele}, Leander and {Jo}, Yongseok and {Philcox}, Oliver H.~E. and {Oppenheimer}, Benjamin D. and {Tillman}, Megan and {Hahn}, ChangHoon and {Kaushal}, Neerav and {Pisani}, Alice and {Gebhardt}, Matthew and {Delgado}, Ana Maria and {Caliendo}, Joyce and {Kreisch}, Christina and {Wong}, Kaze W.~K. and {Coulton}, William R. and {Eickenberg}, Michael and {Parimbelli}, Gabriele and {Ni}, Yueying and {Steinwandel}, Ulrich P. and {La Torre}, Valentina and {Dave}, Romeel and {Battaglia}, Nicholas and {Nagai}, Daisuke and {Spergel}, David N. and {Hernquist}, Lars and {Burkhart}, Blakesley and {Narayanan}, Desika and {Wandelt}, Benjamin and {Somerville}, Rachel S. and {Bryan}, Greg L. and {Viel}, Matteo and {Li}, Yin and {Irsic}, Vid and {Kraljic}, Katarina and {Marinacci}, Federico and {Vogelsberger}, Mark},
        title = "{The CAMELS Project: Public Data Release}",
      journal = {\apjs},
         year = 2023,
        month = apr,
       volume = {265},
       number = {2},
          eid = {54},
        pages = {54},
          doi = {10.3847/1538-4365/acbf47},
archivePrefix = {arXiv},
       eprint = {2201.01300},
 primaryClass = {astro-ph.CO},
       adsurl = {https://ui.adsabs.harvard.edu/abs/2023ApJS..265...54V}
}

@ARTICLE{Roy-2018-KiDS-stellar-mass,
       author = {{Roy}, N. and {Napolitano}, N.~R. and {La Barbera}, F. and {Tortora}, C. and {Getman}, F. and {Radovich}, M. and {Capaccioli}, M. and {Brescia}, M. and {Cavuoti}, S. and {Longo}, G. and {Raj}, M.~A. and {Puddu}, E. and {Covone}, G. and {Amaro}, V. and {Vellucci}, C. and {Grado}, A. and {Kuijken}, K. and {Verdoes Kleijn}, G. and {Valentijn}, E.},
        title = "{Evolution of galaxy size-stellar mass relation from the Kilo-Degree Survey}",
      journal = {\mnras},
         year = 2018,
        month = oct,
       volume = {480},
       number = {1},
        pages = {1057-1080},
          doi = {10.1093/mnras/sty1917},
archivePrefix = {arXiv},
       eprint = {1807.06085},
 primaryClass = {astro-ph.GA},
       adsurl = {https://ui.adsabs.harvard.edu/abs/2018MNRAS.480.1057R}
}

@ARTICLE{thomas-SDSS-III-velocity-dispersion,
       author = {{Thomas}, D. and {Steele}, O. and {Maraston}, C. and {Johansson}, J. and {Beifiori}, A. and {Pforr}, J. and {Str{\"o}mb{\"a}ck}, G. and {Tremonti}, C.~A. and {Wake}, D. and {Bizyaev}, D. and {Bolton}, A. and {Brewington}, H. and {Brownstein}, J.~R. and {Comparat}, J. and {Kneib}, J. -P. and {Malanushenko}, E. and {Malanushenko}, V. and {Oravetz}, D. and {Pan}, K. and {Parejko}, J.~K. and {Schneider}, D.~P. and {Shelden}, A. and {Simmons}, A. and {Snedden}, S. and {Tanaka}, M. and {Weaver}, B.~A. and {Yan}, R.},
        title = "{Stellar velocity dispersions and emission line properties of SDSS-III/BOSS galaxies}",
      journal = {\mnras},
         year = 2013,
        month = may,
       volume = {431},
       number = {2},
        pages = {1383-1397},
          doi = {10.1093/mnras/stt261},
archivePrefix = {arXiv},
       eprint = {1207.6115},
 primaryClass = {astro-ph.CO},
       adsurl = {https://ui.adsabs.harvard.edu/abs/2013MNRAS.431.1383T}
}

@ARTICLE{Lange-2015-GAMA-mass-size-relation,
   author = {{Lange}, R. and {Driver}, S.~P. and {Robotham}, A.~S.~G. and
    {Kelvin}, L.~S. and {Graham}, A.~W. and {Alpaslan}, M. and {Andrews}, S.~K. and
    {Baldry}, I.~K. and {Bamford}, S. and {Bland-Hawthorn}, J. and
    {Brough}, S. and {Cluver}, M.~E. and {Conselice}, C.~J. and
    {Davies}, L.~J.~M. and {Haeussler}, B. and {Konstantopoulos}, I.~S. and
    {Loveday}, J. and {Moffett}, A.~J. and {Norberg}, P. and {Phillipps}, S. and
    {Taylor}, E.~N. and {L{\'o}pez-S{\'a}nchez}, {\'A}.~R. and {Wilkins}, S.~M.
    },
    title = "{Galaxy And Mass Assembly (GAMA): mass-size relations of z $\lt$ 0.1 galaxies subdivided by S{\'e}rsic index, colour and morphology}",
  journal = {\mnras},
archivePrefix = "arXiv",
   eprint = {1411.6355},
     year = 2015,
    month = mar,
   volume = 447,
    pages = {2603-2630},
      doi = {10.1093/mnras/stu2467},
   adsurl = {http://adsabs.harvard.edu/abs/2015MNRAS.447.2603L}
}

@ARTICLE{Romanowsky+03,
   author = {{Romanowsky}, A.~J. and {Douglas}, N.~G. and {Arnaboldi}, M. and
    {Kuijken}, K. and {Merrifield}, M.~R. and {Napolitano}, N.~R. and
    {Capaccioli}, M. and {Freeman}, K.~C.},
    title = "{A Dearth of Dark Matter in Ordinary Elliptical Galaxies}",
  journal = {Science},
   eprint = {arXiv:astro-ph/0308518},
     year = 2003,
    month = sep,
   volume = 301,
    pages = {1696-1698},
      doi = {10.1126/science.1087441},
   adsurl = {http://adsabs.harvard.edu/abs/2003Sci...301.1696R}
}

@ARTICLE{ThomasJ+07_coma,
   author = {{Thomas}, J. and {Saglia}, R.~P. and {Bender}, R. and {Thomas}, D. and
    {Gebhardt}, K. and {Magorrian}, J. and {Corsini}, E.~M. and
    {Wegner}, G.},
    title = "{Dynamical modelling of luminous and dark matter in 17 Coma early-type galaxies}",
  journal = {\mnras},
archivePrefix = "arXiv",
   eprint = {0709.0691},
     year = 2007,
    month = dec,
   volume = 382,
    pages = {657-684},
      doi = {10.1111/j.1365-2966.2007.12434.x},
   adsurl = {http://adsabs.harvard.edu/abs/2007MNRAS.382..657T}
}

@ARTICLE{Kronawitter+00,
   author = {{Kronawitter}, A. and {Saglia}, R.~P. and {Gerhard}, O. and
    {Bender}, R.},
    title = "{Orbital structure and mass distribution in elliptical galaxies}",
  journal = {\aaps},
     year = 2000,
    month = may,
   volume = 144,
    pages = {53-84},
      doi = {10.1051/aas:2000199},
   adsurl = {http://adsabs.harvard.edu/abs/2000A%26AS..144...53K}
}

@ARTICLE{Cappellari-2013-ATLAS3D-XX-2013MNRAS.432.1862C,
       author = {{Cappellari}, Michele and {McDermid}, Richard M. and {Alatalo}, Katherine and {Blitz}, Leo and {Bois}, Maxime and {Bournaud}, Fr{\'e}d{\'e}ric and {Bureau}, M. and {Crocker}, Alison F. and {Davies}, Roger L. and {Davis}, Timothy A. and {de Zeeuw}, P.~T. and {Duc}, Pierre-Alain and {Emsellem}, Eric and {Khochfar}, Sadegh and {Krajnovi{\'c}}, Davor and {Kuntschner}, Harald and {Morganti}, Raffaella and {Naab}, Thorsten and {Oosterloo}, Tom and {Sarzi}, Marc and {Scott}, Nicholas and {Serra}, Paolo and {Weijmans}, Anne-Marie and {Young}, Lisa M.},
        title = "{The ATLAS$^{3D}$ project - XX. Mass-size and mass-{\ensuremath{\sigma}} distributions of early-type galaxies: bulge fraction drives kinematics, mass-to-light ratio, molecular gas fraction and stellar initial mass function}",
      journal = {\mnras},
         year = 2013,
        month = jul,
       volume = {432},
       number = {3},
        pages = {1862-1893},
          doi = {10.1093/mnras/stt644},
archivePrefix = {arXiv},
       eprint = {1208.3523},
 primaryClass = {astro-ph.CO},
       adsurl = {https://ui.adsabs.harvard.edu/abs/2013MNRAS.432.1862C}
}

@ARTICLE{SPIDER-V-stellar-mass-2011AJ....142..118S,
       author = {{Swindle}, R. and {Gal}, R.~R. and {La Barbera}, F. and {de Carvalho}, R.~R.},
        title = "{SPIDER. V. Measuring Systematic Effects in Early-type Galaxy Stellar Masses from Photometric Spectral Energy Distribution Fitting}",
      journal = {\aj},
         year = 2011,
        month = oct,
       volume = {142},
       number = {4},
          eid = {118},
        pages = {118},
          doi = {10.1088/0004-6256/142/4/118},
archivePrefix = {arXiv},
       eprint = {1107.5371},
 primaryClass = {astro-ph.CO},
       adsurl = {https://ui.adsabs.harvard.edu/abs/2011AJ....142..118S}
}

@ARTICLE{Wechsler-connectioin-galaxy-DM-halo-2018ARA&A..56..435W,
       author = {{Wechsler}, Risa H. and {Tinker}, Jeremy L.},
        title = "{The Connection Between Galaxies and Their Dark Matter Halos}",
      journal = {\araa},
         year = 2018,
        month = sep,
       volume = {56},
        pages = {435-487},
          doi = {10.1146/annurev-astro-081817-051756},
archivePrefix = {arXiv},
       eprint = {1804.03097},
 primaryClass = {astro-ph.GA},
       adsurl = {https://ui.adsabs.harvard.edu/abs/2018ARA&A..56..435W}
}

@ARTICLE{ATLAS3D-I-2011MNRAS.413..813C,
       author = {{Cappellari}, Michele and {Emsellem}, Eric and {Krajnovi{\'c}}, Davor and {McDermid}, Richard M. and {Scott}, Nicholas and {Verdoes Kleijn}, G.~A. and {Young}, Lisa M. and {Alatalo}, Katherine and {Bacon}, R. and {Blitz}, Leo and {Bois}, Maxime and {Bournaud}, Fr{\'e}d{\'e}ric and {Bureau}, M. and {Davies}, Roger L. and {Davis}, Timothy A. and {de Zeeuw}, P.~T. and {Duc}, Pierre-Alain and {Khochfar}, Sadegh and {Kuntschner}, Harald and {Lablanche}, Pierre-Yves and {Morganti}, Raffaella and {Naab}, Thorsten and {Oosterloo}, Tom and {Sarzi}, Marc and {Serra}, Paolo and {Weijmans}, Anne-Marie},
        title = "{The ATLAS$^{3D}$ project - I. A volume-limited sample of 260 nearby early-type galaxies: science goals and selection criteria}",
      journal = {\mnras},
         year = 2011,
        month = may,
       volume = {413},
       number = {2},
        pages = {813-836},
          doi = {10.1111/j.1365-2966.2010.18174.x},
archivePrefix = {arXiv},
       eprint = {1012.1551},
 primaryClass = {astro-ph.CO},
       adsurl = {https://ui.adsabs.harvard.edu/abs/2011MNRAS.413..813C}
}

@ARTICLE{4MOST-project-overview-2019Msngr.175....3D,
       author = {{de Jong}, R.~S. and {Agertz}, O. and {Berbel}, A.~A. and {Aird}, J. and {Alexander}, D.~A. and {Amarsi}, A. and {Anders}, F. and {Andrae}, R. and {Ansarinejad}, B. and {Ansorge}, W. and {Antilogus}, P. and {Anwand-Heerwart}, H. and {Arentsen}, A. and {Arnadottir}, A. and {Asplund}, M. and {Auger}, M. and {Azais}, N. and {Baade}, D. and {Baker}, G. and {Baker}, S. and {Balbinot}, E. and {Baldry}, I.~K. and {Banerji}, M. and {Barden}, S. and {Barklem}, P. and {Barth{\'e}l{\'e}my-Mazot}, E. and {Battistini}, C. and {Bauer}, S. and {Bell}, C.~P.~M. and {Bellido-Tirado}, O. and {Bellstedt}, S. and {Belokurov}, V. and {Bensby}, T. and {Bergemann}, M. and {Bestenlehner}, J.~M. and {Bielby}, R. and {Bilicki}, M. and {Blake}, C. and {Bland-Hawthorn}, J. and {Boeche}, C. and {Boland}, W. and {Boller}, T. and {Bongard}, S. and {Bongiorno}, A. and {Bonifacio}, P. and {Boudon}, D. and {Brooks}, D. and {Brown}, M.~J.~I. and {Brown}, R. and {Br{\"u}ggen}, M. and {Brynnel}, J. and {Brzeski}, J. and {Buchert}, T. and {Buschkamp}, P. and {Caffau}, E. and {Caillier}, P. and {Carrick}, J. and {Casagrande}, L. and {Case}, S. and {Casey}, A. and {Cesarini}, I. and {Cescutti}, G. and {Chapuis}, D. and {Chiappini}, C. and {Childress}, M. and {Christlieb}, N. and {Church}, R. and {Cioni}, M. -R.~L. and {Cluver}, M. and {Colless}, M. and {Collett}, T. and {Comparat}, J. and {Cooper}, A. and {Couch}, W. and {Courbin}, F. and {Croom}, S. and {Croton}, D. and {Daguis{\'e}}, E. and {Dalton}, G. and {Davies}, L.~J.~M. and {Davis}, T. and {de Laverny}, P. and {Deason}, A. and {Dionies}, F. and {Disseau}, K. and {Doel}, P. and {D{\"o}scher}, D. and {Driver}, S.~P. and {Dwelly}, T. and {Eckert}, D. and {Edge}, A. and {Edvardsson}, B. and {Youssoufi}, D.~E. and {Elhaddad}, A. and {Enke}, H. and {Erfanianfar}, G. and {Farrell}, T. and {Fechner}, T. and {Feiz}, C. and {Feltzing}, S. and {Ferreras}, I. and {Feuerstein}, D. and {Feuillet}, D. and {Finoguenov}, A. and {Ford}, D. and {Fotopoulou}, S. and {Fouesneau}, M. and {Frenk}, C. and {Frey}, S. and {Gaessler}, W. and {Geier}, S. and {Gentile Fusillo}, N. and {Gerhard}, O. and {Giannantonio}, T. and {Giannone}, D. and {Gibson}, B. and {Gillingham}, P. and {Gonz{\'a}lez-Fern{\'a}ndez}, C. and {Gonzalez-Solares}, E. and {Gottloeber}, S. and {Gould}, A. and {Grebel}, E.~K. and {Gueguen}, A. and {Guiglion}, G. and {Haehnelt}, M. and {Hahn}, T. and {Hansen}, C.~J. and {Hartman}, H. and {Hauptner}, K. and {Hawkins}, K. and {Haynes}, D. and {Haynes}, R. and {Heiter}, U. and {Helmi}, A. and {Aguayo}, C.~H. and {Hewett}, P. and {Hinton}, S. and {Hobbs}, D. and {Hoenig}, S. and {Hofman}, D. and {Hook}, I. and {Hopgood}, J. and {Hopkins}, A. and {Hourihane}, A. and {Howes}, L. and {Howlett}, C. and {Huet}, T. and {Irwin}, M. and {Iwert}, O. and {Jablonka}, P. and {Jahn}, T. and {Jahnke}, K. and {Jarno}, A. and {Jin}, S. and {Jofre}, P. and {Johl}, D. and {Jones}, D. and {J{\"o}nsson}, H. and {Jordan}, C. and {Karovicova}, I. and {Khalatyan}, A. and {Kelz}, A. and {Kennicutt}, R. and {King}, D. and {Kitaura}, F. and {Klar}, J. and {Klauser}, U. and {Kneib}, J. -P. and {Koch}, A. and {Koposov}, S. and {Kordopatis}, G. and {Korn}, A. and {Kosmalski}, J. and {Kotak}, R. and {Kovalev}, M. and {Kreckel}, K. and {Kripak}, Y. and {Krumpe}, M. and {Kuijken}, K. and {Kunder}, A. and {Kushniruk}, I. and {Lam}, M.~I. and {Lamer}, G. and {Laurent}, F. and {Lawrence}, J. and {Lehmitz}, M. and {Lemasle}, B. and {Lewis}, J. and {Li}, B. and {Lidman}, C. and {Lind}, K. and {Liske}, J. and {Lizon}, J. -L. and {Loveday}, J. and {Ludwig}, H. -G. and {McDermid}, R.~M. and {Maguire}, K. and {Mainieri}, V. and {Mali}, S. and {Mandel}, H. and {Mandel}, K. and {Mannering}, L. and {Martell}, S. and {Martinez Delgado}, D. and {Matijevic}, G. and {McGregor}, H. and {McMahon}, R. and {McMillan}, P. and {Mena}, O. and {Merloni}, A. and {Meyer}, M.~J. and {Michel}, C. and {Micheva}, G. and {Migniau}, J. -E. and {Minchev}, I. and {Monari}, G. and {Muller}, R. and {Murphy}, D. and {Muthukrishna}, D. and {Nandra}, K. and {Navarro}, R. and {Ness}, M. and {Nichani}, V. and {Nichol}, R. and {Nicklas}, H. and {Niederhofer}, F. and {Norberg}, P. and {Obreschkow}, D. and {Oliver}, S. and {Owers}, M. and {Pai}, N. and {Pankratow}, S. and {Parkinson}, D. and {Paschke}, J. and {Paterson}, R. and {Pecontal}, A. and {Parry}, I. and {Phillips}, D. and {Pillepich}, A. and {Pinard}, L. and {Pirard}, J. and {Piskunov}, N. and {Plank}, V. and {Pl{\"u}schke}, D. and {Pons}, E. and {Popesso}, P. and {Power}, C. and {Pragt}, J. and {Pramskiy}, A. and {Pryer}, D. and {Quattri}, M. and {Queiroz}, A.~B. d. A. and {Quirrenbach}, A. and {Rahurkar}, S. and {Raichoor}, A. and {Ramstedt}, S. and {Rau}, A. and {Recio-Blanco}, A. and {Reiss}, R. and {Renaud}, F. and {Revaz}, Y. and {Rhode}, P. and {Richard}, J. and {Richter}, A.~D. and {Rix}, H. -W. and {Robotham}, A.~S.~G. and {Roelfsema}, R. and {Romaniello}, M. and {Rosario}, D. and {Rothmaier}, F. and {Roukema}, B. and {Ruchti}, G. and {Rupprecht}, G. and {Rybizki}, J. and {Ryde}, N. and {Saar}, A. and {Sadler}, E. and {Sahl{\'e}n}, M. and {Salvato}, M. and {Sassolas}, B. and {Saunders}, W. and {Saviauk}, A. and {Sbordone}, L. and {Schmidt}, T. and {Schnurr}, O. and {Scholz}, R. -D. and {Schwope}, A. and {Seifert}, W. and {Shanks}, T. and {Sheinis}, A. and {Sivov}, T. and {Sk{\'u}lad{\'o}ttir}, {\'A}. and {Smartt}, S. and {Smedley}, S. and {Smith}, G. and {Smith}, R. and {Sorce}, J. and {Spitler}, L. and {Starkenburg}, E. and {Steinmetz}, M. and {Stilz}, I. and {Storm}, J. and {Sullivan}, M. and {Sutherland}, W. and {Swann}, E. and {Tamone}, A. and {Taylor}, E.~N. and {Teillon}, J. and {Tempel}, E. and {ter Horst}, R. and {Thi}, W. -F. and {Tolstoy}, E. and {Trager}, S. and {Traven}, G. and {Tremblay}, P. -E. and {Tresse}, L. and {Valentini}, M. and {van de Weygaert}, R. and {van den Ancker}, M. and {Veljanoski}, J. and {Venkatesan}, S. and {Wagner}, L. and {Wagner}, K. and {Walcher}, C.~J. and {Waller}, L. and {Walton}, N. and {Wang}, L. and {Winkler}, R. and {Wisotzki}, L. and {Worley}, C.~C. and {Worseck}, G. and {Xiang}, M. and {Xu}, W. and {Yong}, D. and {Zhao}, C. and {Zheng}, J. and {Zscheyge}, F. and {Zucker}, D.},
        title = "{4MOST: Project overview and information for the First Call for Proposals}",
      journal = {The Messenger},
         year = 2019,
        month = mar,
       volume = {175},
        pages = {3-11},
          doi = {10.18727/0722-6691/5117},
archivePrefix = {arXiv},
       eprint = {1903.02464},
 primaryClass = {astro-ph.IM},
       adsurl = {https://ui.adsabs.harvard.edu/abs/2019Msngr.175....3D}
}

@ARTICLE{DESI-whitepaper-2013arXiv1308.0847L,
       author = {{Levi}, Michael and {Bebek}, Chris and {Beers}, Timothy and {Blum}, Robert and {Cahn}, Robert and {Eisenstein}, Daniel and {Flaugher}, Brenna and {Honscheid}, Klaus and {Kron}, Richard and {Lahav}, Ofer and {McDonald}, Patrick and {Roe}, Natalie and {Schlegel}, David and {representing the DESI collaboration}},
        title = "{The DESI Experiment, a whitepaper for Snowmass 2013}",
      journal = {arXiv e-prints},
         year = 2013,
        month = aug,
          eid = {arXiv:1308.0847},
        pages = {arXiv:1308.0847},
          doi = {10.48550/arXiv.1308.0847},
archivePrefix = {arXiv},
       eprint = {1308.0847},
 primaryClass = {astro-ph.CO},
       adsurl = {https://ui.adsabs.harvard.edu/abs/2013arXiv1308.0847L}
}

@ARTICLE{DESI-Scienc-target-survey-design-2016arXiv161100036D,
       author = {{DESI Collaboration} and {Aghamousa}, Amir and {Aguilar}, Jessica and {Ahlen}, Steve and {Alam}, Shadab and {Allen}, Lori E. and {Allende Prieto}, Carlos and {Annis}, James and {Bailey}, Stephen and {Balland}, Christophe and {Ballester}, Otger and {Baltay}, Charles and {Beaufore}, Lucas and {Bebek}, Chris and {Beers}, Timothy C. and {Bell}, Eric F. and {Bernal}, Jos{\'e} Luis and {Besuner}, Robert and {Beutler}, Florian and {Blake}, Chris and {Bleuler}, Hannes and {Blomqvist}, Michael and {Blum}, Robert and {Bolton}, Adam S. and {Briceno}, Cesar and {Brooks}, David and {Brownstein}, Joel R. and {Buckley-Geer}, Elizabeth and {Burden}, Angela and {Burtin}, Etienne and {Busca}, Nicolas G. and {Cahn}, Robert N. and {Cai}, Yan-Chuan and {Cardiel-Sas}, Laia and {Carlberg}, Raymond G. and {Carton}, Pierre-Henri and {Casas}, Ricard and {Castander}, Francisco J. and {Cervantes-Cota}, Jorge L. and {Claybaugh}, Todd M. and {Close}, Madeline and {Coker}, Carl T. and {Cole}, Shaun and {Comparat}, Johan and {Cooper}, Andrew P. and {Cousinou}, M. -C. and {Crocce}, Martin and {Cuby}, Jean-Gabriel and {Cunningham}, Daniel P. and {Davis}, Tamara M. and {Dawson}, Kyle S. and {de la Macorra}, Axel and {De Vicente}, Juan and {Delubac}, Timoth{\'e}e and {Derwent}, Mark and {Dey}, Arjun and {Dhungana}, Govinda and {Ding}, Zhejie and {Doel}, Peter and {Duan}, Yutong T. and {Ealet}, Anne and {Edelstein}, Jerry and {Eftekharzadeh}, Sarah and {Eisenstein}, Daniel J. and {Elliott}, Ann and {Escoffier}, St{\'e}phanie and {Evatt}, Matthew and {Fagrelius}, Parker and {Fan}, Xiaohui and {Fanning}, Kevin and {Farahi}, Arya and {Farihi}, Jay and {Favole}, Ginevra and {Feng}, Yu and {Fernandez}, Enrique and {Findlay}, Joseph R. and {Finkbeiner}, Douglas P. and {Fitzpatrick}, Michael J. and {Flaugher}, Brenna and {Flender}, Samuel and {Font-Ribera}, Andreu and {Forero-Romero}, Jaime E. and {Fosalba}, Pablo and {Frenk}, Carlos S. and {Fumagalli}, Michele and {Gaensicke}, Boris T. and {Gallo}, Giuseppe and {Garcia-Bellido}, Juan and {Gaztanaga}, Enrique and {Pietro Gentile Fusillo}, Nicola and {Gerard}, Terry and {Gershkovich}, Irena and {Giannantonio}, Tommaso and {Gillet}, Denis and {Gonzalez-de-Rivera}, Guillermo and {Gonzalez-Perez}, Violeta and {Gott}, Shelby and {Graur}, Or and {Gutierrez}, Gaston and {Guy}, Julien and {Habib}, Salman and {Heetderks}, Henry and {Heetderks}, Ian and {Heitmann}, Katrin and {Hellwing}, Wojciech A. and {Herrera}, David A. and {Ho}, Shirley and {Holland}, Stephen and {Honscheid}, Klaus and {Huff}, Eric and {Hutchinson}, Timothy A. and {Huterer}, Dragan and {Hwang}, Ho Seong and {Illa Laguna}, Joseph Maria and {Ishikawa}, Yuzo and {Jacobs}, Dianna and {Jeffrey}, Niall and {Jelinsky}, Patrick and {Jennings}, Elise and {Jiang}, Linhua and {Jimenez}, Jorge and {Johnson}, Jennifer and {Joyce}, Richard and {Jullo}, Eric and {Juneau}, St{\'e}phanie and {Kama}, Sami and {Karcher}, Armin and {Karkar}, Sonia and {Kehoe}, Robert and {Kennamer}, Noble and {Kent}, Stephen and {Kilbinger}, Martin and {Kim}, Alex G. and {Kirkby}, David and {Kisner}, Theodore and {Kitanidis}, Ellie and {Kneib}, Jean-Paul and {Koposov}, Sergey and {Kovacs}, Eve and {Koyama}, Kazuya and {Kremin}, Anthony and {Kron}, Richard and {Kronig}, Luzius and {Kueter-Young}, Andrea and {Lacey}, Cedric G. and {Lafever}, Robin and {Lahav}, Ofer and {Lambert}, Andrew and {Lampton}, Michael and {Landriau}, Martin and {Lang}, Dustin and {Lauer}, Tod R. and {Le Goff}, Jean-Marc and {Le Guillou}, Laurent and {Le Van Suu}, Auguste and {Lee}, Jae Hyeon and {Lee}, Su-Jeong and {Leitner}, Daniela and {Lesser}, Michael and {Levi}, Michael E. and {L'Huillier}, Benjamin and {Li}, Baojiu and {Liang}, Ming and {Lin}, Huan and {Linder}, Eric and {Loebman}, Sarah R. and {Luki{\'c}}, Zarija and {Ma}, Jun and {MacCrann}, Niall and {Magneville}, Christophe and {Makarem}, Laleh and {Manera}, Marc and {Manser}, Christopher J. and {Marshall}, Robert and {Martini}, Paul and {Massey}, Richard and {Matheson}, Thomas and {McCauley}, Jeremy and {McDonald}, Patrick and {McGreer}, Ian D. and {Meisner}, Aaron and {Metcalfe}, Nigel and {Miller}, Timothy N. and {Miquel}, Ramon and {Moustakas}, John and {Myers}, Adam and {Naik}, Milind and {Newman}, Jeffrey A. and {Nichol}, Robert C. and {Nicola}, Andrina and {Nicolati da Costa}, Luiz and {Nie}, Jundan and {Niz}, Gustavo and {Norberg}, Peder and {Nord}, Brian and {Norman}, Dara and {Nugent}, Peter and {O'Brien}, Thomas and {Oh}, Minji and {Olsen}, Knut A.~G. and {Padilla}, Cristobal and {Padmanabhan}, Hamsa and {Padmanabhan}, Nikhil and {Palanque-Delabrouille}, Nathalie and {Palmese}, Antonella and {Pappalardo}, Daniel and {P{\^a}ris}, Isabelle and {Park}, Changbom and {Patej}, Anna and {Peacock}, John A. and {Peiris}, Hiranya V. and {Peng}, Xiyan and {Percival}, Will J. and {Perruchot}, Sandrine and {Pieri}, Matthew M. and {Pogge}, Richard and {Pollack}, Jennifer E. and {Poppett}, Claire and {Prada}, Francisco and {Prakash}, Abhishek and {Probst}, Ronald G. and {Rabinowitz}, David and {Raichoor}, Anand and {Ree}, Chang Hee and {Refregier}, Alexandre and {Regal}, Xavier and {Reid}, Beth and {Reil}, Kevin and {Rezaie}, Mehdi and {Rockosi}, Constance M. and {Roe}, Natalie and {Ronayette}, Samuel and {Roodman}, Aaron and {Ross}, Ashley J. and {Ross}, Nicholas P. and {Rossi}, Graziano and {Rozo}, Eduardo and {Ruhlmann-Kleider}, Vanina and {Rykoff}, Eli S. and {Sabiu}, Cristiano and {Samushia}, Lado and {Sanchez}, Eusebio and {Sanchez}, Javier and {Schlegel}, David J. and {Schneider}, Michael and {Schubnell}, Michael and {Secroun}, Aur{\'e}lia and {Seljak}, Uros and {Seo}, Hee-Jong and {Serrano}, Santiago and {Shafieloo}, Arman and {Shan}, Huanyuan and {Sharples}, Ray and {Sholl}, Michael J. and {Shourt}, William V. and {Silber}, Joseph H. and {Silva}, David R. and {Sirk}, Martin M. and {Slosar}, Anze and {Smith}, Alex and {Smoot}, George F. and {Som}, Debopam and {Song}, Yong-Seon and {Sprayberry}, David and {Staten}, Ryan and {Stefanik}, Andy and {Tarle}, Gregory and {Sien Tie}, Suk and {Tinker}, Jeremy L. and {Tojeiro}, Rita and {Valdes}, Francisco and {Valenzuela}, Octavio and {Valluri}, Monica and {Vargas-Magana}, Mariana and {Verde}, Licia and {Walker}, Alistair R. and {Wang}, Jiali and {Wang}, Yuting and {Weaver}, Benjamin A. and {Weaverdyck}, Curtis and {Wechsler}, Risa H. and {Weinberg}, David H. and {White}, Martin and {Yang}, Qian and {Yeche}, Christophe and {Zhang}, Tianmeng and {Zhao}, Gong-Bo and {Zheng}, Yi and {Zhou}, Xu and {Zhou}, Zhimin and {Zhu}, Yaling and {Zou}, Hu and {Zu}, Ying},
        title = "{The DESI Experiment Part I: Science,Targeting, and Survey Design}",
      journal = {arXiv e-prints},
         year = 2016,
        month = oct,
          eid = {arXiv:1611.00036},
        pages = {arXiv:1611.00036},
          doi = {10.48550/arXiv.1611.00036},
archivePrefix = {arXiv},
       eprint = {1611.00036},
 primaryClass = {astro-ph.IM},
       adsurl = {https://ui.adsabs.harvard.edu/abs/2016arXiv161100036D}
}

@ARTICLE{4MOST2011Msngr.145...14D,
       author = {{de Jong}, R.},
        title = "{4MOST {\textemdash} 4-metre Multi-Object Spectroscopic Telescope}",
      journal = {The Messenger},
         year = 2011,
        month = sep,
       volume = {145},
        pages = {14-16},
       adsurl = {https://ui.adsabs.harvard.edu/abs/2011Msngr.145...14D}
}

@ARTICLE{ZhanHu-CSST-2011SSPMA..41.1441Z,
       author = {{Zhan}, Hu},
        title = "{Consideration for a large-scale multi-color imaging and slitless spectroscopy survey on the Chinese space station and its application in dark energy research}",
      journal = {Scientia Sinica Physica, Mechanica \& Astronomica},
         year = 2011,
        month = jan,
       volume = {41},
       number = {12},
        pages = {1441},
          doi = {10.1360/132011-961},
       adsurl = {https://ui.adsabs.harvard.edu/abs/2011SSPMA..41.1441Z}
}

@ARTICLE{Roman-2015arXiv150303757S,
       author = {{Spergel}, D. and {Gehrels}, N. and {Baltay}, C. and {Bennett}, D. and {Breckinridge}, J. and {Donahue}, M. and {Dressler}, A. and {Gaudi}, B.~S. and {Greene}, T. and {Guyon}, O. and {Hirata}, C. and {Kalirai}, J. and {Kasdin}, N.~J. and {Macintosh}, B. and {Moos}, W. and {Perlmutter}, S. and {Postman}, M. and {Rauscher}, B. and {Rhodes}, J. and {Wang}, Y. and {Weinberg}, D. and {Benford}, D. and {Hudson}, M. and {Jeong}, W. -S. and {Mellier}, Y. and {Traub}, W. and {Yamada}, T. and {Capak}, P. and {Colbert}, J. and {Masters}, D. and {Penny}, M. and {Savransky}, D. and {Stern}, D. and {Zimmerman}, N. and {Barry}, R. and {Bartusek}, L. and {Carpenter}, K. and {Cheng}, E. and {Content}, D. and {Dekens}, F. and {Demers}, R. and {Grady}, K. and {Jackson}, C. and {Kuan}, G. and {Kruk}, J. and {Melton}, M. and {Nemati}, B. and {Parvin}, B. and {Poberezhskiy}, I. and {Peddie}, C. and {Ruffa}, J. and {Wallace}, J.~K. and {Whipple}, A. and {Wollack}, E. and {Zhao}, F.},
        title = "{Wide-Field InfrarRed Survey Telescope-Astrophysics Focused Telescope Assets WFIRST-AFTA 2015 Report}",
      journal = {arXiv e-prints},
         year = 2015,
        month = mar,
          eid = {arXiv:1503.03757},
        pages = {arXiv:1503.03757},
          doi = {10.48550/arXiv.1503.03757},
archivePrefix = {arXiv},
       eprint = {1503.03757},
 primaryClass = {astro-ph.IM},
       adsurl = {https://ui.adsabs.harvard.edu/abs/2015arXiv150303757S}
}

@ARTICLE{Euclid-2011,
       author = {{Laureijs}, R. and {Amiaux}, J. and {Arduini}, S. and {Augu{\`e}res}, J. -L. and {Brinchmann}, J. and {Cole}, R. and {Cropper}, M. and {Dabin}, C. and {Duvet}, L. and {Ealet}, A. and {Garilli}, B. and {Gondoin}, P. and {Guzzo}, L. and {Hoar}, J. and {Hoekstra}, H. and {Holmes}, R. and {Kitching}, T. and {Maciaszek}, T. and {Mellier}, Y. and {Pasian}, F. and {Percival}, W. and {Rhodes}, J. and {Saavedra Criado}, G. and {Sauvage}, M. and {Scaramella}, R. and {Valenziano}, L. and {Warren}, S. and {Bender}, R. and {Castander}, F. and {Cimatti}, A. and {Le F{\`e}vre}, O. and {Kurki-Suonio}, H. and {Levi}, M. and {Lilje}, P. and {Meylan}, G. and {Nichol}, R. and {Pedersen}, K. and {Popa}, V. and {Rebolo Lopez}, R. and {Rix}, H. -W. and {Rottgering}, H. and {Zeilinger}, W. and {Grupp}, F. and {Hudelot}, P. and {Massey}, R. and {Meneghetti}, M. and {Miller}, L. and {Paltani}, S. and {Paulin-Henriksson}, S. and {Pires}, S. and {Saxton}, C. and {Schrabback}, T. and {Seidel}, G. and {Walsh}, J. and {Aghanim}, N. and {Amendola}, L. and {Bartlett}, J. and {Baccigalupi}, C. and {Beaulieu}, J. -P. and {Benabed}, K. and {Cuby}, J. -G. and {Elbaz}, D. and {Fosalba}, P. and {Gavazzi}, G. and {Helmi}, A. and {Hook}, I. and {Irwin}, M. and {Kneib}, J. -P. and {Kunz}, M. and {Mannucci}, F. and {Moscardini}, L. and {Tao}, C. and {Teyssier}, R. and {Weller}, J. and {Zamorani}, G. and {Zapatero Osorio}, M.~R. and {Boulade}, O. and {Foumond}, J.~J. and {Di Giorgio}, A. and {Guttridge}, P. and {James}, A. and {Kemp}, M. and {Martignac}, J. and {Spencer}, A. and {Walton}, D. and {Bl{\"u}mchen}, T. and {Bonoli}, C. and {Bortoletto}, F. and {Cerna}, C. and {Corcione}, L. and {Fabron}, C. and {Jahnke}, K. and {Ligori}, S. and {Madrid}, F. and {Martin}, L. and {Morgante}, G. and {Pamplona}, T. and {Prieto}, E. and {Riva}, M. and {Toledo}, R. and {Trifoglio}, M. and {Zerbi}, F. and {Abdalla}, F. and {Douspis}, M. and {Grenet}, C. and {Borgani}, S. and {Bouwens}, R. and {Courbin}, F. and {Delouis}, J. -M. and {Dubath}, P. and {Fontana}, A. and {Frailis}, M. and {Grazian}, A. and {Koppenh{\"o}fer}, J. and {Mansutti}, O. and {Melchior}, M. and {Mignoli}, M. and {Mohr}, J. and {Neissner}, C. and {Noddle}, K. and {Poncet}, M. and {Scodeggio}, M. and {Serrano}, S. and {Shane}, N. and {Starck}, J. -L. and {Surace}, C. and {Taylor}, A. and {Verdoes-Kleijn}, G. and {Vuerli}, C. and {Williams}, O.~R. and {Zacchei}, A. and {Altieri}, B. and {Escudero Sanz}, I. and {Kohley}, R. and {Oosterbroek}, T. and {Astier}, P. and {Bacon}, D. and {Bardelli}, S. and {Baugh}, C. and {Bellagamba}, F. and {Benoist}, C. and {Bianchi}, D. and {Biviano}, A. and {Branchini}, E. and {Carbone}, C. and {Cardone}, V. and {Clements}, D. and {Colombi}, S. and {Conselice}, C. and {Cresci}, G. and {Deacon}, N. and {Dunlop}, J. and {Fedeli}, C. and {Fontanot}, F. and {Franzetti}, P. and {Giocoli}, C. and {Garcia-Bellido}, J. and {Gow}, J. and {Heavens}, A. and {Hewett}, P. and {Heymans}, C. and {Holland}, A. and {Huang}, Z. and {Ilbert}, O. and {Joachimi}, B. and {Jennins}, E. and {Kerins}, E. and {Kiessling}, A. and {Kirk}, D. and {Kotak}, R. and {Krause}, O. and {Lahav}, O. and {van Leeuwen}, F. and {Lesgourgues}, J. and {Lombardi}, M. and {Magliocchetti}, M. and {Maguire}, K. and {Majerotto}, E. and {Maoli}, R. and {Marulli}, F. and {Maurogordato}, S. and {McCracken}, H. and {McLure}, R. and {Melchiorri}, A. and {Merson}, A. and {Moresco}, M. and {Nonino}, M. and {Norberg}, P. and {Peacock}, J. and {Pello}, R. and {Penny}, M. and {Pettorino}, V. and {Di Porto}, C. and {Pozzetti}, L. and {Quercellini}, C. and {Radovich}, M. and {Rassat}, A. and {Roche}, N. and {Ronayette}, S. and {Rossetti}, E. and {Sartoris}, B. and {Schneider}, P. and {Semboloni}, E. and {Serjeant}, S. and {Simpson}, F. and {Skordis}, C. and {Smadja}, G. and {Smartt}, S. and {Spano}, P. and {Spiro}, S. and {Sullivan}, M. and {Tilquin}, A. and {Trotta}, R. and {Verde}, L. and {Wang}, Y. and {Williger}, G. and {Zhao}, G. and {Zoubian}, J. and {Zucca}, E.},
        title = "{Euclid Definition Study Report}",
      journal = {arXiv e-prints},
         year = 2011,
        month = oct,
          eid = {arXiv:1110.3193},
        pages = {arXiv:1110.3193},
          doi = {10.48550/arXiv.1110.3193},
archivePrefix = {arXiv},
       eprint = {1110.3193},
 primaryClass = {astro-ph.CO},
       adsurl = {https://ui.adsabs.harvard.edu/abs/2011arXiv1110.3193L}
}

@ARTICLE{LSST-2019,
       author = {{Ivezi{\'c}}, {\v{Z}}eljko and {Kahn}, Steven M. and {Tyson}, J. Anthony and {Abel}, Bob and {Acosta}, Emily and {Allsman}, Robyn and {Alonso}, David and {AlSayyad}, Yusra and {Anderson}, Scott F. and {Andrew}, John and {Angel}, James Roger P. and {Angeli}, George Z. and {Ansari}, Reza and {Antilogus}, Pierre and {Araujo}, Constanza and {Armstrong}, Robert and {Arndt}, Kirk T. and {Astier}, Pierre and {Aubourg}, {\'E}ric and {Auza}, Nicole and {Axelrod}, Tim S. and {Bard}, Deborah J. and {Barr}, Jeff D. and {Barrau}, Aurelian and {Bartlett}, James G. and {Bauer}, Amanda E. and {Bauman}, Brian J. and {Baumont}, Sylvain and {Bechtol}, Ellen and {Bechtol}, Keith and {Becker}, Andrew C. and {Becla}, Jacek and {Beldica}, Cristina and {Bellavia}, Steve and {Bianco}, Federica B. and {Biswas}, Rahul and {Blanc}, Guillaume and {Blazek}, Jonathan and {Blandford}, Roger D. and {Bloom}, Josh S. and {Bogart}, Joanne and {Bond}, Tim W. and {Booth}, Michael T. and {Borgland}, Anders W. and {Borne}, Kirk and {Bosch}, James F. and {Boutigny}, Dominique and {Brackett}, Craig A. and {Bradshaw}, Andrew and {Brandt}, William Nielsen and {Brown}, Michael E. and {Bullock}, James S. and {Burchat}, Patricia and {Burke}, David L. and {Cagnoli}, Gianpietro and {Calabrese}, Daniel and {Callahan}, Shawn and {Callen}, Alice L. and {Carlin}, Jeffrey L. and {Carlson}, Erin L. and {Chandrasekharan}, Srinivasan and {Charles-Emerson}, Glenaver and {Chesley}, Steve and {Cheu}, Elliott C. and {Chiang}, Hsin-Fang and {Chiang}, James and {Chirino}, Carol and {Chow}, Derek and {Ciardi}, David R. and {Claver}, Charles F. and {Cohen-Tanugi}, Johann and {Cockrum}, Joseph J. and {Coles}, Rebecca and {Connolly}, Andrew J. and {Cook}, Kem H. and {Cooray}, Asantha and {Covey}, Kevin R. and {Cribbs}, Chris and {Cui}, Wei and {Cutri}, Roc and {Daly}, Philip N. and {Daniel}, Scott F. and {Daruich}, Felipe and {Daubard}, Guillaume and {Daues}, Greg and {Dawson}, William and {Delgado}, Francisco and {Dellapenna}, Alfred and {de Peyster}, Robert and {de Val-Borro}, Miguel and {Digel}, Seth W. and {Doherty}, Peter and {Dubois}, Richard and {Dubois-Felsmann}, Gregory P. and {Durech}, Josef and {Economou}, Frossie and {Eifler}, Tim and {Eracleous}, Michael and {Emmons}, Benjamin L. and {Fausti Neto}, Angelo and {Ferguson}, Henry and {Figueroa}, Enrique and {Fisher-Levine}, Merlin and {Focke}, Warren and {Foss}, Michael D. and {Frank}, James and {Freemon}, Michael D. and {Gangler}, Emmanuel and {Gawiser}, Eric and {Geary}, John C. and {Gee}, Perry and {Geha}, Marla and {Gessner}, Charles J.~B. and {Gibson}, Robert R. and {Gilmore}, D. Kirk and {Glanzman}, Thomas and {Glick}, William and {Goldina}, Tatiana and {Goldstein}, Daniel A. and {Goodenow}, Iain and {Graham}, Melissa L. and {Gressler}, William J. and {Gris}, Philippe and {Guy}, Leanne P. and {Guyonnet}, Augustin and {Haller}, Gunther and {Harris}, Ron and {Hascall}, Patrick A. and {Haupt}, Justine and {Hernandez}, Fabio and {Herrmann}, Sven and {Hileman}, Edward and {Hoblitt}, Joshua and {Hodgson}, John A. and {Hogan}, Craig and {Howard}, James D. and {Huang}, Dajun and {Huffer}, Michael E. and {Ingraham}, Patrick and {Innes}, Walter R. and {Jacoby}, Suzanne H. and {Jain}, Bhuvnesh and {Jammes}, Fabrice and {Jee}, M. James and {Jenness}, Tim and {Jernigan}, Garrett and {Jevremovi{\'c}}, Darko and {Johns}, Kenneth and {Johnson}, Anthony S. and {Johnson}, Margaret W.~G. and {Jones}, R. Lynne and {Juramy-Gilles}, Claire and {Juri{\'c}}, Mario and {Kalirai}, Jason S. and {Kallivayalil}, Nitya J. and {Kalmbach}, Bryce and {Kantor}, Jeffrey P. and {Karst}, Pierre and {Kasliwal}, Mansi M. and {Kelly}, Heather and {Kessler}, Richard and {Kinnison}, Veronica and {Kirkby}, David and {Knox}, Lloyd and {Kotov}, Ivan V. and {Krabbendam}, Victor L. and {Krughoff}, K. Simon and {Kub{\'a}nek}, Petr and {Kuczewski}, John and {Kulkarni}, Shri and {Ku}, John and {Kurita}, Nadine R. and {Lage}, Craig S. and {Lambert}, Ron and {Lange}, Travis and {Langton}, J. Brian and {Le Guillou}, Laurent and {Levine}, Deborah and {Liang}, Ming and {Lim}, Kian-Tat and {Lintott}, Chris J. and {Long}, Kevin E. and {Lopez}, Margaux and {Lotz}, Paul J. and {Lupton}, Robert H. and {Lust}, Nate B. and {MacArthur}, Lauren A. and {Mahabal}, Ashish and {Mandelbaum}, Rachel and {Markiewicz}, Thomas W. and {Marsh}, Darren S. and {Marshall}, Philip J. and {Marshall}, Stuart and {May}, Morgan and {McKercher}, Robert and {McQueen}, Michelle and {Meyers}, Joshua and {Migliore}, Myriam and {Miller}, Michelle and {Mills}, David J. and {Miraval}, Connor and {Moeyens}, Joachim and {Moolekamp}, Fred E. and {Monet}, David G. and {Moniez}, Marc and {Monkewitz}, Serge and {Montgomery}, Christopher and {Morrison}, Christopher B. and {Mueller}, Fritz and {Muller}, Gary P. and {Mu{\~n}oz Arancibia}, Freddy and {Neill}, Douglas R. and {Newbry}, Scott P. and {Nief}, Jean-Yves and {Nomerotski}, Andrei and {Nordby}, Martin and {O'Connor}, Paul and {Oliver}, John and {Olivier}, Scot S. and {Olsen}, Knut and {O'Mullane}, William and {Ortiz}, Sandra and {Osier}, Shawn and {Owen}, Russell E. and {Pain}, Reynald and {Palecek}, Paul E. and {Parejko}, John K. and {Parsons}, James B. and {Pease}, Nathan M. and {Peterson}, J. Matt and {Peterson}, John R. and {Petravick}, Donald L. and {Libby Petrick}, M.~E. and {Petry}, Cathy E. and {Pierfederici}, Francesco and {Pietrowicz}, Stephen and {Pike}, Rob and {Pinto}, Philip A. and {Plante}, Raymond and {Plate}, Stephen and {Plutchak}, Joel P. and {Price}, Paul A. and {Prouza}, Michael and {Radeka}, Veljko and {Rajagopal}, Jayadev and {Rasmussen}, Andrew P. and {Regnault}, Nicolas and {Reil}, Kevin A. and {Reiss}, David J. and {Reuter}, Michael A. and {Ridgway}, Stephen T. and {Riot}, Vincent J. and {Ritz}, Steve and {Robinson}, Sean and {Roby}, William and {Roodman}, Aaron and {Rosing}, Wayne and {Roucelle}, Cecille and {Rumore}, Matthew R. and {Russo}, Stefano and {Saha}, Abhijit and {Sassolas}, Benoit and {Schalk}, Terry L. and {Schellart}, Pim and {Schindler}, Rafe H. and {Schmidt}, Samuel and {Schneider}, Donald P. and {Schneider}, Michael D. and {Schoening}, William and {Schumacher}, German and {Schwamb}, Megan E. and {Sebag}, Jacques and {Selvy}, Brian and {Sembroski}, Glenn H. and {Seppala}, Lynn G. and {Serio}, Andrew and {Serrano}, Eduardo and {Shaw}, Richard A. and {Shipsey}, Ian and {Sick}, Jonathan and {Silvestri}, Nicole and {Slater}, Colin T. and {Smith}, J. Allyn and {Smith}, R. Chris and {Sobhani}, Shahram and {Soldahl}, Christine and {Storrie-Lombardi}, Lisa and {Stover}, Edward and {Strauss}, Michael A. and {Street}, Rachel A. and {Stubbs}, Christopher W. and {Sullivan}, Ian S. and {Sweeney}, Donald and {Swinbank}, John D. and {Szalay}, Alexander and {Takacs}, Peter and {Tether}, Stephen A. and {Thaler}, Jon J. and {Thayer}, John Gregg and {Thomas}, Sandrine and {Thornton}, Adam J. and {Thukral}, Vaikunth and {Tice}, Jeffrey and {Trilling}, David E. and {Turri}, Max and {Van Berg}, Richard and {Vanden Berk}, Daniel and {Vetter}, Kurt and {Virieux}, Francoise and {Vucina}, Tomislav and {Wahl}, William and {Walkowicz}, Lucianne and {Walsh}, Brian and {Walter}, Christopher W. and {Wang}, Daniel L. and {Wang}, Shin-Yawn and {Warner}, Michael and {Wiecha}, Oliver and {Willman}, Beth and {Winters}, Scott E. and {Wittman}, David and {Wolff}, Sidney C. and {Wood-Vasey}, W. Michael and {Wu}, Xiuqin and {Xin}, Bo and {Yoachim}, Peter and {Zhan}, Hu},
        title = "{LSST: From Science Drivers to Reference Design and Anticipated Data Products}",
      journal = {\apj},
         year = 2019,
        month = mar,
       volume = {873},
       number = {2},
          eid = {111},
        pages = {111},
          doi = {10.3847/1538-4357/ab042c},
archivePrefix = {arXiv},
       eprint = {0805.2366},
 primaryClass = {astro-ph},
       adsurl = {https://ui.adsabs.harvard.edu/abs/2019ApJ...873..111I}
}

@ARTICLE{Fornax-size-mag-catalog-2018,
       author = {{Venhola}, Aku and {Peletier}, Reynier and {Laurikainen}, Eija and {Salo}, Heikki and {Iodice}, Enrichetta and {Mieske}, Steffen and {Hilker}, Michael and {Wittmann}, Carolin and {Lisker}, Thorsten and {Paolillo}, Maurizio and {Cantiello}, Michele and {Janz}, Joachim and {Spavone}, Marilena and {D'Abrusco}, Raffaele and {Ven}, Glennvande and {Napolitano}, Nicola and {Kleijn}, GijsVerdoes and {Maddox}, Natasha and {Capaccioli}, Massimo and {Grado}, Aniello and {Valentijn}, Edwin and {Falc{\'o}n-Barroso}, Jes{\'u}s and {Limatola}, Luca},
        title = "{The Fornax Deep Survey with the VST. IV. A size and magnitude limited catalog of dwarf galaxies in the area of the Fornax cluster}",
      journal = {\aap},
         year = 2018,
        month = dec,
       volume = {620},
          eid = {A165},
        pages = {A165},
          doi = {10.1051/0004-6361/201833933},
archivePrefix = {arXiv},
       eprint = {1810.00550},
 primaryClass = {astro-ph.GA},
       adsurl = {https://ui.adsabs.harvard.edu/abs/2018A&A...620A.165V}
}

@ARTICLE{SAMI-Fornax-I-Scott2020,
       author = {{Scott}, Nicholas and {Eftekhari}, F. Sara and {Peletier}, Reynier F. and {Bryant}, Julia J. and {Bland-Hawthorn}, Joss and {Capaccioli}, Massimo and {Croom}, Scott M. and {Drinkwater}, Michael and {Falc{\'o}n-Barroso}, J{\'e}sus and {Hilker}, Michael and {Iodice}, Enrichetta and {Lorente}, Nuria F.~P. and {Mieske}, Steffen and {Spavone}, Marilena and {van de Ven}, Glenn and {Venhola}, Aku},
        title = "{The SAMI-Fornax Dwarfs Survey I: sample, observations, and the specific stellar angular momentum of dwarf elliptical galaxies}",
      journal = {\mnras},
         year = 2020,
        month = sep,
       volume = {497},
       number = {2},
        pages = {1571-1582},
          doi = {10.1093/mnras/staa2042},
archivePrefix = {arXiv},
       eprint = {2007.04492},
 primaryClass = {astro-ph.GA},
       adsurl = {https://ui.adsabs.harvard.edu/abs/2020MNRAS.497.1571S}
}

@article{Fornax-stellar-mass,
    author = {Taylor, Edward N. and Hopkins, Andrew M. and Baldry, Ivan K. and Brown, Michael J. I. and Driver, Simon P. and Kelvin, Lee S. and Hill, David T. and Robotham, Aaron S. G. and Bland-Hawthorn, Joss and Jones, D. H. and Sharp, R. G. and Thomas, Daniel and Liske, Jochen and Loveday, Jon and Norberg, Peder and Peacock, J. A. and Bamford, Steven P. and Brough, Sarah and Colless, Matthew and Cameron, Ewan and Conselice, Christopher J. and Croom, Scott M. and Frenk, C. S. and Gunawardhana, Madusha and Kuijken, Konrad and Nichol, R. C. and Parkinson, H. R. and Phillipps, S. and Pimbblet, K. A. and Popescu, C. C. and Prescott, Matthew and Sutherland, W. J. and Tuffs, R. J. and van Kampen, Eelco and Wijesinghe, D.},
    title = "{Galaxy And Mass Assembly (GAMA): stellar mass estimates}",
    journal = {\mnras},
    volume = {418},
    number = {3},
    pages = {1587-1620},
    year = {2011},
    month = {12},
    issn = {0035-8711},
    doi = {10.1111/j.1365-2966.2011.19536.x},
    url = {https://doi.org/10.1111/j.1365-2966.2011.19536.x},
    eprint = {https://academic.oup.com/mnras/article-pdf/418/3/1587/18437265/mnras0418-1587.pdf},
}

@ARTICLE{2023MNRAS.522.3912Ceverino,
       author = {{Ceverino}, Daniel and {Mandelker}, Nir and {Snyder}, Gregory F. and {Lapiner}, Sharon and {Dekel}, Avishai and {Primack}, Joel and {Ginzburg}, Omri and {Larkin}, Sean},
        title = "{Effects of feedback on galaxies in the VELA simulations: elongation, clumps, and compaction}",
      journal = {\mnras},
         year = 2023,
        month = jul,
       volume = {522},
       number = {3},
        pages = {3912-3925},
          doi = {10.1093/mnras/stad1255},
archivePrefix = {arXiv},
       eprint = {2210.15372},
 primaryClass = {astro-ph.GA},
       adsurl = {https://ui.adsabs.harvard.edu/abs/2023MNRAS.522.3912C}
}

@ARTICLE{2022_one_galaxy_camels,
       author = {{Villaescusa-Navarro}, Francisco and {Ding}, Jupiter and {Genel}, Shy and {Tonnesen}, Stephanie and {La Torre}, Valentina and {Spergel}, David N. and {Teyssier}, Romain and {Li}, Yin and {Heneka}, Caroline and {Lemos}, Pablo and {Angl{\'e}s-Alc{\'a}zar}, Daniel and {Nagai}, Daisuke and {Vogelsberger}, Mark},
        title = "{Cosmology with One Galaxy?}",
      journal = {\apj},
         year = 2022,
        month = apr,
       volume = {929},
       number = {2},
          eid = {132},
        pages = {132},
          doi = {10.3847/1538-4357/ac5d3f},
archivePrefix = {arXiv},
       eprint = {2201.02202},
 primaryClass = {astro-ph.CO},
       adsurl = {https://ui.adsabs.harvard.edu/abs/2022ApJ...929..132V}
}

@ARTICLE{Bernardi2003a,
       author = {{Bernardi}, Mariangela and {Sheth}, Ravi K. and {Annis}, James and {Burles}, Scott and {Eisenstein}, Daniel J. and {Finkbeiner}, Douglas P. and {Hogg}, David W. and {Lupton}, Robert H. and {Schlegel}, David J. and {SubbaRao}, Mark and {Bahcall}, Neta A. and {Blakeslee}, John P. and {Brinkmann}, J. and {Castander}, Francisco J. and {Connolly}, Andrew J. and {Csabai}, Istv{\'a}n and {Doi}, Mamoru and {Fukugita}, Masataka and {Frieman}, Joshua and {Heckman}, Timothy and {Hennessy}, Gregory S. and {Ivezi{\'c}}, {\v{Z}}eljko and {Knapp}, G.~R. and {Lamb}, Don Q. and {McKay}, Timothy and {Munn}, Jeffrey A. and {Nichol}, Robert and {Okamura}, Sadanori and {Schneider}, Donald P. and {Thakar}, Aniruddha R. and {York}, Donald G.},
        title = "{Early-Type Galaxies in the Sloan Digital Sky Survey. I. The Sample}",
      journal = {\aj},
         year = 2003,
        month = apr,
       volume = {125},
       number = {4},
        pages = {1817-1848},
          doi = {10.1086/367776},
archivePrefix = {arXiv},
       eprint = {astro-ph/0301631},
 primaryClass = {astro-ph},
       adsurl = {https://ui.adsabs.harvard.edu/abs/2003AJ....125.1817B}
}

@ARTICLE{Cardelli1989-extinction-law-ApJ...345..245C,
       author = {{Cardelli}, Jason A. and {Clayton}, Geoffrey C. and {Mathis}, John S.},
        title = "{The Relationship between Infrared, Optical, and Ultraviolet Extinction}",
      journal = {\apj},
         year = 1989,
        month = oct,
       volume = {345},
        pages = {245},
          doi = {10.1086/167900},
       adsurl = {https://ui.adsabs.harvard.edu/abs/1989ApJ...345..245C}
}

@article{Lephare-Ilbert2006,
	author = {{Ilbert, O.} and {Arnouts, S.} and {McCracken, H. J.} and {Bolzonella, M.} and {Bertin, E.} and {Le Fèvre, O.} and {Mellier, Y.} and {Zamorani, G.} and {Pellò, R.} and {Iovino, A.} and {Tresse, L.} and {Le Brun, V.} and {Bottini, D.} and {Garilli, B.} and {Maccagni, D.} and {Picat, J. P.} and {Scaramella, R.} and {Scodeggio, M.} and {Vettolani, G.} and {Zanichelli, A.} and {Adami, C.} and {Bardelli, S.} and {Cappi, A.} and {Charlot, S.} and {Ciliegi, P.} and {Contini, T.} and {Cucciati, O.} and {Foucaud, S.} and {Franzetti, P.} and {Gavignaud, I.} and {Guzzo, L.} and {Marano, B.} and {Marinoni, C.} and {Mazure, A.} and {Meneux, B.} and {Merighi, R.} and {Paltani, S.} and {Pollo, A.} and {Pozzetti, L.} and {Radovich, M.} and {Zucca, E.} and {Bondi, M.} and {Bongiorno, A.} and {Busarello, G.} and {De La Torre, S.} and {Gregorini, L.} and {Lamareille, F.} and {Mathez, G.} and {Merluzzi, P.} and {Ripepi, V.} and {Rizzo, D.} and {Vergani, D.}},
	title = {Accurate photometric redshifts for the CFHT legacy survey calibrated using the VIMOS VLT deep survey},
	DOI= "10.1051/0004-6361:20065138",
	url= "https://doi.org/10.1051/0004-6361:20065138",
	journal = {A\&A},
	year = 2006,
	volume = 457,
	number = 3,
	pages = "841-856",
}

@ARTICLE{BC03-2003MNRAS.344.1000B,
       author = {{Bruzual}, G. and {Charlot}, S.},
        title = "{Stellar population synthesis at the resolution of 2003}",
      journal = {\mnras},
         year = 2003,
        month = oct,
       volume = {344},
       number = {4},
        pages = {1000-1028},
          doi = {10.1046/j.1365-8711.2003.06897.x},
archivePrefix = {arXiv},
       eprint = {astro-ph/0309134},
 primaryClass = {astro-ph},
       adsurl = {https://ui.adsabs.harvard.edu/abs/2003MNRAS.344.1000B}
}

@article{Mstar-NSA-Blanton-2011,
doi = {10.1088/0004-6256/142/1/31},
url = {https://dx.doi.org/10.1088/0004-6256/142/1/31},
year = {2011},
month = {jun},
publisher = {\aj},
volume = {142},
number = {1},
pages = {31},
author = {Michael R. Blanton and Eyal Kazin and Demitri Muna and Benjamin A. Weaver and Adrian Price-Whelan},
title = {IMPROVED BACKGROUND SUBTRACTION FOR THE SLOAN DIGITAL SKY SURVEY IMAGES},
journal = {The Astronomical Journal}
}

@ARTICLE{Mstar-NSA-Blanton2007,
       author = {{Blanton}, Michael R. and {Roweis}, Sam},
        title = "{K-Corrections and Filter Transformations in the Ultraviolet, Optical, and Near-Infrared}",
      journal = {\aj},
         year = 2007,
        month = feb,
       volume = {133},
       number = {2},
        pages = {734-754},
          doi = {10.1086/510127},
archivePrefix = {arXiv},
       eprint = {astro-ph/0606170},
 primaryClass = {astro-ph},
       adsurl = {https://ui.adsabs.harvard.edu/abs/2007AJ....133..734B}
}

@article{MaNGA-DL-morph-2022,
    author = {Domínguez Sánchez, H and Margalef, B and Bernardi, M and Huertas-Company, M},
    title = "{SDSS-IV DR17: final release of MaNGA PyMorph photometric and deep-learning morphological catalogues}",
    journal = {\mnras},
    volume = {509},
    number = {3},
    pages = {4024-4036},
    year = {2021},
    month = {10},
    issn = {0035-8711},
    doi = {10.1093/mnras/stab3089},
    url = {https://doi.org/10.1093/mnras/stab3089},
    eprint = {https://academic.oup.com/mnras/article-pdf/509/3/4024/41475368/stab3089.pdf},
}

@ARTICLE{labarbera2010SPIDER-I/MNRAS.408.1313L,
       author = {{La Barbera}, F. and {de Carvalho}, R.~R. and {de La Rosa}, I.~G. and {Lopes}, P.~A.~A. and {Kohl-Moreira}, J.~L. and {Capelato}, H.~V.},
        title = "{SPIDER - I. Sample and galaxy parameters in the grizYJHK wavebands}",
      journal = {\mnras},
         year = 2010,
        month = nov,
       volume = {408},
       number = {3},
        pages = {1313-1334},
          doi = {10.1111/j.1365-2966.2010.16850.x},
       adsurl = {https://ui.adsabs.harvard.edu/abs/2010MNRAS.408.1313L}
}

@BOOK{Binney_Tremaine,
       author = {{Binney}, James and {Tremaine}, Scott},
        title = "{Galactic dynamics}",
         year = 1987,
       adsurl = {https://ui.adsabs.harvard.edu/abs/1987gady.book.....B}
}

@BOOK{Sersic1968adga.book.....S,
       author = {{Sersic}, Jose Luis},
        title = "{Atlas de Galaxias Australes}",
         year = 1968,
       adsurl = {https://ui.adsabs.harvard.edu/abs/1968adga.book.....S}
}

@ARTICLE{TNG_LF_2022,
       author = {{Tr{\v{c}}ka}, Ana and {Baes}, Maarten and {Camps}, Peter and {Kapoor}, Anand Utsav and {Nelson}, Dylan and {Pillepich}, Annalisa and {Barrientos}, Daniela and {Hernquist}, Lars and {Marinacci}, Federico and {Vogelsberger}, Mark},
        title = "{UV to submillimetre luminosity functions of TNG50 galaxies}",
      journal = {\mnras},
         year = 2022,
        month = nov,
       volume = {516},
       number = {3},
        pages = {3728-3749},
          doi = {10.1093/mnras/stac2277},
archivePrefix = {arXiv},
       eprint = {2208.06424},
 primaryClass = {astro-ph.GA},
       adsurl = {https://ui.adsabs.harvard.edu/abs/2022MNRAS.516.3728T}
}

@ARTICLE{Pillepich2018MNRAS.475..648P,
       author = {{Pillepich}, Annalisa and {Nelson}, Dylan and {Hernquist}, Lars and {Springel}, Volker and {Pakmor}, R{\"u}diger and {Torrey}, Paul and {Weinberger}, Rainer and {Genel}, Shy and {Naiman}, Jill P. and {Marinacci}, Federico and {Vogelsberger}, Mark},
        title = "{First results from the IllustrisTNG simulations: the stellar mass content of groups and clusters of galaxies}",
      journal = {\mnras},
         year = 2018,
        month = mar,
       volume = {475},
       number = {1},
        pages = {648-675},
          doi = {10.1093/mnras/stx3112},
archivePrefix = {arXiv},
       eprint = {1707.03406},
 primaryClass = {astro-ph.GA},
       adsurl = {https://ui.adsabs.harvard.edu/abs/2018MNRAS.475..648P}
}

@article{Weinberger2016/10.1093/mnras/stw2944,
    author = {Weinberger, Rainer and Springel, Volker and Hernquist, Lars and Pillepich, Annalisa and Marinacci, Federico and Pakmor, Rüdiger and Nelson, Dylan and Genel, Shy and Vogelsberger, Mark and Naiman, Jill and Torrey, Paul},
    title = "{Simulating galaxy formation with black hole driven thermal and kinetic feedback}",
    journal = {\mnras},
    volume = {465},
    number = {3},
    pages = {3291-3308},
    year = {2016},
    month = {11},
    issn = {0035-8711},
    doi = {10.1093/mnras/stw2944},
    url = {https://doi.org/10.1093/mnras/stw2944},
    eprint = {https://academic.oup.com/mnras/article-pdf/465/3/3291/8518280/stw2944.pdf},
}

@ARTICLE{Nelson2018MNRAS.475..624N,
       author = {{Nelson}, Dylan and {Pillepich}, Annalisa and {Springel}, Volker and {Weinberger}, Rainer and {Hernquist}, Lars and {Pakmor}, R{\"u}diger and {Genel}, Shy and {Torrey}, Paul and {Vogelsberger}, Mark and {Kauffmann}, Guinevere and {Marinacci}, Federico and {Naiman}, Jill},
        title = "{First results from the IllustrisTNG simulations: the galaxy colour bimodality}",
      journal = {\mnras},
         year = 2018,
        month = mar,
       volume = {475},
       number = {1},
        pages = {624-647},
          doi = {10.1093/mnras/stx3040},
archivePrefix = {arXiv},
       eprint = {1707.03395},
 primaryClass = {astro-ph.GA},
       adsurl = {https://ui.adsabs.harvard.edu/abs/2018MNRAS.475..624N}
}

@article{SAMI-Fornax-Dwarfs-Survey-II,
    author = {Eftekhari, F Sara and Peletier, Reynier F and Scott, Nicholas and Mieske, Steffen and Bland-Hawthorn, Joss and Bryant, Julia J and Cantiello, Michele and Croom, Scott M and Drinkwater, Michael J and Falcón-Barroso, Jésus and Hilker, Michael and Iodice, Enrichetta and Napolitano, Nicola R and Spavone, Marilena and Valentijn, Edwin A and van de Ven, Glenn and Venhola, Aku},
    title = "{The SAMI–Fornax Dwarfs Survey – II. The Stellar Mass Fundamental Plane and the dark matter fraction of dwarf galaxies}",
    journal = {\mnras},
    volume = {517},
    number = {4},
    pages = {4714-4735},
    year = {2022},
    month = {09},
    issn = {0035-8711},
    doi = {10.1093/mnras/stac2606},
    url = {https://doi.org/10.1093/mnras/stac2606},
    eprint = {https://academic.oup.com/mnras/article-pdf/517/4/4714/46800081/stac2606.pdf},
}

@ARTICLE{2001ApJ..gNFW,
       author = {{Wyithe}, J.~S.~B. and {Turner}, E.~L. and {Spergel}, D.~N.},
        title = "{Gravitational Lens Statistics for Generalized NFW Profiles: Parameter Degeneracy and Implications for Self-Interacting Cold Dark Matter}",
      journal = {\apj},
         year = 2001,
        month = jul,
       volume = {555},
       number = {1},
        pages = {504-523},
          doi = {10.1086/321437},
archivePrefix = {arXiv},
       eprint = {astro-ph/0007354},
 primaryClass = {astro-ph},
       adsurl = {https://ui.adsabs.harvard.edu/abs/2001ApJ...555..504W}
}

@ARTICLE{1996ApJ...NFW,
       author = {{Navarro}, Julio F. and {Frenk}, Carlos S. and {White}, Simon D.~M.},
        title = "{The Structure of Cold Dark Matter Halos}",
      journal = {\apj},
         year = 1996,
        month = may,
       volume = {462},
        pages = {563},
          doi = {10.1086/177173},
archivePrefix = {arXiv},
       eprint = {astro-ph/9508025},
 primaryClass = {astro-ph},
       adsurl = {https://ui.adsabs.harvard.edu/abs/1996ApJ...462..563N}
}

@ARTICLE{2020MNRAS..Shetty,
       author = {{Shetty}, Shravan and {Cappellari}, Michele and {McDermid}, Richard M. and {Krajnovi{\'c}}, Davor and {de Zeeuw}, P.~T. and {Davies}, Roger L. and {Kobayashi}, Chiaki},
        title = "{A precise benchmark for cluster scaling relations: Fundamental Plane, Mass Plane, and IMF in the Coma cluster from dynamical models}",
      journal = {\mnras},
         year = 2020,
        month = jun,
       volume = {494},
       number = {4},
        pages = {5619-5635},
          doi = {10.1093/mnras/staa1043},
archivePrefix = {arXiv},
       eprint = {2004.07449},
 primaryClass = {astro-ph.GA},
       adsurl = {https://ui.adsabs.harvard.edu/abs/2020MNRAS.494.5619S}
}

@ARTICLE{2012Cappellari..Nature,
       author = {{Cappellari}, Michele and {McDermid}, Richard M. and {Alatalo}, Katherine and {Blitz}, Leo and {Bois}, Maxime and {Bournaud}, Fr{\'e}d{\'e}ric and {Bureau}, M. and {Crocker}, Alison F. and {Davies}, Roger L. and {Davis}, Timothy A. and {de Zeeuw}, P.~T. and {Duc}, Pierre-Alain and {Emsellem}, Eric and {Khochfar}, Sadegh and {Krajnovi{\'c}}, Davor and {Kuntschner}, Harald and {Lablanche}, Pierre-Yves and {Morganti}, Raffaella and {Naab}, Thorsten and {Oosterloo}, Tom and {Sarzi}, Marc and {Scott}, Nicholas and {Serra}, Paolo and {Weijmans}, Anne-Marie and {Young}, Lisa M.},
        title = "{Systematic variation of the stellar initial mass function in early-type galaxies}",
      journal = {\nat},
         year = 2012,
        month = apr,
       volume = {484},
       number = {7395},
        pages = {485-488},
          doi = {10.1038/nature10972},
archivePrefix = {arXiv},
       eprint = {1202.3308},
 primaryClass = {astro-ph.CO},
       adsurl = {https://ui.adsabs.harvard.edu/abs/2012Natur.484..485C}
}

@article{2023MaNGADynPop,
    author = {Zhu, Kai and Lu, Shengdong and Cappellari, Michele and Li, Ran and Mao, Shude and Gao, Liang},
    title = "{MaNGA DynPop – I. Quality-assessed stellar dynamical modelling from integral-field spectroscopy of 10K nearby galaxies: a catalogue of masses, mass-to-light ratios, density profiles, and dark matter}",
    journal = {\mnras},
    volume = {522},
    number = {4},
    pages = {6326-6353},
    year = {2023},
    month = {04},
    issn = {0035-8711},
    doi = {10.1093/mnras/stad1299},
    url = {https://doi.org/10.1093/mnras/stad1299},
    eprint = {https://academic.oup.com/mnras/article-pdf/522/4/6326/50426281/stad1299.pdf},
}

@ARTICLE{2020Cappellari..JAM,
       author = {{Cappellari}, Michele},
        title = "{Efficient solution of the anisotropic spherically aligned axisymmetric Jeans equations of stellar hydrodynamics for galactic dynamics}",
      journal = {\mnras},
         year = 2020,
        month = jun,
       volume = {494},
       number = {4},
        pages = {4819-4837},
          doi = {10.1093/mnras/staa959},
archivePrefix = {arXiv},
       eprint = {1907.09894},
 primaryClass = {astro-ph.GA},
       adsurl = {https://ui.adsabs.harvard.edu/abs/2020MNRAS.494.4819C}
}

@ARTICLE{2008Cappellari..JAM,
       author = {{Cappellari}, Michele},
        title = "{Measuring the inclination and mass-to-light ratio of axisymmetric galaxies via anisotropic Jeans models of stellar kinematics}",
      journal = {\mnras},
         year = 2008,
        month = oct,
       volume = {390},
       number = {1},
        pages = {71-86},
          doi = {10.1111/j.1365-2966.2008.13754.x},
archivePrefix = {arXiv},
       eprint = {0806.0042},
 primaryClass = {astro-ph},
       adsurl = {https://ui.adsabs.harvard.edu/abs/2008MNRAS.390...71C}
}

@ARTICLE{2015OverviewMaNGA,
       author = {{Bundy}, Kevin and {Bershady}, Matthew A. and {Law}, David R. and {Yan}, Renbin and {Drory}, Niv and {MacDonald}, Nicholas and {Wake}, David A. and {Cherinka}, Brian and {S{\'a}nchez-Gallego}, Jos{\'e} R. and {Weijmans}, Anne-Marie and {Thomas}, Daniel and {Tremonti}, Christy and {Masters}, Karen and {Coccato}, Lodovico and {Diamond-Stanic}, Aleksandar M. and {Arag{\'o}n-Salamanca}, Alfonso and {Avila-Reese}, Vladimir and {Badenes}, Carles and {Falc{\'o}n-Barroso}, J{\'e}sus and {Belfiore}, Francesco and {Bizyaev}, Dmitry and {Blanc}, Guillermo A. and {Bland-Hawthorn}, Joss and {Blanton}, Michael R. and {Brownstein}, Joel R. and {Byler}, Nell and {Cappellari}, Michele and {Conroy}, Charlie and {Dutton}, Aaron A. and {Emsellem}, Eric and {Etherington}, James and {Frinchaboy}, Peter M. and {Fu}, Hai and {Gunn}, James E. and {Harding}, Paul and {Johnston}, Evelyn J. and {Kauffmann}, Guinevere and {Kinemuchi}, Karen and {Klaene}, Mark A. and {Knapen}, Johan H. and {Leauthaud}, Alexie and {Li}, Cheng and {Lin}, Lihwai and {Maiolino}, Roberto and {Malanushenko}, Viktor and {Malanushenko}, Elena and {Mao}, Shude and {Maraston}, Claudia and {McDermid}, Richard M. and {Merrifield}, Michael R. and {Nichol}, Robert C. and {Oravetz}, Daniel and {Pan}, Kaike and {Parejko}, John K. and {Sanchez}, Sebastian F. and {Schlegel}, David and {Simmons}, Audrey and {Steele}, Oliver and {Steinmetz}, Matthias and {Thanjavur}, Karun and {Thompson}, Benjamin A. and {Tinker}, Jeremy L. and {van den Bosch}, Remco C.~E. and {Westfall}, Kyle B. and {Wilkinson}, David and {Wright}, Shelley and {Xiao}, Ting and {Zhang}, Kai},
        title = "{Overview of the SDSS-IV MaNGA Survey: Mapping nearby Galaxies at Apache Point Observatory}",
      journal = {\apj},
         year = 2015,
        month = jan,
       volume = {798},
       number = {1},
          eid = {7},
        pages = {7},
          doi = {10.1088/0004-637X/798/1/7},
archivePrefix = {arXiv},
       eprint = {1412.1482},
 primaryClass = {astro-ph.GA},
       adsurl = {https://ui.adsabs.harvard.edu/abs/2015ApJ...798....7B}
}

@ARTICLE{amaro+19,
       author = {{Amaro}, V. and {Cavuoti}, S. and {Brescia}, M. and {Vellucci}, C. and {Longo}, G. and {Bilicki}, M. and {de Jong}, J.~T.~A. and {Tortora}, C. and {Radovich}, M. and {Napolitano}, N.~R. and {Buddelmeijer}, H.},
        title = "{Statistical analysis of probability density functions for photometric redshifts through the KiDS-ESO-DR3 galaxies}",
      journal = {\mnras},
         year = 2019,
        month = jan,
       volume = {482},
       number = {3},
        pages = {3116-3134},
          doi = {10.1093/mnras/sty2922},
archivePrefix = {arXiv},
       eprint = {1810.09777},
 primaryClass = {astro-ph.IM},
       adsurl = {https://ui.adsabs.harvard.edu/abs/2019MNRAS.482.3116A}
}

@article{Photometric-BASS-10.1093/mnras/stab3165,
    author = {Li, Changhua and Zhang, Yanxia and Cui, Chenzhou and Fan, Dongwei and Zhao, Yongheng and Wu, Xue-Bing and Zhang, Jing-Yi and Han, Jun and Xu, Yunfei and Tao, Yihan and Li, Shanshan and He, Boliang},
    title = "{Photometric redshift estimation of BASS DR3 quasars by machine learning}",
    journal = {\mnras},
    volume = {509},
    number = {2},
    pages = {2289-2303},
    year = {2021},
    month = {11},
    issn = {0035-8711},
    doi = {10.1093/mnras/stab3165},
    url = {https://doi.org/10.1093/mnras/stab3165},
    eprint = {https://academic.oup.com/mnras/article-pdf/509/2/2289/41199823/stab3165.pdf},
}

@ARTICLE{pulsoni20202_TNG_ETG,
       author = {{Pulsoni}, C. and {Gerhard}, O. and {Arnaboldi}, M. and {Pillepich}, A. and {Nelson}, D. and {Hernquist}, L. and {Springel}, V.},
        title = "{The stellar halos of ETGs in the IllustrisTNG simulations: The photometric and kinematic diversity of galaxies at large radii}",
      journal = {\aap},
         year = 2020,
        month = sep,
       volume = {641},
          eid = {A60},
        pages = {A60},
          doi = {10.1051/0004-6361/202038253},
archivePrefix = {arXiv},
       eprint = {2004.13042},
 primaryClass = {astro-ph.GA},
       adsurl = {https://ui.adsabs.harvard.edu/abs/2020A&A...641A..60P}
}

@ARTICLE{Rodriguez-Gomez_TNG100spin,
       author = {{Rodriguez-Gomez}, Vicente and {Genel}, Shy and {Fall}, S. Michael and {Pillepich}, Annalisa and {Huertas-Company}, Marc and {Nelson}, Dylan and {P{\'e}rez-Monta{\~n}o}, Luis Enrique and {Marinacci}, Federico and {Pakmor}, R{\"u}diger and {Springel}, Volker and {Vogelsberger}, Mark and {Hernquist}, Lars},
        title = "{Galactic angular momentum in the IllustrisTNG simulation - I. Connection to morphology, halo spin, and black hole mass}",
      journal = {\mnras},
         year = 2022,
        month = jun,
       volume = {512},
       number = {4},
        pages = {5978-5994},
          doi = {10.1093/mnras/stac806},
archivePrefix = {arXiv},
       eprint = {2203.10098},
 primaryClass = {astro-ph.GA},
       adsurl = {https://ui.adsabs.harvard.edu/abs/2022MNRAS.512.5978R}
}

@ARTICLE{T09-CentralDM-2009MNRAS.396.1132Tortora09,
       author = {{Tortora}, C. and {Napolitano}, N.~R. and {Romanowsky}, A.~J. and {Capaccioli}, M. and {Covone}, G.},
        title = "{Central mass-to-light ratios and dark matter fractions in early-type galaxies}",
      journal = {\mnras},
         year = 2009,
        month = jun,
       volume = {396},
       number = {2},
        pages = {1132-1150},
          doi = {10.1111/j.1365-2966.2009.14789.x},
archivePrefix = {arXiv},
       eprint = {0901.3781},
 primaryClass = {astro-ph.CO},
       adsurl = {https://ui.adsabs.harvard.edu/abs/2009MNRAS.396.1132T}
}

@ARTICLE{Wolf10_2D_3D,
       author = {{Wolf}, Joe and {Martinez}, Gregory D. and {Bullock}, James S. and {Kaplinghat}, Manoj and {Geha}, Marla and {Mu{\~n}oz}, Ricardo R. and {Simon}, Joshua D. and {Avedo}, Frank F.},
        title = "{Accurate masses for dispersion-supported galaxies}",
      journal = {\mnras},
         year = 2010,
        month = aug,
       volume = {406},
       number = {2},
        pages = {1220-1237},
          doi = {10.1111/j.1365-2966.2010.16753.x},
archivePrefix = {arXiv},
       eprint = {0908.2995},
 primaryClass = {astro-ph.CO},
       adsurl = {https://ui.adsabs.harvard.edu/abs/2010MNRAS.406.1220W}
}

@ARTICLE{Suess19_MLgrad_halfrad_high-z,
       author = {{Suess}, Katherine A. and {Kriek}, Mariska and {Price}, Sedona H. and {Barro}, Guillermo},
        title = "{Half-mass Radii for {\ensuremath{\sim}}7000 Galaxies at 1.0 {\ensuremath{\leq}} z {\ensuremath{\leq}} 2.5: Most of the Evolution in the Mass-Size Relation Is Due to Color Gradients}",
      journal = {\apj},
         year = 2019,
        month = jun,
       volume = {877},
       number = {2},
          eid = {103},
        pages = {103},
          doi = {10.3847/1538-4357/ab1bda},
archivePrefix = {arXiv},
       eprint = {1904.10992},
 primaryClass = {astro-ph.GA},
       adsurl = {https://ui.adsabs.harvard.edu/abs/2019ApJ...877..103S}
}

@ARTICLE{Bernardi23_MLgrad_halfmassrad,
       author = {{Bernardi}, M. and {Sheth}, R.~K. and {Dom{\'\i}nguez S{\'a}nchez}, H. and {Margalef-Bentabol}, B. and {Bizyaev}, D. and {Lane}, R.~R.},
        title = "{The half-mass radius of MaNGA galaxies: effect of IMF gradients}",
      journal = {\mnras},
         year = 2023,
        month = jan,
       volume = {518},
       number = {3},
        pages = {3494-3508},
          doi = {10.1093/mnras/stac3361},
archivePrefix = {arXiv},
       eprint = {2201.07810},
 primaryClass = {astro-ph.GA},
       adsurl = {https://ui.adsabs.harvard.edu/abs/2023MNRAS.518.3494B}
}

@ARTICLE{2022FrASS...8..197T&N22,
       author = {{Tortora}, C. and {Napolitano}, N.~R.},
        title = "{Central dark matter in early-type galaxies}",
      journal = {Frontiers in Astronomy and Space Sciences},
         year = 2022,
        month = jan,
       volume = {8},
          eid = {197},
        pages = {197},
          doi = {10.3389/fspas.2021.704419},
archivePrefix = {arXiv},
       eprint = {2201.00842},
 primaryClass = {astro-ph.GA},
       adsurl = {https://ui.adsabs.harvard.edu/abs/2022FrASS...8..197N}
}

@ARTICLE{NFW-Burkert1995,
       author = {{Burkert}, A.},
        title = "{The Structure of Dark Matter Halos in Dwarf Galaxies}",
      journal = {\apjl},
         year = 1995,
        month = jul,
       volume = {447},
        pages = {L25-L28},
          doi = {10.1086/309560},
archivePrefix = {arXiv},
       eprint = {astro-ph/9504041},
 primaryClass = {astro-ph},
       adsurl = {https://ui.adsabs.harvard.edu/abs/1995ApJ...447L..25B}
}

@ARTICLE{Qiu2023_cosmo_cl_ML,
       author = {{Qiu}, Lanlan and {Napolitano}, Nicola R. and {Borgani}, Stefano and {Zhong}, Fucheng and {Li}, Xiaodong and {Radovich}, Mario and {Lin}, Weipeng and {Dolag}, Klaus and {Tortora}, Crescenzo and {Wang}, Yang and {Remus}, Rhea-Silvia and {Longo}, Giuseppe},
        title = "{Cosmology with Galaxy Cluster Properties using Machine Learning}",
      journal = {arXiv e-prints},
         year = 2023,
        month = apr,
          eid = {arXiv:2304.09142},
        pages = {arXiv:2304.09142},
          doi = {10.48550/arXiv.2304.09142},
archivePrefix = {arXiv},
       eprint = {2304.09142},
 primaryClass = {astro-ph.CO},
       adsurl = {https://ui.adsabs.harvard.edu/abs/2023arXiv230409142Q}
}

@ARTICLE{2023A&A...677A.102F,
       author = {{Fortuni}, Flaminia and {Merlin}, Emiliano and {Fontana}, Adriano and {Giocoli}, Carlo and {Romelli}, Erik and {Graziani}, Luca and {Santini}, Paola and {Castellano}, Marco and {Charlot}, St{\'e}phane and {Chevallard}, Jacopo},
        title = "{FORECAST: A flexible software to forward model cosmological hydrodynamical simulations mimicking real observations}",
      journal = {\aap},
         year = 2023,
        month = sep,
       volume = {677},
          eid = {A102},
        pages = {A102},
          doi = {10.1051/0004-6361/202346725},
archivePrefix = {arXiv},
       eprint = {2305.19166},
 primaryClass = {astro-ph.IM},
       adsurl = {https://ui.adsabs.harvard.edu/abs/2023A&A...677A.102F}
}

@ARTICLE{Navarro1997,
       author = {{Navarro}, Julio F. and {Frenk}, Carlos S. and {White}, Simon D.~M.},
        title = "{A Universal Density Profile from Hierarchical Clustering}",
      journal = {\apj},
         year = 1997,
        month = dec,
       volume = {490},
       number = {2},
        pages = {493-508},
          doi = {10.1086/304888},
archivePrefix = {arXiv},
       eprint = {astro-ph/9611107},
 primaryClass = {astro-ph},
       adsurl = {https://ui.adsabs.harvard.edu/abs/1997ApJ...490..493N}
}

@ARTICLE{FJ_96,
       author = {{Faber}, S.~M. and {Jackson}, R.~E.},
        title = "{Velocity dispersions and mass-to-light ratios for elliptical galaxies.}",
      journal = {\apj},
         year = 1976,
        month = mar,
       volume = {204},
        pages = {668-683},
          doi = {10.1086/154215},
       adsurl = {https://ui.adsabs.harvard.edu/abs/1976ApJ...204..668F}
}

@ARTICLE{napolitano-lamost2020-central-velocity-dispersion,
       author = {{Napolitano}, Nicola R. and {D'Ago}, Giuseppe and {Tortora}, Crescenzo and {Zhao}, Gang and {Luo}, A. -Li and {Tang}, Baitian and {Zhang}, Wei and {Zhang}, Yong and {Li}, Rui},
        title = "{Central velocity dispersion catalogue of LAMOST-DR7 galaxies}",
      journal = {\mnras},
         year = 2020,
        month = nov,
       volume = {498},
       number = {4},
        pages = {5704-5719},
          doi = {10.1093/mnras/staa2409},
archivePrefix = {arXiv},
       eprint = {2007.07823},
 primaryClass = {astro-ph.GA},
       adsurl = {https://ui.adsabs.harvard.edu/abs/2020MNRAS.498.5704N}
}

@ARTICLE{Napolitano09-NGC4494-planetary-nebulae,
       author = {{Napolitano}, N.~R. and {Romanowsky}, A.~J. and {Coccato}, L. and {Capaccioli}, M. and {Douglas}, N.~G. and {Noordermeer}, E. and {Gerhard}, O. and {Arnaboldi}, M. and {de Lorenzi}, F. and {Kuijken}, K. and {Merrifield}, M.~R. and {O'Sullivan}, E. and {Cortesi}, A. and {Das}, P. and {Freeman}, K.~C.},
        title = "{The Planetary Nebula Spectrograph elliptical galaxy survey: the dark matter in NGC 4494}",
      journal = {\mnras},
         year = 2009,
        month = feb,
       volume = {393},
       number = {2},
        pages = {329-353},
          doi = {10.1111/j.1365-2966.2008.14053.x},
archivePrefix = {arXiv},
       eprint = {0810.1291},
 primaryClass = {astro-ph},
       adsurl = {https://ui.adsabs.harvard.edu/abs/2009MNRAS.393..329N}
}

@ARTICLE{Coccato09,
       author = {{Coccato}, L. and {Gerhard}, O. and {Arnaboldi}, M. and {Das}, P. and {Douglas}, N.~G. and {Kuijken}, K. and {Merrifield}, M.~R. and {Napolitano}, N.~R. and {Noordermeer}, E. and {Romanowsky}, A.~J. and {Capaccioli}, M. and {Cortesi}, A. and {De Lorenzi}, F. and {Freeman}, K.~C.},
        title = "{Kinematic properties of early-type galaxy haloes using planetary nebulae*}",
      journal = {\mnras},
         year = 2009,
        month = apr,
       volume = {394},
       number = {3},
        pages = {1249-1283},
          doi = {10.1111/j.1365-2966.2009.14417.x},
archivePrefix = {arXiv},
       eprint = {0811.3203},
 primaryClass = {astro-ph},
       adsurl = {https://ui.adsabs.harvard.edu/abs/2009MNRAS.394.1249C}
}

@ARTICLE{Pulsoni_ePNS,
       author = {{Pulsoni}, C. and {Gerhard}, O. and {Arnaboldi}, M. and {Coccato}, L. and {Longobardi}, A. and {Napolitano}, N.~R. and {Moylan}, E. and {Narayan}, C. and {Gupta}, V. and {Burkert}, A. and {Capaccioli}, M. and {Chies-Santos}, A.~L. and {Cortesi}, A. and {Freeman}, K.~C. and {Kuijken}, K. and {Merrifield}, M.~R. and {Romanowsky}, A.~J. and {Tortora}, C.},
        title = "{The extended Planetary Nebula Spectrograph (ePN.S) early-type galaxy survey: The kinematic diversity of stellar halos and the relation between halo transition scale and stellar mass}",
      journal = {\aap},
         year = 2018,
        month = oct,
       volume = {618},
          eid = {A94},
        pages = {A94},
          doi = {10.1051/0004-6361/201732473},
archivePrefix = {arXiv},
       eprint = {1712.05833},
 primaryClass = {astro-ph.GA},
       adsurl = {https://ui.adsabs.harvard.edu/abs/2018A&A...618A..94P}
}

@ARTICLE{Napolitano2011,
       author = {{Napolitano}, N.~R. and {Romanowsky}, A.~J. and {Capaccioli}, M. and {Douglas}, N.~G. and {Arnaboldi}, M. and {Coccato}, L. and {Gerhard}, O. and {Kuijken}, K. and {Merrifield}, M.~R. and {Bamford}, S.~P. and {Cortesi}, A. and {Das}, P. and {Freeman}, K.~C.},
        title = "{The PN.S Elliptical Galaxy Survey: a standard {\ensuremath{\Lambda}}CDM halo around NGC 4374?}",
      journal = {\mnras},
         year = 2011,
        month = mar,
       volume = {411},
       number = {3},
        pages = {2035-2053},
          doi = {10.1111/j.1365-2966.2010.17833.x},
archivePrefix = {arXiv},
       eprint = {1010.1533},
 primaryClass = {astro-ph.CO},
       adsurl = {https://ui.adsabs.harvard.edu/abs/2011MNRAS.411.2035N}
}

@ARTICLE{C06-cappellari2006,
       author = {{Cappellari}, Michele and {Bacon}, R. and {Bureau}, M. and {Damen}, M.~C. and {Davies}, Roger L. and {de Zeeuw}, P.~T. and {Emsellem}, Eric and {Falc{\'o}n-Barroso}, Jes{\'u}s and {Krajnovi{\'c}}, Davor and {Kuntschner}, Harald and {McDermid}, Richard M. and {Peletier}, Reynier F. and {Sarzi}, Marc and {van den Bosch}, Remco C.~E. and {van de Ven}, Glenn},
        title = "{The SAURON project - IV. The mass-to-light ratio, the virial mass estimator and the Fundamental Plane of elliptical and lenticular galaxies}",
      journal = {\mnras},
         year = 2006,
        month = mar,
       volume = {366},
       number = {4},
        pages = {1126-1150},
          doi = {10.1111/j.1365-2966.2005.09981.x},
archivePrefix = {arXiv},
       eprint = {astro-ph/0505042},
 primaryClass = {astro-ph},
       adsurl = {https://ui.adsabs.harvard.edu/abs/2006MNRAS.366.1126C}
}

@ARTICLE{2003MNRAS.343..978Shen03,
       author = {{Shen}, Shiyin and {Mo}, H.~J. and {White}, Simon D.~M. and {Blanton}, Michael R. and {Kauffmann}, Guinevere and {Voges}, Wolfgang and {Brinkmann}, J. and {Csabai}, Istvan},
        title = "{The size distribution of galaxies in the Sloan Digital Sky Survey}",
      journal = {\mnras},
         year = 2003,
        month = aug,
       volume = {343},
       number = {3},
        pages = {978-994},
          doi = {10.1046/j.1365-8711.2003.06740.x},
archivePrefix = {arXiv},
       eprint = {astro-ph/0301527},
 primaryClass = {astro-ph},
       adsurl = {https://ui.adsabs.harvard.edu/abs/2003MNRAS.343..978S}
}

@ARTICLE{Li+20_KiDS,
       author = {{Li}, R. and {Napolitano}, N.~R. and {Tortora}, C. and {Spiniello}, C. and {Koopmans}, L.~V.~E. and {Huang}, Z. and {Roy}, N. and {Vernardos}, G. and {Chatterjee}, S. and {Giblin}, B. and {Getman}, F. and {Radovich}, M. and {Covone}, G. and {Kuijken}, K.},
        title = "{New High-quality Strong Lens Candidates with Deep Learning in the Kilo-Degree Survey}",
      journal = {\apj},
         year = 2020,
        month = aug,
       volume = {899},
       number = {1},
          eid = {30},
        pages = {30},
          doi = {10.3847/1538-4357/ab9dfa},
archivePrefix = {arXiv},
       eprint = {2004.02715},
 primaryClass = {astro-ph.GA},
       adsurl = {https://ui.adsabs.harvard.edu/abs/2020ApJ...899...30L}
}

@article{BASS-identification-10.1093/mnras/stab1650,
    author = {Li, Changhua and Zhang, Yanxia and Cui, Chenzhou and Fan, Dongwei and Zhao, Yongheng and Wu, Xue-Bing and He, Boliang and Xu, Yunfei and Li, Shanshan and Han, Jun and Tao, Yihan and Mi, Linying and Yang, Hanxi and Yang, Sisi},
    title = "{Identification of BASS DR3 sources as stars, galaxies, and quasars by XGBoost}",
    journal = {\mnras},
    volume = {506},
    number = {2},
    pages = {1651-1664},
    year = {2021},
    month = {07},
    issn = {0035-8711},
    doi = {10.1093/mnras/stab1650},
    url = {https://doi.org/10.1093/mnras/stab1650},
    eprint = {https://academic.oup.com/mnras/article-pdf/506/2/1651/39200449/stab1650.pdf},
}

@ARTICLE{Li-CNN-gal-structure-params,
       author = {{Li}, R. and {Napolitano}, N.~R. and {Roy}, N. and {Tortora}, C. and {La Barbera}, F. and {Sonnenfeld}, A. and {Qiu}, C. and {Liu}, S.},
        title = "{Galaxy Light Profile Convolutional Neural Networks (GaLNets). I. Fast and Accurate Structural Parameters for Billion-galaxy Samples}",
      journal = {\apj},
         year = 2022,
        month = apr,
       volume = {929},
       number = {2},
          eid = {152},
        pages = {152},
          doi = {10.3847/1538-4357/ac5ea0},
archivePrefix = {arXiv},
       eprint = {2111.05434},
 primaryClass = {astro-ph.GA},
       adsurl = {https://ui.adsabs.harvard.edu/abs/2022ApJ...929..152L}
}

@ARTICLE{2022subhalo-galaxy-MLP,
       author = {{Shao}, Helen and {Villaescusa-Navarro}, Francisco and {Genel}, Shy and {Spergel}, David N. and {Angl{\'e}s-Alc{\'a}zar}, Daniel and {Hernquist}, Lars and {Dav{\'e}}, Romeel and {Narayanan}, Desika and {Contardo}, Gabriella and {Vogelsberger}, Mark},
        title = "{Finding Universal Relations in Subhalo Properties with Artificial Intelligence}",
      journal = {\apj},
         year = 2022,
        month = mar,
       volume = {927},
       number = {1},
          eid = {85},
        pages = {85},
          doi = {10.3847/1538-4357/ac4d30},
archivePrefix = {arXiv},
       eprint = {2109.04484},
 primaryClass = {astro-ph.CO},
       adsurl = {https://ui.adsabs.harvard.edu/abs/2022ApJ...927...85S}
}

@ARTICLE{2021halo-galaxy-GNN,
       author = {{Villanueva-Domingo}, Pablo and {Villaescusa-Navarro}, Francisco and {Angl{\'e}s-Alc{\'a}zar}, Daniel and {Genel}, Shy and {Marinacci}, Federico and {Spergel}, David N. and {Hernquist}, Lars and {Vogelsberger}, Mark and {Dave}, Romeel and {Narayanan}, Desika},
        title = "{Inferring Halo Masses with Graph Neural Networks}",
      journal = {\apj},
         year = 2022,
        month = aug,
       volume = {935},
       number = {1},
          eid = {30},
        pages = {30},
          doi = {10.3847/1538-4357/ac7aa3},
archivePrefix = {arXiv},
       eprint = {2111.08683},
 primaryClass = {astro-ph.CO},
       adsurl = {https://ui.adsabs.harvard.edu/abs/2022ApJ...935...30V}
}

@ARTICLE{LCDM-cit_planck,
       author = {{Planck Collaboration} and {Ade}, P.~A.~R. and {Aghanim}, N. and {Arnaud}, M. and {Ashdown}, M. and {Aumont}, J. and {Baccigalupi}, C. and {Banday}, A.~J. and {Barreiro}, R.~B. and {Bartlett}, J.~G. and {Bartolo}, N. and {Battaner}, E. and {Battye}, R. and {Benabed}, K. and {Beno{\^\i}t}, A. and {Benoit-L{\'e}vy}, A. and {Bernard}, J. -P. and {Bersanelli}, M. and {Bielewicz}, P. and {Bock}, J.~J. and {Bonaldi}, A. and {Bonavera}, L. and {Bond}, J.~R. and {Borrill}, J. and {Bouchet}, F.~R. and {Boulanger}, F. and {Bucher}, M. and {Burigana}, C. and {Butler}, R.~C. and {Calabrese}, E. and {Cardoso}, J. -F. and {Catalano}, A. and {Challinor}, A. and {Chamballu}, A. and {Chary}, R. -R. and {Chiang}, H.~C. and {Chluba}, J. and {Christensen}, P.~R. and {Church}, S. and {Clements}, D.~L. and {Colombi}, S. and {Colombo}, L.~P.~L. and {Combet}, C. and {Coulais}, A. and {Crill}, B.~P. and {Curto}, A. and {Cuttaia}, F. and {Danese}, L. and {Davies}, R.~D. and {Davis}, R.~J. and {de Bernardis}, P. and {de Rosa}, A. and {de Zotti}, G. and {Delabrouille}, J. and {D{\'e}sert}, F. -X. and {Di Valentino}, E. and {Dickinson}, C. and {Diego}, J.~M. and {Dolag}, K. and {Dole}, H. and {Donzelli}, S. and {Dor{\'e}}, O. and {Douspis}, M. and {Ducout}, A. and {Dunkley}, J. and {Dupac}, X. and {Efstathiou}, G. and {Elsner}, F. and {En{\ss}lin}, T.~A. and {Eriksen}, H.~K. and {Farhang}, M. and {Fergusson}, J. and {Finelli}, F. and {Forni}, O. and {Frailis}, M. and {Fraisse}, A.~A. and {Franceschi}, E. and {Frejsel}, A. and {Galeotta}, S. and {Galli}, S. and {Ganga}, K. and {Gauthier}, C. and {Gerbino}, M. and {Ghosh}, T. and {Giard}, M. and {Giraud-H{\'e}raud}, Y. and {Giusarma}, E. and {Gjerl{\o}w}, E. and {Gonz{\'a}lez-Nuevo}, J. and {G{\'o}rski}, K.~M. and {Gratton}, S. and {Gregorio}, A. and {Gruppuso}, A. and {Gudmundsson}, J.~E. and {Hamann}, J. and {Hansen}, F.~K. and {Hanson}, D. and {Harrison}, D.~L. and {Helou}, G. and {Henrot-Versill{\'e}}, S. and {Hern{\'a}ndez-Monteagudo}, C. and {Herranz}, D. and {Hildebrandt}, S.~R. and {Hivon}, E. and {Hobson}, M. and {Holmes}, W.~A. and {Hornstrup}, A. and {Hovest}, W. and {Huang}, Z. and {Huffenberger}, K.~M. and {Hurier}, G. and {Jaffe}, A.~H. and {Jaffe}, T.~R. and {Jones}, W.~C. and {Juvela}, M. and {Keih{\"a}nen}, E. and {Keskitalo}, R. and {Kisner}, T.~S. and {Kneissl}, R. and {Knoche}, J. and {Knox}, L. and {Kunz}, M. and {Kurki-Suonio}, H. and {Lagache}, G. and {L{\"a}hteenm{\"a}ki}, A. and {Lamarre}, J. -M. and {Lasenby}, A. and {Lattanzi}, M. and {Lawrence}, C.~R. and {Leahy}, J.~P. and {Leonardi}, R. and {Lesgourgues}, J. and {Levrier}, F. and {Lewis}, A. and {Liguori}, M. and {Lilje}, P.~B. and {Linden-V{\o}rnle}, M. and {L{\'o}pez-Caniego}, M. and {Lubin}, P.~M. and {Mac{\'\i}as-P{\'e}rez}, J.~F. and {Maggio}, G. and {Maino}, D. and {Mandolesi}, N. and {Mangilli}, A. and {Marchini}, A. and {Maris}, M. and {Martin}, P.~G. and {Martinelli}, M. and {Mart{\'\i}nez-Gonz{\'a}lez}, E. and {Masi}, S. and {Matarrese}, S. and {McGehee}, P. and {Meinhold}, P.~R. and {Melchiorri}, A. and {Melin}, J. -B. and {Mendes}, L. and {Mennella}, A. and {Migliaccio}, M. and {Millea}, M. and {Mitra}, S. and {Miville-Desch{\^e}nes}, M. -A. and {Moneti}, A. and {Montier}, L. and {Morgante}, G. and {Mortlock}, D. and {Moss}, A. and {Munshi}, D. and {Murphy}, J.~A. and {Naselsky}, P. and {Nati}, F. and {Natoli}, P. and {Netterfield}, C.~B. and {N{\o}rgaard-Nielsen}, H.~U. and {Noviello}, F. and {Novikov}, D. and {Novikov}, I. and {Oxborrow}, C.~A. and {Paci}, F. and {Pagano}, L. and {Pajot}, F. and {Paladini}, R. and {Paoletti}, D. and {Partridge}, B. and {Pasian}, F. and {Patanchon}, G. and {Pearson}, T.~J. and {Perdereau}, O. and {Perotto}, L. and {Perrotta}, F. and {Pettorino}, V. and {Piacentini}, F. and {Piat}, M. and {Pierpaoli}, E. and {Pietrobon}, D. and {Plaszczynski}, S. and {Pointecouteau}, E. and {Polenta}, G. and {Popa}, L. and {Pratt}, G.~W. and {Pr{\'e}zeau}, G. and {Prunet}, S. and {Puget}, J. -L. and {Rachen}, J.~P. and {Reach}, W.~T. and {Rebolo}, R. and {Reinecke}, M. and {Remazeilles}, M. and {Renault}, C. and {Renzi}, A. and {Ristorcelli}, I. and {Rocha}, G. and {Rosset}, C. and {Rossetti}, M. and {Roudier}, G. and {Rouill{\'e} d'Orfeuil}, B. and {Rowan-Robinson}, M. and {Rubi{\~n}o-Mart{\'\i}n}, J.~A. and {Rusholme}, B. and {Said}, N. and {Salvatelli}, V. and {Salvati}, L. and {Sandri}, M. and {Santos}, D. and {Savelainen}, M. and {Savini}, G. and {Scott}, D. and {Seiffert}, M.~D. and {Serra}, P. and {Shellard}, E.~P.~S. and {Spencer}, L.~D. and {Spinelli}, M. and {Stolyarov}, V. and {Stompor}, R. and {Sudiwala}, R. and {Sunyaev}, R. and {Sutton}, D. and {Suur-Uski}, A. -S. and {Sygnet}, J. -F. and {Tauber}, J.~A. and {Terenzi}, L. and {Toffolatti}, L. and {Tomasi}, M. and {Tristram}, M. and {Trombetti}, T. and {Tucci}, M. and {Tuovinen}, J. and {T{\"u}rler}, M. and {Umana}, G. and {Valenziano}, L. and {Valiviita}, J. and {Van Tent}, F. and {Vielva}, P. and {Villa}, F. and {Wade}, L.~A. and {Wandelt}, B.~D. and {Wehus}, I.~K. and {White}, M. and {White}, S.~D.~M. and {Wilkinson}, A. and {Yvon}, D. and {Zacchei}, A. and {Zonca}, A.},
        title = "{Planck 2015 results. XIII. Cosmological parameters}",
      journal = {\aap},
         year = 2016,
        month = sep,
       volume = {594},
          eid = {A13},
        pages = {A13},
          doi = {10.1051/0004-6361/201525830},
archivePrefix = {arXiv},
       eprint = {1502.01589},
 primaryClass = {astro-ph.CO},
       adsurl = {https://ui.adsabs.harvard.edu/abs/2016A&A...594A..13P}
}

@article{TNG-data-nelson2019illustristng,
  title={The IllustrisTNG simulations: public data release},
  author={Nelson, Dylan and Springel, Volker and Pillepich, Annalisa and Rodriguez-Gomez, Vicente and Torrey, Paul and Genel, Shy and Vogelsberger, Mark and Pakmor, Ruediger and Marinacci, Federico and Weinberger, Rainer and others},
  journal={Computational Astrophysics and Cosmology},
  volume={6},
  number={1},
  pages={1--29},
  year={2019},
  publisher={Springer}
}

@article{tortora2012spider,
  title={SPIDER—VI. The central dark matter content of luminous early-type galaxies: Benchmark correlations with mass, structural parameters and environment},
  author={Tortora, C and La Barbera, F and Napolitano, NR and de Carvalho, RR and Romanowsky, Aaron J},
  journal={\mnras},
  volume={425},
  number={1},
  pages={577--594},
  year={2012},
  publisher={Blackwell Science Ltd Oxford, UK}
}

@ARTICLE{2019MNRAS.490.5390Bottrell,
       author = {{Bottrell}, Connor and {Hani}, Maan H. and {Teimoorinia}, Hossen and {Ellison}, Sara L. and {Moreno}, Jorge and {Torrey}, Paul and {Hayward}, Christopher C. and {Thorp}, Mallory and {Simard}, Luc and {Hernquist}, Lars},
        title = "{Deep learning predictions of galaxy merger stage and the importance of observational realism}",
      journal = {\mnras},
         year = 2019,
        month = dec,
       volume = {490},
       number = {4},
        pages = {5390-5413},
          doi = {10.1093/mnras/stz2934},
archivePrefix = {arXiv},
       eprint = {1910.07031},
 primaryClass = {astro-ph.GA},
       adsurl = {https://ui.adsabs.harvard.edu/abs/2019MNRAS.490.5390B}
}

@ARTICLE{2021MNRAS.508.3321Tang,
       author = {{Tang}, Lin and {Lin}, Weipeng and {Wang}, Yang and {Napolitano}, N.~R.},
        title = "{The importance of mock observations in validating galaxy properties for cosmological simulations}",
      journal = {\mnras},
         year = 2021,
        month = dec,
       volume = {508},
       number = {3},
        pages = {3321-3336},
          doi = {10.1093/mnras/stab2722},
archivePrefix = {arXiv},
       eprint = {2109.08838},
 primaryClass = {astro-ph.CO},
       adsurl = {https://ui.adsabs.harvard.edu/abs/2021MNRAS.508.3321T}
}

@article{von2022inferring,
  title={Inferring galaxy dark halo properties from visible matter with machine learning},
  author={von Marttens, Rodrigo and Casarini, Luciano and Napolitano, Nicola R and Wu, Sirui and Amaro, Valeria and Li, Rui and Tortora, Crescenzo and Canabarro, Askery and Wang, Yang},
  journal={\mnras},
  volume={516},
  number={3},
  pages={3924--3943},
  year={2022},
  publisher={Oxford University Press}
}

@ARTICLE{2021JCAP...08..062N,
       author = {{Newton}, Oliver and {Leo}, Matteo and {Cautun}, Marius and {Jenkins}, Adrian and {Frenk}, Carlos S. and {Lovell}, Mark R. and {Helly}, John C. and {Benson}, Andrew J. and {Cole}, Shaun},
        title = "{Constraints on the properties of warm dark matter using the satellite galaxies of the Milky Way}",
      journal = {\jcap},
         year = 2021,
        month = aug,
       volume = {2021},
       number = {8},
          eid = {062},
        pages = {062},
          doi = {10.1088/1475-7516/2021/08/062},
archivePrefix = {arXiv},
       eprint = {2011.08865},
 primaryClass = {astro-ph.CO},
       adsurl = {https://ui.adsabs.harvard.edu/abs/2021JCAP...08..062N}
}

@ARTICLE{2020JCAP...02..024B,
       author = {{Banerjee}, Arka and {Adhikari}, Susmita and {Dalal}, Neal and {More}, Surhud and {Kravtsov}, Andrey},
        title = "{Signatures of self-interacting dark matter on cluster density profile and subhalo distributions}",
      journal = {\jcap},
         year = 2020,
        month = feb,
       volume = {2020},
       number = {2},
          eid = {024},
        pages = {024},
          doi = {10.1088/1475-7516/2020/02/024},
archivePrefix = {arXiv},
       eprint = {1906.12026},
 primaryClass = {astro-ph.CO},
       adsurl = {https://ui.adsabs.harvard.edu/abs/2020JCAP...02..024B}
}

\begin{appendix}
\section{Dark Matter predictions of the dynamical sample}
\label{app:mass_DM}
In this Appendix, we show the dark matter predictions of the real dynamical sample, both the ``augmented'' one, $\tilde{M}_{\rm DM}(\re)$ defined in \S\ref{sec:preamble}, and the standard one, \mdre. As discussed in \S\ref{sec:preamble}, the former corresponds to the standard definition in observational data, when there is no explicit estimate of the internal the gas content of galaxies. This is expected to produce a minimal impact on the ETG estimates, but it can give a larger bias in the LTGs. For this same reason, since we miss the information on the hidden baryons of the real galaxies, we cannot test whether the true DM mass inside the \re\ predicted by \mla\ is consistent with the dynamical inferences. Hence, as introduced in \S\ref{sec:pred_real}, we will argue that if the augmented mass is correctly predicted, the pure DM content of the galaxy is correctly predicted too, despite the fact that this has to be feedback-dependent. In fact, as extensively discussed across this paper, both stars and gas content of galaxies depend on the baryonic physics adopted in TNG100.

\begin{figure*}[ht]
\centering
\includegraphics[width=1.8\columnwidth]{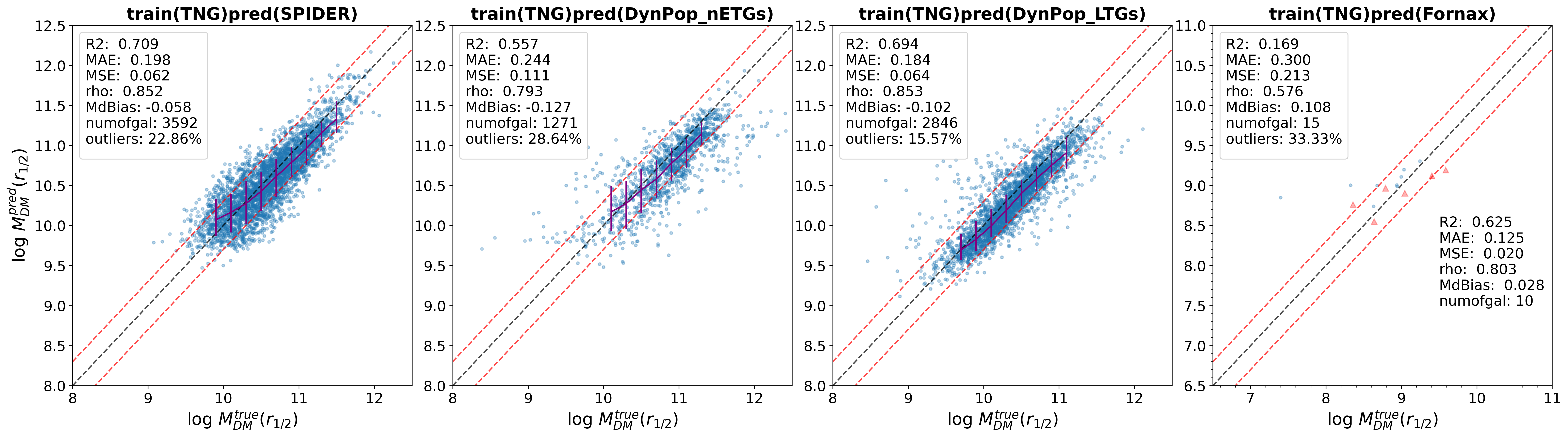}
\includegraphics[width=1.8\columnwidth]{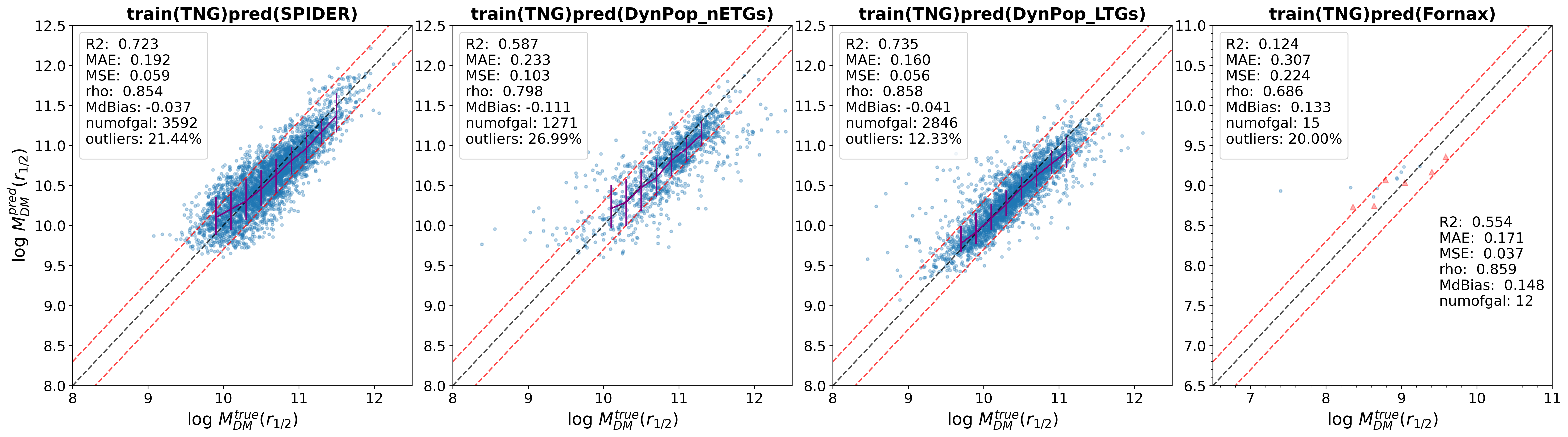}
\caption{\mlaall\ predictions of the central dark matter (\mdre, top row) and augmented DM ($\tilde{M}_{\rm DM}(\re$), bottom row) for the real galaxy dynamical samples. The different definition of the target is discussed in \S\ref{sec:preamble}. The optimal feature combination was used (i.e., \re, \ms\ and \sig) as indicated in \S\ref{sec:self_pred} and Table~\ref{tab: DM-number of features and accuracy}. As for Fig.~\ref{fig: train-simu-pred-real_1}, the dynamical model used in MaNGA Dynpop is $\rm JAM_{sph}$+generalized NFW profile, as mentioned in \S\ref{sec:dynpop}. For the DSAMI sample, the red triangle represents data point from the secondary test sample (1kpc<\re\ <2$R_{\rm p}$). The legend shows the statistical estimators. The bottom right corner inset of the DSAMI sample gives the estimators after  excluding the outliers, as previously done in Fig.~\ref{fig: train-simu-pred-real_ReMRe}.}
\label{fig: train-simu-pred-real-DM}
\end{figure*}

In Fig.~\ref{fig: train-simu-pred-real-DM} we show the $\tilde{M}_{\rm DM}(\re)$ and the \mdre\ predictions, using the minimal feature set as reference result and the \mlaall\ configuration. We compare them with the only DM estimate we have from observations in both cases, although we expect the \mdre\ should miss the baryon part and show a negative offset. The first thing to note is that all the statistical estimators are degraded with respect to the analogous predictions of the total mass as in Fig.~\ref{fig: train-simu-pred-real_1}. This has, again, to do with the larger scatter of the scaling relations involving \mdre\ discussed in Appendix~\ref{app:errors}. However, we see a better accuracy of the $\tilde{M}_{\rm DM}(\re)$ predictions with respect to the \mdre, as measured by the larger $R^2$ and the smaller MAE and MSE, although the improvements in $R^2$ are never larger than 10\%, while the $\rho$ is almost unchanged in all cases. This latter effect is due to the small offset of the predicted \mdre\ with respect to the 1-to-1 relation, due to the missing baryons mentioned above and slightly visible in the plot, especially for the DynPop/LTG sample, which, by definition, is gas richer. We stress though that for the DynPop/ETG sample the $R^2$ is particularly poor, due to the combination of a higher outlier fraction and, also a larger offset, that seems not to be resolved even in the  $\tilde{M}_{\rm DM}(\re)$ prediction. In fact, the residual offset has to be tracked back to the shift of the scaling relations discussed for the \mtotre\ in \S\ref{sec:systematics}, which is rather understood. We conclude this appendix with the claim that the accuracy of the augmented dark mass obtained for \mla\ predictions is in line with the one found for the total mass, \mtotre, in \S\ref{sec:pred_real}, except that the statistical estimators are poorer due to the larger scatter in the estimates, coming from the noisier DM scaling relations. As postulated above, we also claim that we provide the estimate of the ``true'' DM content of galaxies in the TNG100 cosmology+feedback framework  (i.e., deprived of the gas content that is hidden in the dynamical inferences). These parameters will also be part of the catalog of \mla\ predictions which we make public (see \S\ref{sec:conclusions}).

\section{Self-predictions with no measurement errors}
\label{app:errors}
As mentioned in \S\ref{sec:self_pred} we present here the results of the self-predictions excluding the measurement errors (i.e., taking the catalogs of the features and targets of the TNG100 at face value). This is a standard approach when comparing simulation and observations, but it might be eventually too idealized if one wants to use simulations to obtain realistic forecasts for real applications. In order to check the net impact of this idealized approach and to directly compare our results with similar literature analyses (e.g., \citealt{2022subhalo-galaxy-MLP}) and our previous results, in vM+22. The accuracy of both the \mdre\ and \mtotre\ are shown in Tables~\ref{tab:feature_importance_app_1} and~\ref{tab:feature_importance_app_2}, in a similar way to what is shown in Tables~\ref{tab: DM-number of features and accuracy} and~\ref{tab: DYN-number of features and accuracy}, for the case with measurement errors included. From the comparison of these pairs of Tables, we see the same features noticed for the case of errors included, in particular, the growth of accuracy with the increasing number of features and also the same order in the ``feature importance''. Also, the inclusions of the errors do not impact the number of features needed to reach the ``plateau'', with \re, \ms, and \sig\ remaining the most important features. The only difference we see is that all accuracies increase at every step, meaning that the only impact of the errors, as expected, is to increase the scatter and then the overall accuracy of the predictions. With no errors, \mla\ reaches up to 98\% accuracy for \mdre\ and up to 99\% for \mtotre.
In particular, we can compare the new results with the one of vM+22, by looking at the TNG\_all sample predictions for the DM mass in the \re. The closer experiments to compare are then the ones considering all the features in our self-predictions (last line in both Tables~\ref{tab:feature_importance_app_1} and~\ref{tab:feature_importance_app_2}), showing $R^2$ values of 0.98 and the result from vM+22 also reporting a $R^2\sim$ 0.981 (see their Table~1). 
In Fig.~\ref{fig:self-prediction_no_error} we also show the one-to-one plot of the same targets for the case of \mlaall\ configuration using the full-count training sample, which is again the closest experiment to vM+22 (e.g., their Fig.~4). Compared to our Fig.~\ref{fig: self-prediction-simu-mela_all}, we can visually see that the net effect of the absence of measurement errors is the extremely tight correlation with a much smaller scatter and a tiny outlier fraction. This turns out to be consistent with what we have previously found in vM+22, where the DM mass inside \re. We can also see that the scatter of the total mass is systematically smaller than the one of the DM. This is also seen in Fig.~\ref{fig: self-prediction-simu-mela_all}, although it does not show up as evident as it is in this no-error case. Overall, this reveals that the total mass is slightly better predicted by \mla\ than the dark mass. The reason for that can be the smaller scatter in the scaling relations shown by the \mtotre\ with respect to the \mdre, for example, \mtotre--\msre\ or \mtotre--\sig\ versus the same relations of the \mdre\ in Fig.~\ref{fig:correlations}.  

\begin{table}
\centering
\small
\begin{tabular}{clc}
\toprule
$R^2$ & features & num \\ 
\hline
\multicolumn{3}{c}{nETGs}\\
\hline
0.9244 & \re & 1 \\ 
0.9837 & \re, $M_\star$ & 2 \\ 
0.9886 & \re, $M_\star$, $\sigma$ & 3 \\ 
0.9901 & r, \re, $M_\star$, $\sigma$ & 4 \\ 
0.9900 & g, r, \re, $M_\star$, $\sigma$ & 5 \\ 
\hline
\multicolumn{3}{c}{dETGs}\\
\hline
0.2207 & $\sigma$ & 1 \\ 
0.8882 & \re, $\sigma$ & 2 \\ 
0.9249 & \re, $M_\star$, $\sigma$ & 3 \\ 
0.9463 & r, \re, $M_\star$, $\sigma$ & 4 \\ 
0.9461 & g, r, \re, $M_\star$, $\sigma$ & 5 \\ 
\hline
\multicolumn{3}{c}{LTGs}\\
\hline
0.6514 & \re & 1 \\ 
0.9596 & \re, $M_\star$ & 2 \\ 
0.9707 & \re, $M_\star$, $\sigma$ & 3 \\ 
0.9753 & r, \re, $M_\star$, $\sigma$ & 4 \\ 
0.9757 & g, r, \re, $M_\star$, $\sigma$ & 5 \\ 
\hline
\multicolumn{3}{c}{TNG\_all}\\
\hline
0.6400 & $M_\star$ & 1 \\ 
0.9580 & \re, $\sigma$ & 2 \\ 
0.9755 & \re, $M_\star$, $\sigma$ & 3 \\ 
0.9810 & r, \re, $M_\star$, $\sigma$ & 4 \\ 
0.9813 & g, r, \re, $M_\star$, $\sigma$ & 5 \\
\bottomrule
\end{tabular}
\caption{Accuracy as a function of the number of features for the \mdre, excluding measurement errors. Each row displays the optimal feature combination for the corresponding feature count. The result considering measurement errors can be found in Table~\ref{tab: DM-number of features and accuracy}.}
\label{tab:feature_importance_app_1}
\end{table}

\begin{table}
\centering
\small
\begin{tabular}{clc}
\toprule
$R^2$ & features & num \\ 
\hline
\multicolumn{3}{c}{nETGs}\\
\hline
0.9501 & $M_\star$ & 1 \\ 
0.9959 & \re, $M_\star$ & 2 \\ 
0.9973 & \re, $M_\star$, $\sigma$ & 3 \\ 
0.9976 & g, \re, $M_\star$, $\sigma$ & 4 \\ 
0.9976 & g, r, \re, $M_\star$, $\sigma$ & 5 \\  
\hline
\multicolumn{3}{c}{dETGs}\\
\hline
0.5985 & $\sigma$ & 1 \\ 
0.8923 & \re, $\sigma$ & 2 \\ 
0.9578 & \re, $M_\star$, $\sigma$ & 3 \\ 
0.9701 & r, \re, $M_\star$, $\sigma$ & 4 \\ 
0.9700 & g, r, \re, $M_\star$, $\sigma$ & 5 \\  
\hline
\multicolumn{3}{c}{LTGs}\\
\hline
0.8105 & $M_\star$ & 1 \\ 
0.9799 & \re, $M_\star$ & 2 \\ 
0.9862 & \re, $M_\star$, $\sigma$ & 3 \\ 
0.9877 & r, \re, $M_\star$, $\sigma$ & 4 \\ 
0.9879 & g, r, \re, $M_\star$, $\sigma$ & 5 \\  
\hline
\multicolumn{3}{c}{TNG\_all}\\
\hline
0.8520 & $M_\star$ & 1 \\ 
0.9729 & \re, $M_\star$ & 2 \\ 
0.9895 & \re, $M_\star$, $\sigma$ & 3 \\ 
0.9911 & r, \re, $M_\star$, $\sigma$ & 4 \\ 
0.9912 & g, r, \re, $M_\star$, $\sigma$ & 5 \\ 
\bottomrule
\end{tabular}
\caption{Accuracy as a function of the number of features for the \mtotre, excluding measurement errors. Each row displays the optimal feature combination for the corresponding feature count. The result considering measurement errors can be found in Table~\ref{tab: DYN-number of features and accuracy}.}
\label{tab:feature_importance_app_2}
\end{table}

\section{Different settings of DynPop}
\label{app:dynpop}
\begin{figure*}[ht]
\centering
\includegraphics[width=1.8\columnwidth]{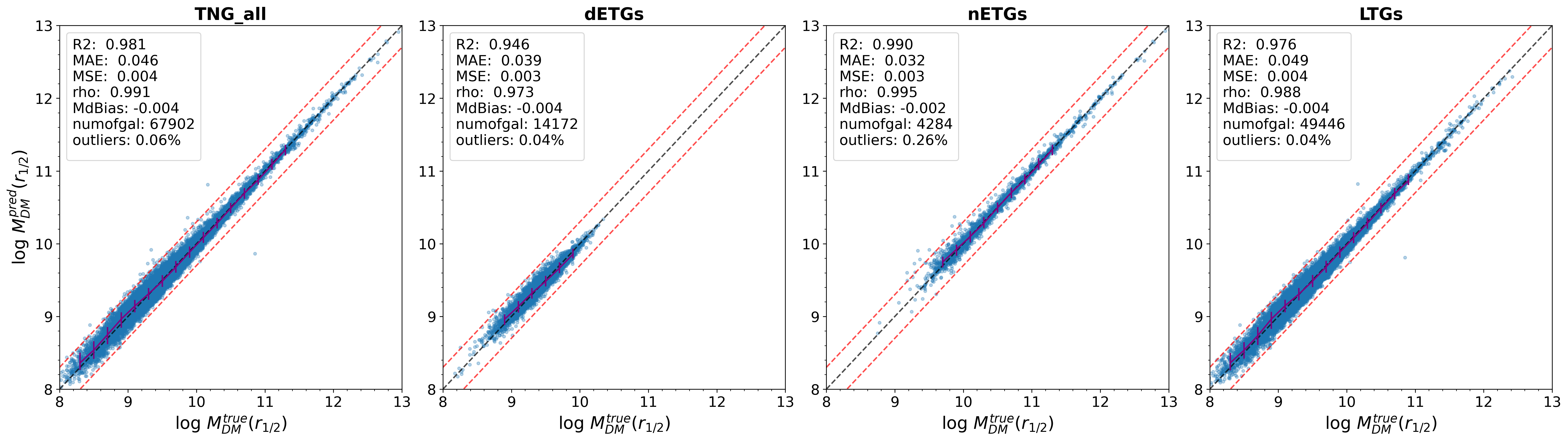}
\includegraphics[width=1.8\columnwidth]{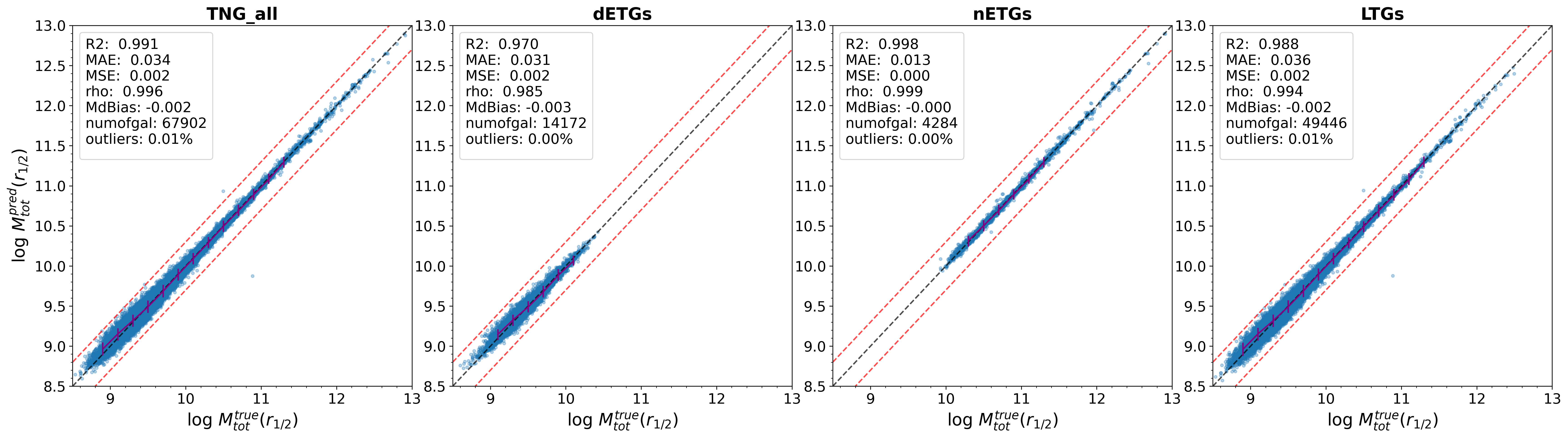}
\caption{Self-prediction test using full features as indicated in Table~\ref{tab: features and targets}, with the full-counts training sample incorporating excluding measurement errors, as described in \S\ref{sec:meas_err}. Top row: Target is \mdre. Bottom row: Target is \mtotre\ . The results including measurement errors are presented in   Fig.~\ref{fig: self-prediction-simu}.}
\label{fig:self-prediction_no_error}
\end{figure*}

\begin{figure}[htp]
\centering
\includegraphics[width=1.03\columnwidth]{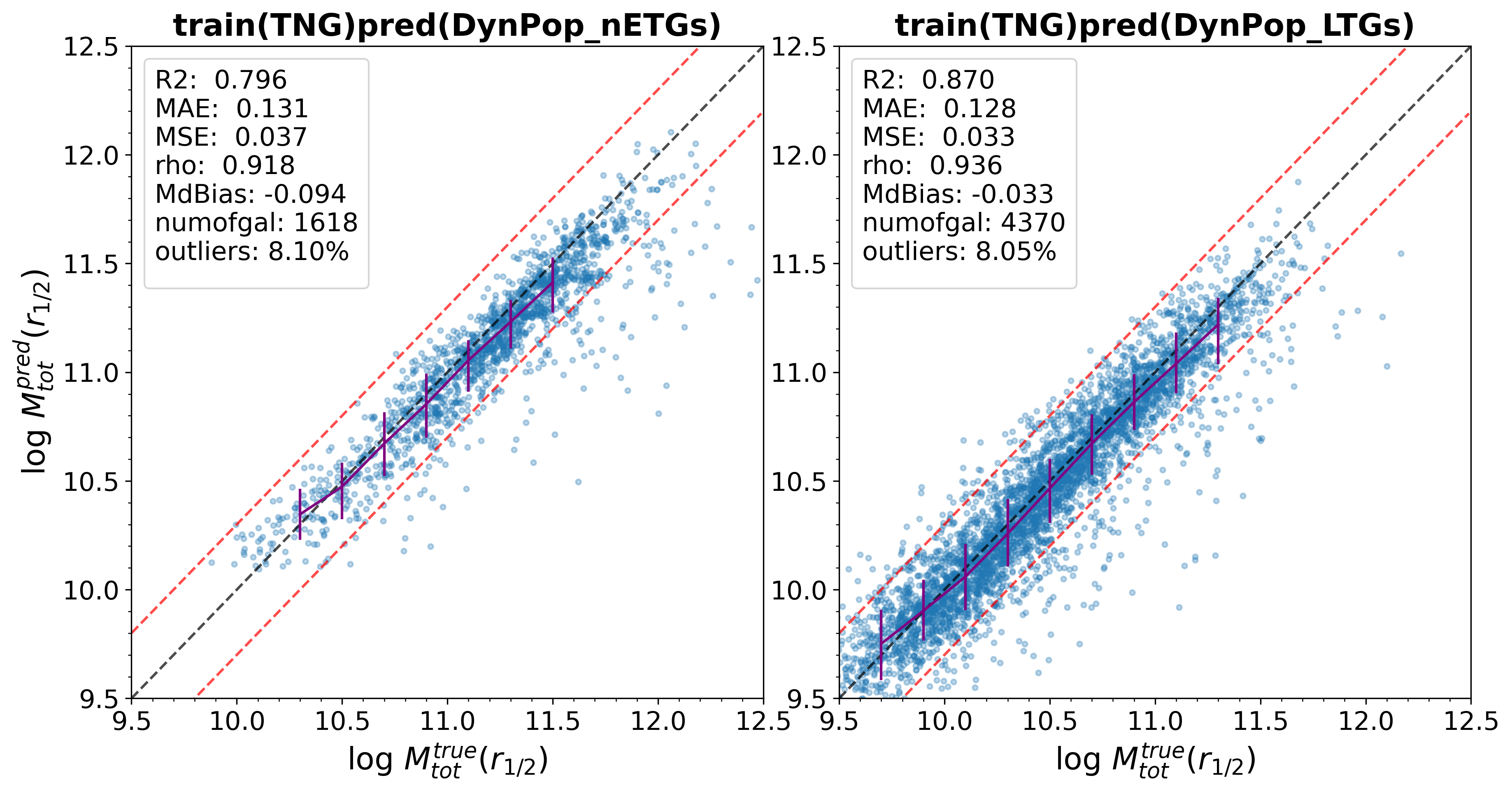}
\caption{\mlaall\ predictions of the central total mass, \mtotre, for the DynPop sample, but using quality flag Qual$\geq 0$. This can be compared to the corresponding result of Qual$\geq 1$ in Fig.~\ref{fig: train-simu-pred-real_1}.}
\label{fig:MaNGA-Qual0}
\end{figure}
In this Appendix, we briefly collect the results of all available settings (i.e., different Qual of galaxies and different assumptions of JAM analysis) of the DynPop dynamical sample. In Fig.~\ref{fig:MaNGA-Qual0} we show the prediction of DynPop sample with Qual $\geq0$ for the reference dynamical analysis, i.e, the one based on the generalized NFW, that can be directly compared with the results based on the Qual$\geq1$ of the same model, as in Fig.~\ref{fig: train-simu-pred-real_1}. As expected, we see a degradation of all the statistical estimators for both nETGs and, especially, LTGs. The indicators having more affected are the ones related to the scatter (MAE and MSE) and the outlier fraction, indeed suggesting a larger scatter of the poorer quality dynamical analysis estimates used as ground-truth. It is interesting that the nETG sample has $R^2$ overall consistent with the one found for the  
Qual$\geq1$ predictions ($R^2$=0.796 vs. 0.790, outliers fraction 8.10\% vs. 7.47\%.). We finally stress that galaxies with Qual=0 are mainly concentrated in the small mass end and mainly in the LTG sample. This selection effect is rather obvious as because of their smaller masses (and luminosities), they are harder to observe and therefore have poorer image quality. 

In Fig.~\ref{fig:diff-MaNGA-result} we finally show the \mlaall\ predictions (see URL in the foornote~\ref{foot:release}) against all other mass estimates from the models as in K+23. The overall impression is that the \mlaall\ predictions are rather robust with respect to the majority of the dynamical models adopted. We stress here that in the different plots, the \mla\ estimates stay the same and only the DynPop, along the x-axis, change. In this respect this figure measures the level of fidelity of the DynPop model with respect to the TNG100 predictions, assuming that \mla\ produces realistic, observationally consistent estimates of the total mass of galaxies. In this perspective, LTGs are also the sample showing minimal differences as a function of the models, while nETGs show quite a large variation with the worse case provided by the sph/fNFW, while the Mass-Follows-Light (MLF) models, seem to be surprisingly consistent with TNG100, performing even better than our reference gNFW models. This might suggest that the DM in nETGs is steeper than typical NFW cusps (e.g., because of adiabatic contraction). This might not be a surprising result, as the dynamical models (consistently with simulations) make use of a Chabrier IMF, and lower normalization IMF has been found to require adiabatic contraction to fit both the central (e.g., \citealt{2010MNRAS.405.2351N}) and extended kinematics (e.g., \citealt{Napolitano2011}) of ETGs. 

\begin{figure*}[htp]
\centering
\includegraphics[width=1.8\columnwidth]{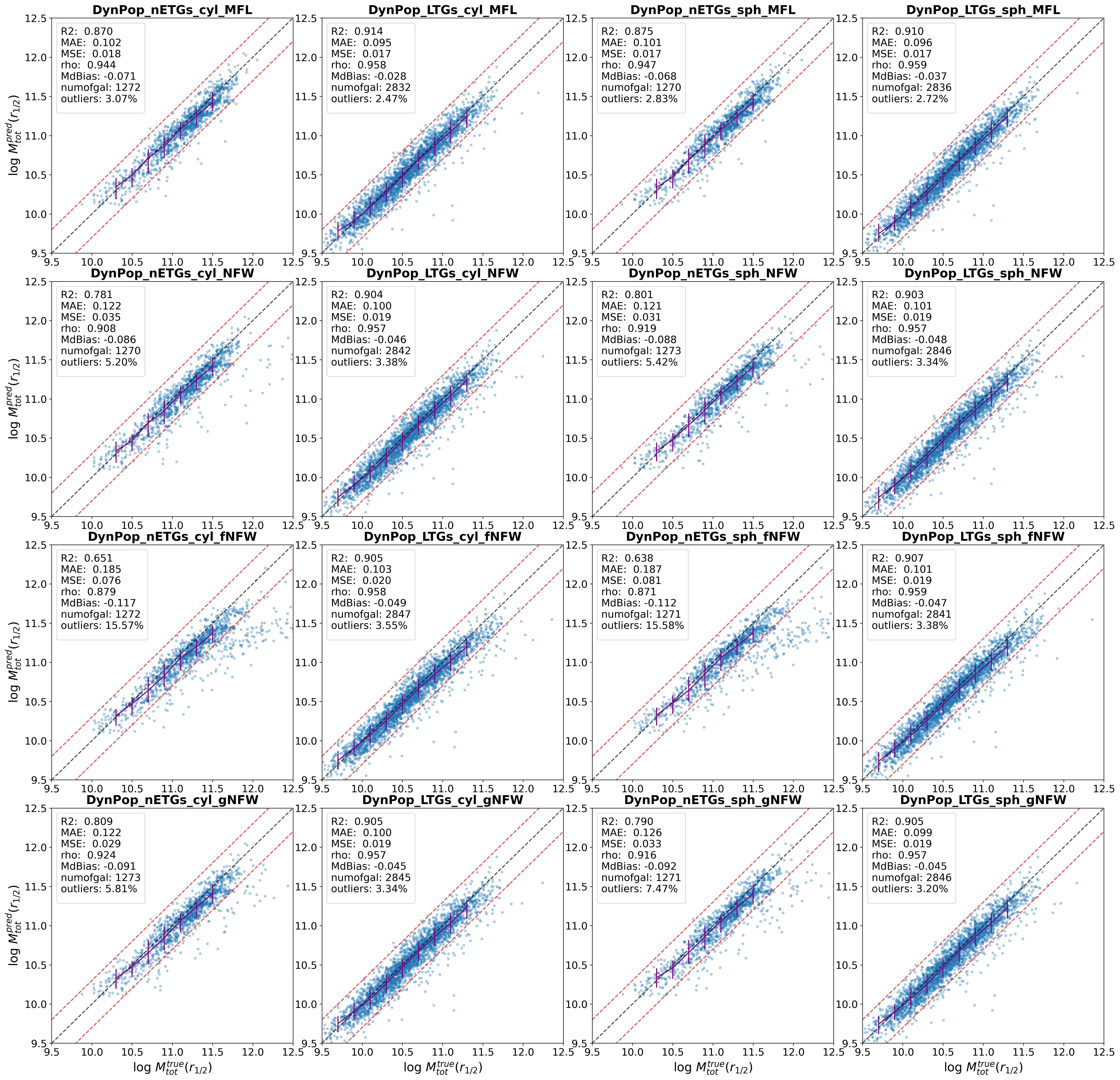}
\caption{\mlaall\ predictions of the central total mass, \mtotre, for all the JAM models from K+23, as mentioned in \S\ref{sec:dynpop}. These scenarios encompass different orientations of the velocity ellipsoid and the assumptions about the dark vs. luminous matter distribution. In detail, they combine the following assumptions:  1) cyl and sph make use of the orientations of the velocity ellipsoid along cylindrical and spherical coordinates, respectively; 2) MFL corresponds to the mass-follows-light assumption; 3) NFW corresponds to the free NFW dark halo assumption; 4) fNFW is the fixed NFW, using the cosmologically constrained NFW halo assumption; 5)  gNFW is the generalized NFW dark halo. Our reference model is sph\_gNFW, which are the last two plots (see also Fig.~\ref{fig: train-simu-pred-real_1}).}
\label{fig:diff-MaNGA-result}
\end{figure*}

\end{appendix}

\end{document}